\documentclass{jfm}
\usepackage{graphicx}
\usepackage{amsmath}
\usepackage{epstopdf, epsfig}
\usepackage{subfig}
\usepackage{soul}
\usepackage{xcolor}

\shorttitle{ML Building-block-flow wall model for LES}
\shortauthor{A. Lozano-Dur\'an and H. J. Bae}

\title{Machine learning building-block-flow wall model for large-eddy simulation}

\author{Adri\'an Lozano-Dur\'an\aff{1}
  \corresp{\email{adrianld@mit.edu}}
 \and H. Jane  Bae\aff{2}}

\affiliation{\aff{1}Department of Aeronautics and Astronautics, Massachusetts Institute of Technology, Cambridge, MA 02139, USA
\aff{2}Graduate Aerospace Laboratories, California Institute of Technology, Pasadena, California 91125, USA}

\begin{document}

\maketitle

\newcommand{\corr}[1]{\textcolor{black}{#1}}

\begin{abstract}
A wall model for large-eddy simulation (LES) is proposed by devising
the flow as a combination of building blocks.  The core assumption of
the model is that a finite set of simple canonical flows contains the
essential physics to predict the wall-shear stress in more complex
scenarios.  The model is constructed to predict
zero/favourable/adverse mean pressure gradient wall turbulence,
separation, statistically unsteady turbulence with mean flow
three-dimensionality, and laminar flow.  The approach is implemented
using two types of artificial neural networks: a classifier, which
identifies the contribution of each building block in the flow, and a
predictor, which estimates the wall-shear stress via combination of
the building-block flows. The training data are directly obtained from
wall-modelled LES (WMLES) optimised to reproduce the correct mean
quantities. This approach guarantees the consistency of the training
data with the numerical discretisation and the gridding strategy of
the flow solver. The output of the model is accompanied by a
confidence score in the prediction that aids the detection of
regions where the model underperforms. The model is validated in canonical flows (e.g. laminar/turbulent
boundary layers, turbulent channels, turbulent Poiseuille--Couette
flow, turbulent pipe) and two realistic aircraft configurations:
the NASA Common Research Model High-lift and NASA Juncture Flow
experiment.  It is shown that the building-block-flow wall model
outperforms (or matches) the predictions by an equilibrium wall
model. It is also concluded that further improvements in WMLES should  
incorporate advances in subgrid-scale modelling to minimise
error propagation to the wall model.
\end{abstract}


\section{Introduction}

The use of computational fluid dynamics (CFD) for external aerodynamic
applications has been a key tool for aircraft design in the modern
aerospace industry~\citep{casey2000}. However, flow predictions from
state-of-the-art solvers are still unable to comply with the stringent
accuracy requirements and computational efficiency demanded by the
industry~\citep{Mauery2021}. In recent years, wall-modelled large-eddy
simulation (WMLES) has gained momentum as a high-fidelity tool for
routine industrial design~\citep{Goc2021}.  In WMLES, only the
large-scale motions in the outer region of the boundary layer are
resolved, which enables a competitive computational cost compared with
other CFD approaches~\citep{Chapman1979, Choi2012, Yang2021}. As such,
NASA has recognised WMLES as an important pacing item for ``developing
a visionary CFD capability required by the notional year
2030''~\citep{Slotnick2014}. In the present work, we introduce a wall
model based on flow-state classification applicable to a wide variety
of flow regimes that also provides a confidence score for the
prediction.

Several strategies for modelling the near-wall region have been
explored in the literature, and comprehensive reviews can be found in
\citet{Cabot2000}, \citet{Piomelli2002}, \citet{Spalart2009},
\citet{Larsson2015} and \citet{Bose2018}. One of the most widely used
approaches for wall modelling is the wall flux approach (or
approximate boundary conditions), where the no-slip and thermal wall
boundary conditions are replaced with shear stress and heat flux
boundary conditions provided by the wall model. This category of wall
models utilises the large-eddy simulation (LES) solution at a given
location in the domain as input and returns the wall fluxes needed by
the solver as boundary conditions.  Examples of the most popular
approaches are those computing the wall shear stress using the
law of the wall \citep{Deardorff1970, Schumann1975, Piomelli1989}, the
full/simplified Reynolds-averaged Navier-Stokes equations
\citep{Balaras1996, Wang2002, Bodart2011, Kawai2013,
  Bermejo-Moreno2014, Park2014, Yang2015}, structural vortex
models~\citep{Chung2009} or dynamic wall models~\citep{Bose2014,
Bae2019}. Despite the progress, recent results from the American 
Institute of Aeronautics and Astronautics Workshop on high-lift 
prediction~\citep{Kiris2022} have evidenced the deficiencies of
state-of-the-art WMLES in realistic aircraft.  Even simulations with
over 350 million degrees of freedom, which are too costly for
routine industrial design cycle, are unable to accurately match the
experimental results~\citep{Rumsey2019, Lozano2022, Goc2022,
Cetin2022}.

The need for improved predictions has incited the adoption of machine
learning (ML) tools to complement and enhance existing turbulence
models.  The reader is referred to the multiple reviews in the
literature for a comprehensive overview on ML for fluid
mechanics~\citep{Duraisamy2019, Brunton2019, Brenner2019, Pandey2020,
  Duraisamy2021, Beck2021, Vinuesa2022, Vinuesa2022b}.  Most models
follow the supervised learning paradigm, i.e., the ML task of learning
a function that maps an input to an output based on known input--output
pairs. The first ML-based models for LES were introduced in the form
of subgrid-scale (SGS) models. Early approaches used artificial neural
networks (ANNs) to emulate and speed up a conventional, but
computationally expensive, SGS model~\citep{Sarghini2003}. More
recently, SGS models have been trained to predict the (so-called)
perfect SGS terms using data from filtered direct numerical simulation
(DNS) \citep{Gamahara2017, Xie2019}.  Other approaches include
deriving SGS terms from optimal estimator theory~\citep{Vollant2017},
deconvolution operators \citep[e.g.][]{Hickel2004, Maulik2017,
  Fukami2019} or optimised SGS tensor accounting for numerical
errors~\citep{Ling2022}.

One of the first attempts at using supervised learning for wall models
in LES can be found in \citet{Yang2019}. The authors noted that a
model trained on turbulent channel flow data at a single Reynolds
number could be extrapolated to higher Reynolds numbers in the same
configuration.
Similar approaches for data-driven wall models using supervised
learning were developed for various flow configurations, such as a
spanwise rotating channel flow \citep{huang2019wall}, flow over periodic
hills \citep{zhou2021wall}, turbulent flows with separation
\citep{zangeneh2021data}, and boundary layer flow in the presence of
shock--boundary layer interaction \citep{bhaskaran2021science}, with
mixed results in \emph{a posteriori} testing. The first attempt at
semi-supervised learning for WMLES can be found in
\citet{bae2022scientific}, where the authors used reinforcement
learning to train on turbulent channel flow data at relatively low
Reynolds numbers. The model was able to extrapolate to higher Reynolds
numbers for turbulent channel flows and zero pressure gradient
turbulent boundary layers. The reinforcement learning wall model has
recently been extended to account for pressure gradient
effects~\citep{Zhou2022}.
Nonetheless, most of the models cited above rely on information about
the flow that is typically inaccessible in real-world applications,
such as the boundary layer thickness, and are limited to simple flow
configurations.  One exception is the ML wall model introduced by
\citet{Lozano_brief_2020}, which is is directly applicable to
arbitrary complex geometries and provides the foundations for the
present modelling effort.

Currently, one major challenge for WMLES of realistic external
aerodynamic applications is achieving robustness and accuracy
necessary to model the myriad of different flow regimes that are
characteristic of these problems. Examples include turbulence with
mean flow three-dimensionality, laminar-to-turbulent transition, flow
separation, secondary flow motions at corners, and shock wave
formation, to name a few. The wall stress generation mechanisms in
these complex scenarios differ from those in flat-plate turbulence.
However, the most widespread wall models are built upon the assumption
of statistically in equilibrium wall-bounded turbulence without
mean flow three-dimensionality, which only applies to a handful number
of flows. The latter raises the question of how to devise models
capable of seamlessly accounting for such a vast and rich collection
of flow physics in a single unified approach. \corr{Another important
  consideration is the data required for training the ML models and
  consistency with the numerical schemes and grid generation
  strategy.}

In this work, we develop a wall model for LES using building-block
flows. The model is formulated to account for various flow
configurations, such as wall-attached turbulence, wall turbulence
under favourable/adverse pressure gradients, separated turbulence,
statistically unsteady turbulence, and laminar flow. The model
comprises two components: a classifier and a predictor. The classifier
is trained to place the flows into separate categories along with a
confidence score, while the predictor outputs the modelled wall stress
based on the likelihood of each category.  The training data are
directly obtained from WMLES with an `exact' model for
mean quantities to guarantee consistency with the numerical
discretisation and grid structure. The model is validated in canonical
flows outside the training set and complex flows. The latter includes
two realistic aircraft configurations, namely, the NASA Common
Research Model High-lift and NASA Juncture Flow Experiment.

This paper is organised as follows.  The formulation of the new model
is discussed in \S\ref{sec:formulation}. The numerical approach and
traditional wall models that serves as a comparison point are
introduced in \S\ref{sec:methods}.  The model is validated in
\S\ref{sec:validation} and compared with an equilibrium wall
model. The model limitations are discussed in
\S\ref{sec:limitations}. Finally, conclusions are offered in
\S\ref{sec:conclusions}.

\section{Model formulation}\label{sec:formulation}

\begin{figure}
      \centering
      \includegraphics[width=1\textwidth]{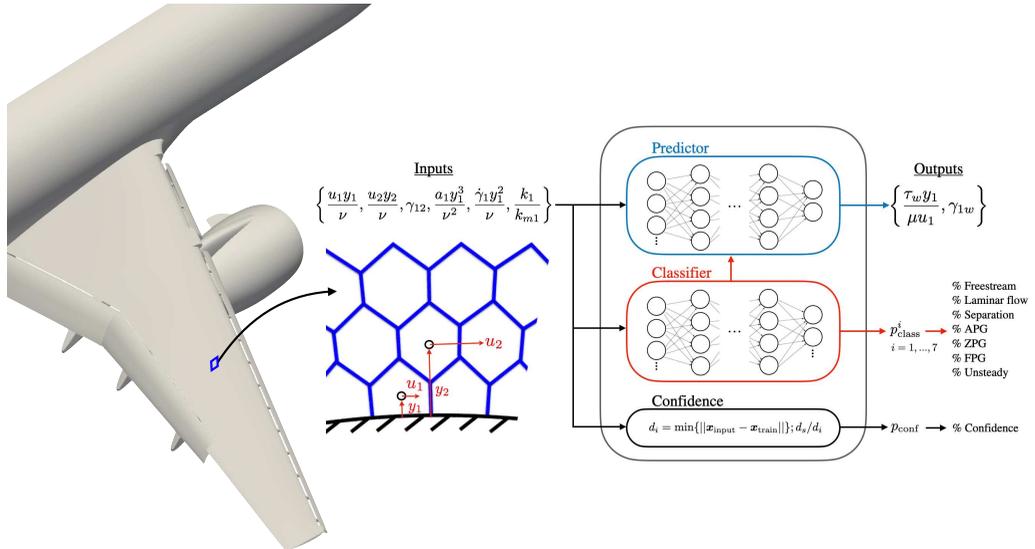}
      \caption{\corr{Schematic of the building-block flow wall model
        (BFWM). Details of the formulation are provided in \S\ref{sec:formulation}.}
      \label{fig:model_sketch}}
\end{figure}

The working principle of the proposed model is summarised in
figure~\ref{fig:model_sketch}. The model is referred to as the
building-block-flow wall model (BFWM) and was first introduced by
\citet{Lozano_brief_2020}.  The BFWM is comprised of two elements: a
classifier and a predictor.  First, the classifier is fed data from
the LES solver and quantifies the similarities of the input with a
collection of known building-block flows. The predictor leverages the
information of the classifier together with the input to generate the
wall-shear stress prediction via combination of the building-block
flows from the database. Each building block is dedicated to modelling
different flow physics (e.g. wall-attached turbulence,
adverse/favourable pressure-gradient effects, separation, laminar
flow), and the model provides a blending between flows using
information from the classifier. A confidence score is generated based
on the similarity between the input data and the building-block flows.
If the input data look extraneous, then the model prompts a low confidence
score, which essentially means that the flow is unknown and does not
match any knowledge from the database. In the following, we elaborate
on the model requirements, assumptions, input/output data, training
data, and ANN architecture.

\subsection{Model requirements}

We consider the following model requirements.
\begin{enumerate}
\item The wall model should be able to account for different flow
  physics (e.g. laminar flow, wall-attached turbulence, separated
  flow) in a unified manner (i.e., the input and output structure
  of the model must be identical regardless of the case).
\item  The model must be scalable to incorporate additional
  building-block flows if needed in future versions.
\item  The model must provide a confidence score in the
  prediction at each point at the wall.
\item The model formulation must be directly applicable to complex
  geometries (e.g., realistic aircraft configuration) without any
  additional modifications. This requirement will constrain the
  allowable input variables and the parameters used for their
  non-dimensionalisation (i.e., they will need to be
  local in space).
\item The model must account for the numerical errors of the schemes
  employed to integrate the LES equations. This implies that the input
  data used to train the model must be consistent with the data 
  from the LES solver rather than from filtered DNS.
\item  The inputs and outputs of the model must be given in
  non-dimensional form to comply with dimensional consistency.
\item  The model must be invariant under constant space/time
  translations and rotations of the frame of reference.
\item The model must be Galilean invariant.
\end{enumerate}

\subsection{Model assumptions}\label{subsec:overview}

The main modelling assumptions are as follows.
\begin{enumerate}
\item There is a finite set of simple flows (referred to as
  building-block flows) that contains the essential flow physics to
  formulate generalisable wall models.
\item The effect of the (missing) near-wall SGS in WMLES of
  complex flows is representable by a linear combination of the
  near-wall behaviour of the building-block flows.
\item \corr{A set of N non-dimensional model inputs based on local
  flow quantities is enough to discern among building-block flows,
  where N is the minimum number of parameters characterising the
  building-block flow collection.}
\item The non-dimensional form of the model inputs/outputs that
  provides the best predictive capabilities (i.e.,
  interpolation/extrapolation between training cases) is obtained by
  scaling the variables with the kinematic viscosity and the distance
  to the wall.
\item The flow information from two contiguous wall-normal locations is
  enough to predict the wall-shear stress in the vicinity of those
  locations.
\item History effects for the SGS flow are captured using
  instantaneous accelerations without the need for additional
  information from past times.
\end{enumerate}

Assumptions (i) and (ii) imply that the BFWM would provide accurate
  predictions as long as the (non-universal) large-scale flow motions
  are resolved by the WMLES grid, and the (smaller-scale) near-wall
  dynamics resemble the building-block flows or a combination of
  them.
Assumption (iii) stems from the fact that the number of input
  variables to distinguish among different cases in the building-block
  flow collection must be, at least, equal to the number of
  non-dimensional groups required to completely characterise the
  wall-shear stress across all the building-block flows. For the
  building-block flows chosen in this work, six parameters are needed
  to unambiguously identify the wall-shear stress from one particular
  case (as will be discussed in \S\ref{subsec:buildingblock}). These
  parameters are global quantities not available to the
  model. Instead, six non-dimensional inputs using local flow
  information are fed into the BFWM to predict the wall-shear
  stress. However, there is no guarantee that these inputs can be used to
  discern among all possible building-block flows in an univocal
  manner at all times, and from there comes the need for assumption (iii).

In assumption (iv), it is assumed that the scaling proposed will be
  the best in all the flow scenarios that the model may encounter, which is
  not true in general. For example, the current scaling choice will
  not provide the best performance in the presence of strong
  compressibility effects, chemically reacting flows or multiphase
  flows.

\corr{Assumptions (v) and (vi) are adopted for the sake of model
  simplicity. The choice is informed by our previous work in
  \cite{Lozano_brief_2020}, where a seven-point stencil was used for the
  input variables compared to the simpler, two-point stencil selected in
  the present work. It was noted that the seven-point stencil greatly
  complicated the model implementation without providing important
  benefits in terms of model performance. There are additional
  modelling assumptions that are not explicitly stated
  above. Nonetheless, points (i) to (vi) are the most critical
  assumptions affecting the model performance.}

\subsection{Building-block flows} \label{subsec:buildingblock}

\begin{figure}
  \centering
  \includegraphics[width=0.9\textwidth]{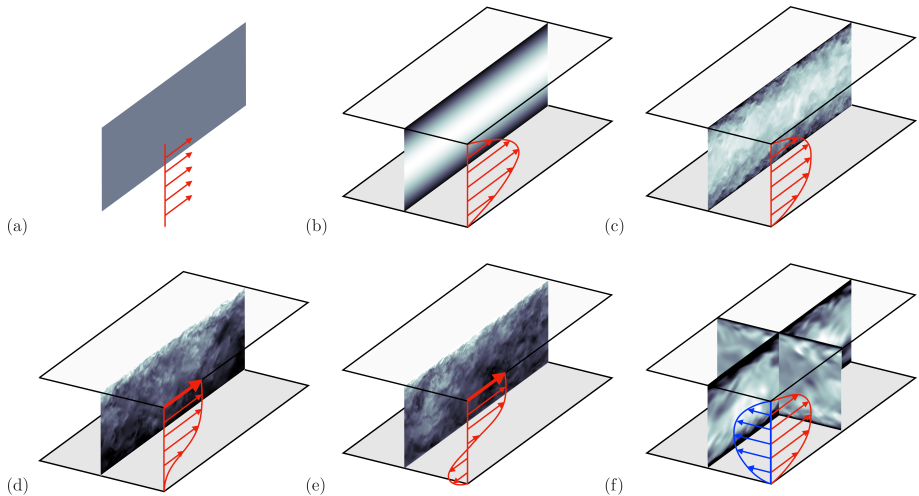}
  \caption{Examples of building-block units taken as
    representative of different flow regimes: (a) Freestream, (b)
    laminar channel flow, (c) turbulent channel flow, (d) turbulent
    Poiseuille--Couette flow for zero wall stress, (e) turbulent
    Poiseuille--Couette flow with a strong adverse mean pressure
    gradient, and (f) turbulent channel flow with sudden imposition of
    spanwise mean pressure gradient. \label{fig:flow_units}}
\end{figure}

Seven types of building-block flows are considered, and examples are
shown in figure \ref{fig:flow_units}. All the cases entail an
incompressible flow confined between two parallel walls, with the
exception of the freestream flow. Cases with additional complexity,
such as airfoils, wings, bumps, are intentionally avoided. The
rationale behind this choice is that the building blocks should encode
the key flow physics to predict more complex scenarios. Hence we
intent to avoid case overfitting, i.e. correctly predicting the flow
over a wing merely because the model was also trained on similar wings
instead of faithfully capturing the flow physics.

For the building blocks, the streamwise, wall-normal and spanwise
spatial coordinates are denoted by $x$, $y$ and $z$, respectively.
The walls are separated by a distance $2h$. The density is $\rho$, and
the dynamic and kinematic viscosities of the fluid are $\mu$ and
$\nu$, respectively.  In the following, we discuss the configuration
and physical motivation behind each building block. We also include in
parentheses the label for each building-block flow.
\begin{itemize}
 \item \underline{Freestream} (Freestream). A uniform and constant
   velocity field is used as representative of freestream flows. The
   main role of this building-block flow is to identify near-wall
   regions with zero points per boundary layer thickness. In these
   situations, the BFWM applies a ZPG model (defined below) to estimate
   the wall stress.  The prediction is accompanied by a low confidence
   score as there is not enough information to provide reliable
   predictions.
\item \underline{Wall-bounded laminar flow} (Laminar). The near-wall
  region of a laminar flow is modelled using Poiseuille flow (i.e.,
  parabolic mean velocity profile). The goal of this building block is
  to allow the prediction of the wall stress in wall-attached laminar
  scenarios. The case is characterised by the friction Reynolds number
  $\Rey_\tau= u_\tau h /\nu$, where $u_\tau$ is the friction velocity
  at the wall, and $h$ is the channel half-height.  The range of
  Reynolds numbers considered is from $\Rey_\tau= 5$ to $\Rey_tau = 10^4$.  A
  parabolic mean velocity profile is used to analytically predict the
  stress at the wall.
\item \underline{Wall-bounded turbulence under zero
  mean-pressure-gradient} (ZPG). Canonical wall turbulence without
  mean pressure gradient effects is modelled using turbulent channel
  flows as a building block. The wall stress predictions are provided
  by an ANN trained for $\Rey_\tau = 100$ to $\Rey_\tau=10,000$ using
  numerical data~\citep{Lozano2014a, Hoyas2022}.
\item \ul{Wall-bounded turbulence under favourable/adverse
  mean-pressure-gradient and separation} (FPG, APG, and Separation,
  respectively). We utilise the turbulent Poiseuille--Couette flow as a
  simplified representation of wall-bounded turbulence subject to
  favourable and adverse mean pressure gradient effects. The bottom
  wall is set at rest, and the top wall is set at a constant velocity equal to
  $U_t>0$. A streamwise mean pressure gradient, denoted by
  $\mathrm{d}P/\mathrm{d}x$, is applied to the flow. Favourable
  pressure gradient effects are obtained for values of
  $\mathrm{d}P/\mathrm{d}x$ accelerating the flow in the same
  direction as the top wall. Adverse pressure gradient conditions are
  achieved for values of $\mathrm{d}P/\mathrm{d}x$ that accelerate the
  flow in the opposite direction as the top wall. Flow separation is
  represented by values of $\mathrm{d}P/\mathrm{d}x$ at which the wall
  stress at the bottom wall is zero. The different flow regimes are
  characterised by the two Reynolds numbers $\Rey_P=
  \pm\sqrt{|\mathrm{d}P/\mathrm{d}x| / \rho} h/\nu$ and $\Rey_{U}=
  U_th/\nu$. The wall stress predictions are provided by an ANN
  trained for $\Rey_P$ from $-1.2 \times 10^3$ to $1.2 \times 10^3$,
  and $\Rey_{U}$ from $5 \times 10^3$ to $2 \times 10^4$. New DNS runs were
  conducted for these cases using the same numerical solver as in
  previous investigations by our group~\citep{Lozano2018a,Lozano2019}.
\item \underline{Statistically unsteady wall turbulence with 3-D mean
  flow} (Unsteady). The last building-block flow considered is a
  turbulent channel flow subject to a sudden spanwise mean pressure
  gradient $\mathrm{d}P/\mathrm{d}z$.  The role of this case is to
  capture out-of-equilibrium effects due to strong unsteadiness and
  mean flow three-dimensionality, such as the decrease in the
  magnitude of the wall stress and the misalignment between the
  wall stress vector and mean shear vector.  Only the initial
  transient of the flow, where non-equilibrium effects manifest, is
  considered. The different flow regimes are characterised by the
  streamwise friction Reynolds number $\Rey_\tau = u_{\tau} h /\nu$
  before the imposition of $\mathrm{d}P/\mathrm{d}z$ and the ratio of
  streamwise to spanwise mean pressure gradients $\Pi =
  (\mathrm{d}P/\mathrm{d}z)/(\mathrm{d}P/\mathrm{d}x)$. The
  wall stress predictions are provided by an ANN trained for
  $\Rey_\tau = 100$ to $1000$ and $\Pi = 5$ to $100$ using data from
  \citet{Lozano2020}.
\end{itemize}

\corr{The wall-shear stress from each case in the building-block flow
  collection considered is completely determined by the specification
  of six parameters: Re$_\tau$ (or Re$_P$), $Re_U$, $\Pi$, $t
  u_\tau/h$ (non-dimensional time for Unsteady cases),
  turbulent/non-turbulent flow, and laminar/freestream. Not all
  parameters are relevant for each case, yet they are required to
  avoid ambiguities.}

\begin{figure}
    \vspace{0.5cm}
  \begin{center}
    \subfloat[]{\includegraphics[width=0.43\textwidth]{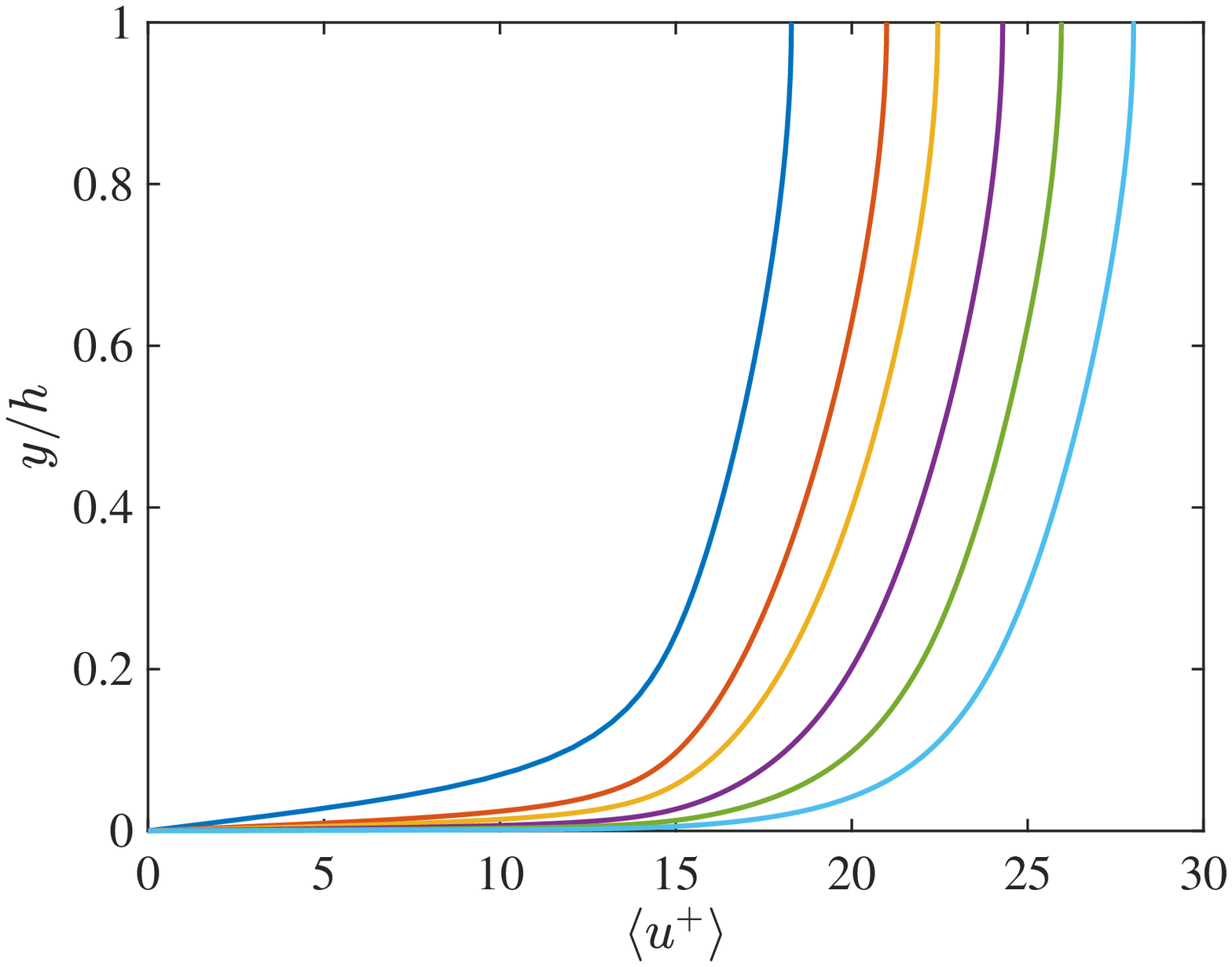}}
    \hspace{0.5cm}
    \subfloat[]{\includegraphics[width=0.43\textwidth]{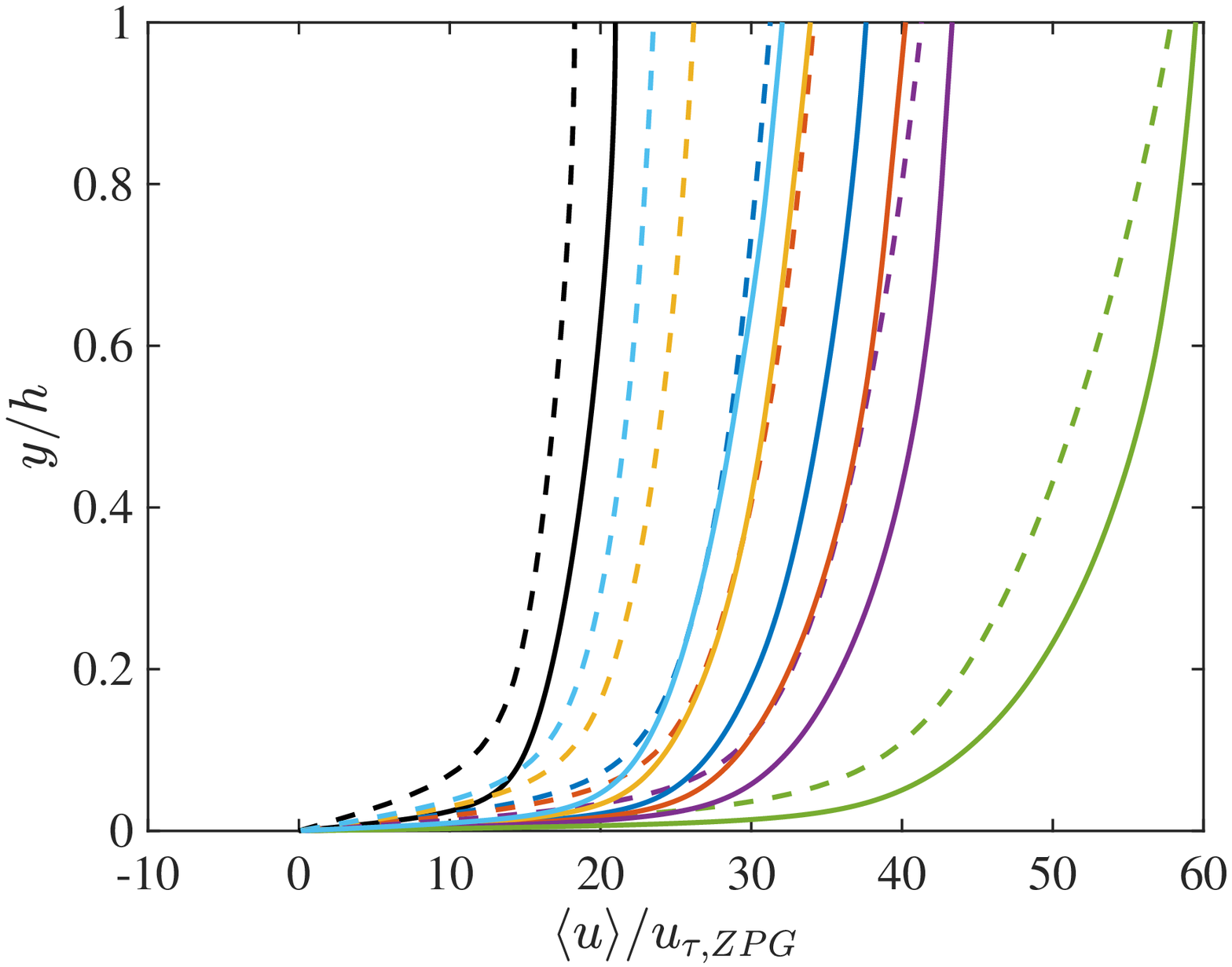}}
  \end{center}
  \begin{center}
    \subfloat[]{\includegraphics[width=0.43\textwidth]{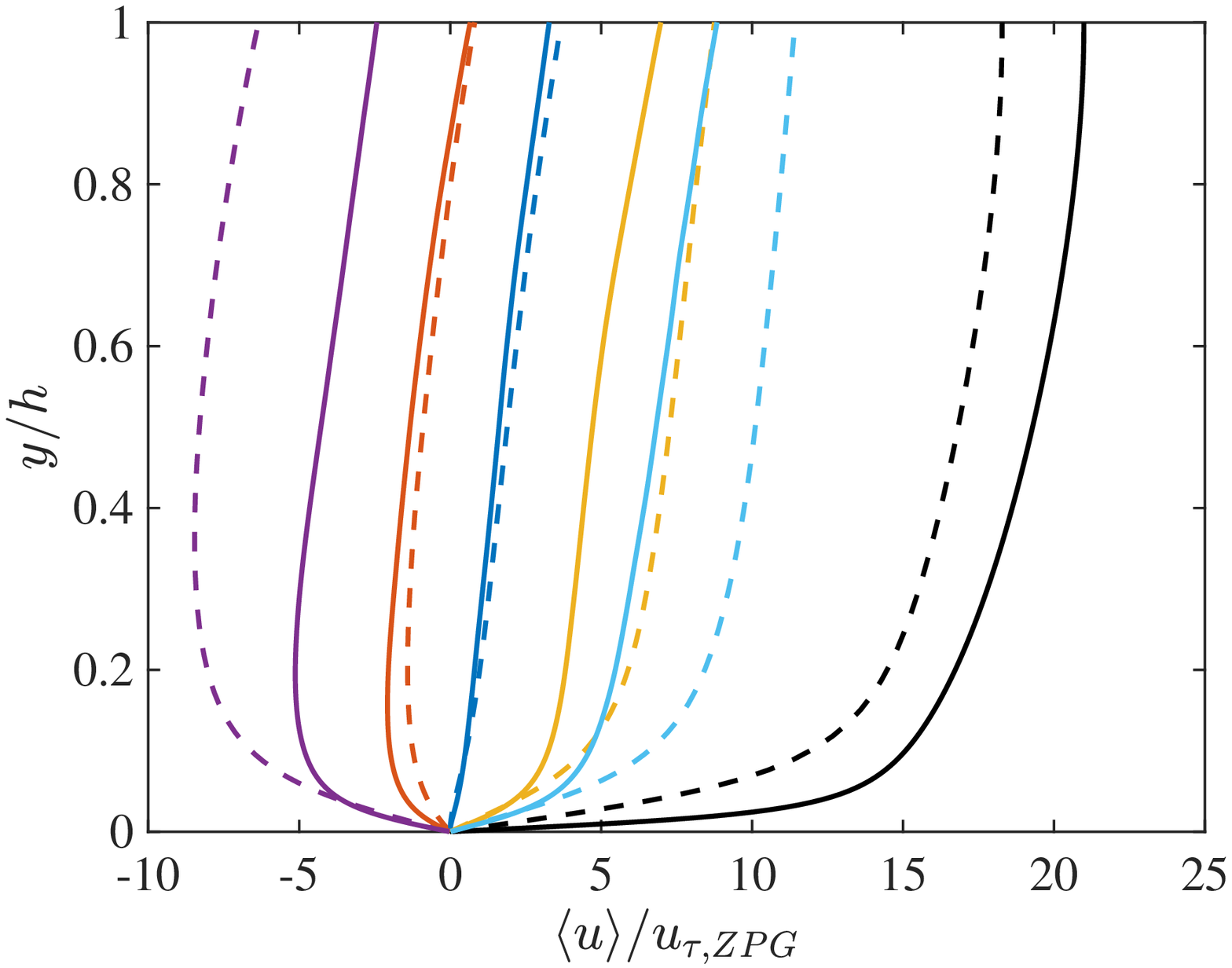}}
    \hspace{0.5cm}
    \subfloat[]{\includegraphics[width=0.43\textwidth]{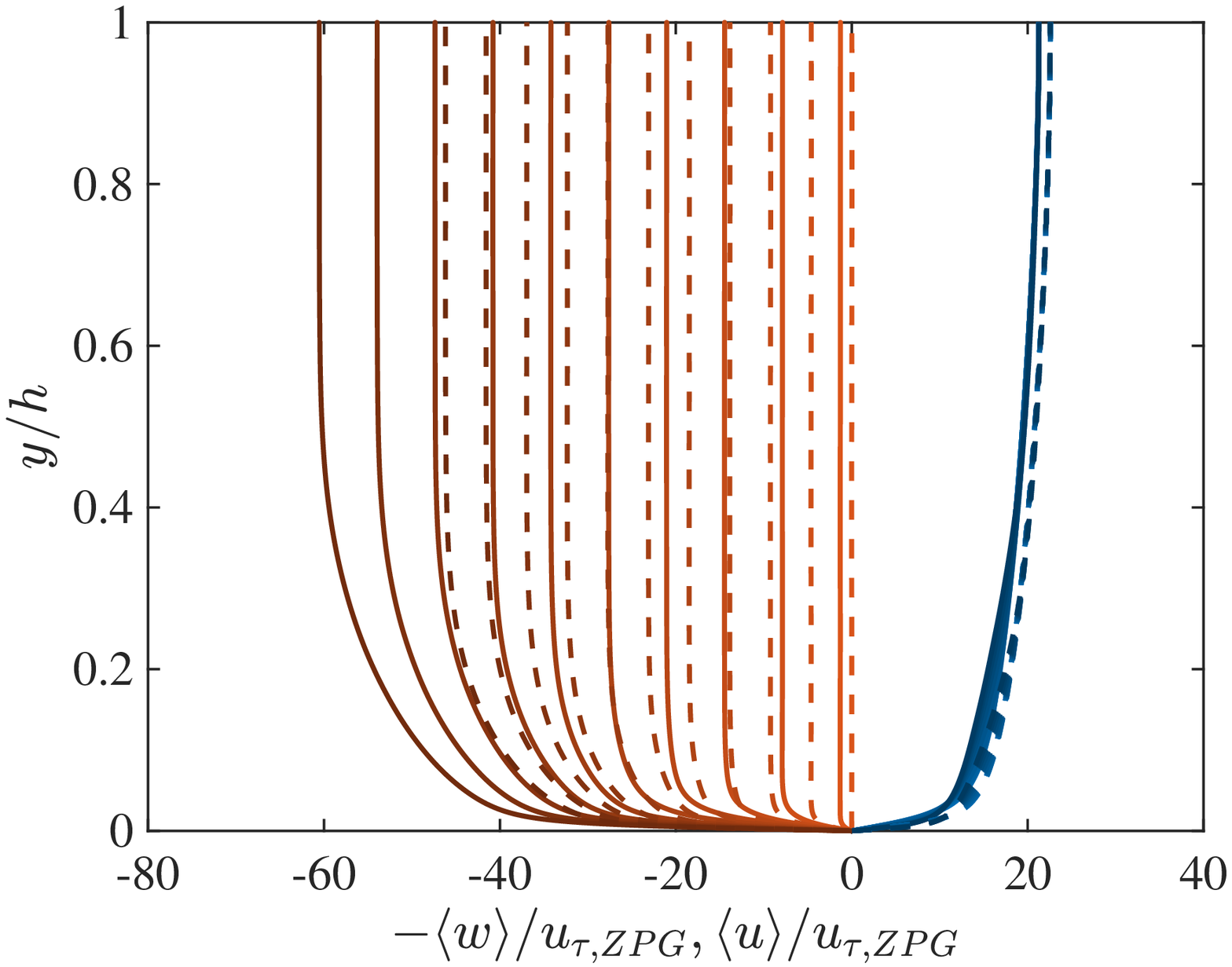}}
  \end{center}  
  \caption{ Mean velocity profiles for a selection of building-block
    flows. (a) Turbulent channel flows for (from left to right)
    $\Rey_\tau=180, 550, 950, 2000, 4200$ and $10,000$ (ZPG).  (b)
    Turbulent Poiseuille--Couette with FPG at $\Rey_U=6500$ (dashed) $\Rey_P=0,
    100, 150, 200, 250, 300$ (from left to right) and $\Rey_U=22,360$
    (solid) for $\Rey_P= 0, 380, 550, 750, 800, 1000$ (from left to
    right). (c) Turbulent Poiseuille--Couette with APG and separation at $\Rey_U=6500$
    (dashed) $\Rey_P=0, -100, -150, -200, -250, -300$ (from right to
    left) and $\Rey_U=22,360$ (solid) for $\Rey_P= 0, -380, -550,
    -750, -800, -1000$ (from right to left). (d) Turbulent channel
    flows with the sudden imposition of spanwise mean pressure gradient
    for increasing time (from light to dark colour) for the streamwise
    (in blue) and spanwise (in red) mean velocity profiles at
    $\Rey_\tau=550$ (solid) and $\Rey_\tau=950$ (dashed) for $\Pi =
    60$. In all cases, $u_{\tau,ZPG}$ is the friction velocity of the
    same case without adverse/favourable/spanwise mean pressure
    gradient. \label{fig:BF_Umean}}
\end{figure}

\corr{The DNS database for ZPG, FPG, APG, Separation, and Unsteady
  contains roughly 500 simulations}. Figure \ref{fig:BF_Umean}
contains examples of the DNS mean velocity profiles for a selection of
building-block flows.  An advantage of the present building-block
set-up is that it allows the generation of contiguous data from one
flow regime to another without modifying the geometry but only varying
the non-dimensional numbers defining the case (e.g., from FPG to ZPG
to APG to Separation by just changing the mean pressure
gradient). This facilitates the generation of training data filling
the non-dimensional space of inputs, which translates into a more
reliable model. The latter is an important requirement, as the role of
a model should not be fitting one particular dataset but to learn the
scaling of the non-dimensional inputs and outputs controlling the
problem at hand.
%

\subsection{Input and output variables}\label{subsec:IO}

The input variables are acquired using a two-point stencil as shown in
figure \ref{fig:model_sketch}.  The stencil contains the centre of the
control volume attached to the wall (where the wall-shear stress is to
be predicted) and the second control volume off the wall along the
wall-normal direction. Given that the model is intended to be used in
complex geometries, the centres do not need to align perfectly along
the wall-normal direction. This misalignment is considered during the
model training. The information collected from the flow is
\begin{equation}
  \left\{ u_1, u_2, \gamma_{12}, a_1, \dot\gamma_1, k_1 , k_{m1} \right\},
  \label{eq:input_variables_raw}
\end{equation}
where $u_1 = |\vec u_1|$ and $u_2=|\vec u_2|$ are the magnitude of the
wall-parallel velocities relative to the wall at the first and second
control volumes, respectively, $\gamma_{12}$ is the angle between
$\vec u_1$ and $\vec u_2$, $a_1$ is the magnitude of the acceleration
at the first control volume, $\dot\gamma_1$ is the time derivative of
the angle of $\vec u_1$ in the wall-parallel direction, and $k_1$ and
$k_{m1}$ are the turbulent kinetic energy and mean kinetic energy,
respectively, at the first control volume and relative to the wall. The
mean velocity to compute $k_1$ and $k_{m1}$ is obtained via an
exponential average in time (denoted by $(\bar{\cdot})$) with time
scale $10\Delta /\sqrt{k_{m1}}$, where $\Delta$ is the characteristic
grid size based on the cube root of the control volume. The quantities
predicted by the model are
\begin{equation}
  \left\{\tau_w, \gamma_{1w}, p_\mathrm{class}^i, p_{\mathrm{conf}} \right\}, 
\end{equation}
where $\tau_w = |\vec \tau_w|$ is the magnitude of the wall stress
vector, $\gamma_{1w}$ is the angle between $\vec u_1$ and $\vec
\tau_w$, $p_\mathrm{class}^i \in [0,1]$ for $i=1,...,7$ is the
probability of the flow corresponding to each building-block category,
and $p_{\mathrm{conf}} \in [0,1]$ is the model confidence score.

The input to the wall model comprises the non-dimensional groups
formed by the set
\begin{equation}
\left\{ \frac{ u_1 y_1}{\nu} , \frac{ u_2 y_2}{\nu}, \gamma_{12}, \frac{a_1 y_1^3}{\nu^2}, \frac{{\dot{\gamma}}_{1} y_1^2}{\nu}, \frac{ k_1}{ k_{m1}} \right\}.\label{eq:input}
\end{equation}
The model applies the exponential averaged $(\bar\cdot)$ to the input
variables, and this is accounted for in the training.  The
non-dimensional output of the wall model is
\begin{equation}
  \left\{ \frac{\tau_w y_1}{\mu \bar u_1}, \gamma_{1w} , p_\mathrm{class}^i, p_{\mathrm{conf}} \right\}.
\end{equation}
We have explicitly avoided the use of flow parameters such as
freestream velocity, and bulk velocity flow, as they are not
unambiguously identifiable for arbitrary geometries.

The non-dimensional groups in (\ref{eq:input}) are devised to enable
the classification and wall stress prediction according to the
building-block flows considered. All the inputs contribute to the
prediction of the outputs to some degree. However, it is possible to
highlight the main role played by each input variable. The first two
inputs, $u_1 y_1/\nu$ and $u_2 y_2/\nu$, represent the local Reynolds
numbers at the first and second control volumes, respectively. They
are key drivers for the prediction of the wall-shear stress in all
cases by detecting the shape of the mean velocity profile. They aid
the distinction between turbulent channel flow, turbulence with
favourable/adverse mean pressure gradient, and separation. The
relative angle $\gamma_{12}$ is used to detect three-dimensionality in
the mean velocity profile. The non-dimensional acceleration $a_1
y_1^3/\nu^2$ and the angular rate of change $\dot{\gamma}_{1}
y_1^2/\nu$ facilitate the prediction of the magnitude and direction of
statistically unsteady effects in the wall-shear stress. Finally,
$k_1/k_{m1}$ is leveraged to discern between turbulent flows (ZPG,
APG, FPG, etc.)  and non-turbulent flows (freestream and laminar).  The wall stress at the output is non-dimensionalised using the
pseudo-wall-stress $\mu \bar u_1/y_1$, as it was found to minimise the
spread of the data between cases. This scaling improved the
generalisability of the model and facilitated learning the trends in
the data by the ANNs.

\subsection{WMLES with optimised SGS/wall model}
\label{subsec:training_data}

The training data (discussed in \S\ref{subsec:training_steps}) are
generated using WMLES with SGS/wall models optimised to obtain the
exact values for the mean velocity profiles and wall stress
distribution in order to attain consistency with the numerical
discretisation and gridding strategy of the solver.  This approach was
preferred over filtered DNS data, as it is known that the SGS tensor
in implicitly filtered LES ($\tau^\mathrm{SGS}_{ij}$) does not
coincide with the Reynolds stress terms resulting from filtering the
Navier-Stokes equations. The ambiguity in the filter operator renders
DNS data inadequate for the development of SGS models because of
inconsistent governing equations~\citep{Lund1995, Lund2003,
  bae_brief_2017, bae_brief_2018, bae2022numerical}.  This limitation
is particularly relevant in the present work, as the typical grid
sizes utilised in WMLES of external aerodynamics are orders of
magnitude larger than the characteristic Kolmogorov length scale of
the near-wall turbulence. Consequently, numerical errors are
comparable to modelling errors, and the former must be accounted for
in order to yield accurate predictions.

We introduce an exact-for-the-mean SGS (ESGS) model given by
\begin{equation}
  \tau^\mathrm{ESGS}_{ij} = \tau^\mathrm{base}_{ij} + \Delta \tau^\mathrm{SGS}_{ij},
\end{equation}
where $\tau^\mathrm{base}_{ij}$ is the SGS stress tensor provided by
the baseline (imperfect) SGS model (in this case, the dynamic
Smagorinsky model, albeit another model could have been selected), and
$\Delta \tau^\mathrm{SGS}_{ij}$ is the SGS model correction such that
the WMLES mean velocity profiles match the DNS counterparts.  In this
approach, instantaneous DNS flow fields are not needed, and neither is
a specific LES filter shape.  \corr{The correction $\Delta
  \tau^\mathrm{SGS}_{ij}$ is calculated on the fly as the
  instantaneous force, $\partial \Delta
  \tau^\mathrm{SGS}_{ij}/\partial x_j$, required for $\langle u_i
  \rangle_{xz} = \langle u_i^\mathrm{DNS} \rangle_{xzt}$, where
  $\langle \cdot \rangle_{xz}$ denotes average over the homogeneous
  directions $x$ and $z$, and $\langle u_i^\mathrm{DNS} \rangle_{xzt}$
  is the DNS mean velocity profile for the $i$-th velocity component
  averaged over $x$, $z$ and $t$.  If the DNS flow is not
  statistically stationary (e.g., the Unsteady building-block flow),
  then the averaged is taken over $x$, $z$, and statistically
  equivalent realisations. At each wall-normal location, the force
  $\partial \Delta \tau^\mathrm{SGS}_{ij}/\partial x_j$ is numerically
  calculated as the value required in the right-hand side of the
  equations to match the DNS mean velocity profile in the next time
  step}.

The boundary condition at the wall is also modified to reproduce the
probability density function (p.d.f.) of the wall-shear stress from
DNS, and it is referred to as the exact boundary condition (EBC). \corr{This was achieved by using an inverse probability
  integral transform, which generates random numbers from an arbitrary
  probability distribution given its cumulative distribution
  function~\citep{Devroye2006}. During runtime of WMLES with EGSG,
  random numbers are sampled from the uniform distribution and mapped
  onto the p.d.f. of the wall-shear stress previously obtained from
  DNS. The process is performed for all wall locations at each time
  step.}

WMLES of all the building-block flows was conducted using ESGS with
EBC (hereafter, E-WMLES). The simulations were carried out in the same
flow solver later used to implement the BFWM. This guarantees
consistency of the training data with the actual numerical errors of
the solver and gridding structure under the assumption of an SGS model
able to accurately predict the mean velocity profiles.

\subsection{E-WMLES training data, ANN architecture, and training method}
\label{subsec:training_steps}

\begin{figure}
  \centering
  \vspace{0.5cm}
\includegraphics[width=1.0\textwidth]{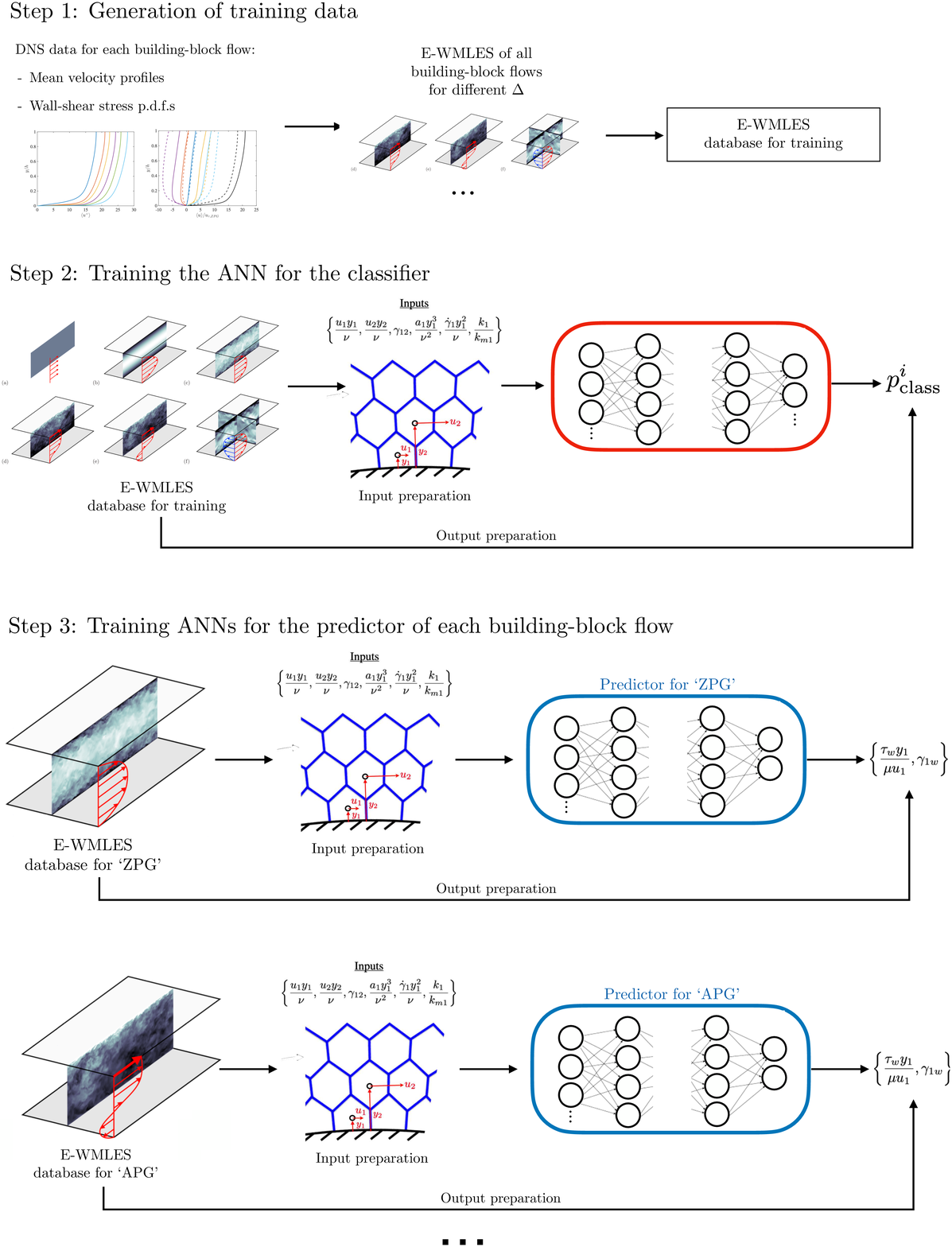}
\caption{ \corr{Overview of the training workflow for the BFWM. The details are discussed in \S\ref{subsec:training_steps}} \label{fig:training_overview}}
\end{figure}

\corr{Figure \ref{fig:training_overview} offers an overview of the training workflow, which is divided into three steps.}
%
%
\begin{itemize}
\item[-] \corr{Step 1: Generation of E-WMLES training database. The
  mean velocity profiles and the p.d.f.s of the wall-shear stress from
  DNS are used to generate the training database using E-WMLES as
  described in \S\ref{subsec:training_data}.  The E-WMLES database
  comprise the building-block flows cases discussed in
  \S\ref{subsec:buildingblock} at isotropic grid resolutions $\Delta =
  h/N$ with $N=5,10,20,40,80,160$ and $380$, which cover (by a wide
  margin) the grid resolutions encountered in external aerodynamic
  applications.  For grid resolutions finer than $\Delta =h/380$, the BFWM
  was trained with DNS data to ensure convergence to the no-slip
  boundary condition.
For each grid resolution, the E-WMLES data generated are time-resolved
with a varying time step such that the Courant--Friedrichs--Lewy number
is equal to 0.5. Time-resolved data were required to compute
accelerations and time averages of the input variables that are
representative of WMLES. The E-WMLES cases were run for 10
eddy-turnover times (defined by $h/u_\tau$). This time was sufficient
to capture the statistical trends of the smallest flow scales in the
E-WMLES, which have lifetimes of the order of
$\Delta/u_\tau$~\citep{Lozano2014b}. It was found that reducing the
time length of the training data below 3 eddy-turnover times
significantly reduced the performance of the ANNs. The Unsteady cases
were only simulated for the period along which non-equilibrium effects
are relevant (i.e., between 0.5 and 2 eddy-turnover times). Each
E-WMLES contains of the order of 1,000 to 10,000 snapshots depending
on the case.
The training set was augmented by performing E-WMLES in which the $x$
direction (i.e., mean flow direction) was rotated parallel to the wall
by -45, -40, -35,..., 35, 40 and 45 degrees.}
    \item[-] \corr{Step 2: Training the ANN for classifier. The
      classifier is trained using the full E-WMLES database.} %
      classifier The classifier is an ANN with 5 hidden layers and 20
      neurons per layer. The layers are connected with rectified
      linear units (ReLUs) as the activation function. The input to
      the classifier is the set of non-dimensional groups from
      Eq. (\ref{eq:input}). The last layer of the classifier is
      followed by the softmax activation function, which provides the
      probabilities of belonging to each building-block flow category
      ($p_\mathrm{class}^i$, $i=1,...,7$) such that $\sum
      p_\mathrm{class}^i = 1$.  The network is trained using the
      Broyden-Fletcher-Goldfarb-Shanno quasi-Newton
      algorithm~\citep{broyden1970convergence, fletcher1970new,
        goldfarb1970family, shanno1970conditioning,
        fletcher2013practical}.  The performance of the classifier
      after training is evaluated in figure \ref{fig:confusion}, which
      shows the normalised confusion matrix. The diagonal of the
      matrix contains the percentages of inputs that are correctly
      classified, whereas off-diagonal values show the amount of
      misclassification among cases.  The matrix shows that most
      samples are correctly classified. There are a few
      misclassifications (off-diagonal values) which are intentional
      to ensure a smooth transition between building blocks. This is
      caused by the overlap between contiguous building blocks, such
      as Separation and strong APG, mild APG and ZPG, ZPG and mild
      FPG, and ZPG and Unsteady. The only unintentional
      misclassification is the larger number of ZPG samples classified
      as FPG. However, this did not impact the performance of the BFWM, as
      the wall stress predictions for FPG and ZPG are comparable. Note
      that there is no ambiguity in the distinction between
      Freestream/Laminar and turbulent cases, which is important as
      their predictions vary considerably.  A confidence score
      $p_\mathrm{conf} \in [0,1]$ is inferred from the distance of the
      input to its nearest neighbour in the training set.  In
      particular, the confidence is based on the ratio $d_s/d_i$
      (clipped to be less than 1), where $d_i$ is the distance of
      the input to the closest sample in the training set, and $d_s$
      is the average distance of the latter to its ten closest samples
      in the training set.
%
\begin{figure}
  \centering
  \vspace{0.5cm}
\includegraphics[width=0.8\textwidth]{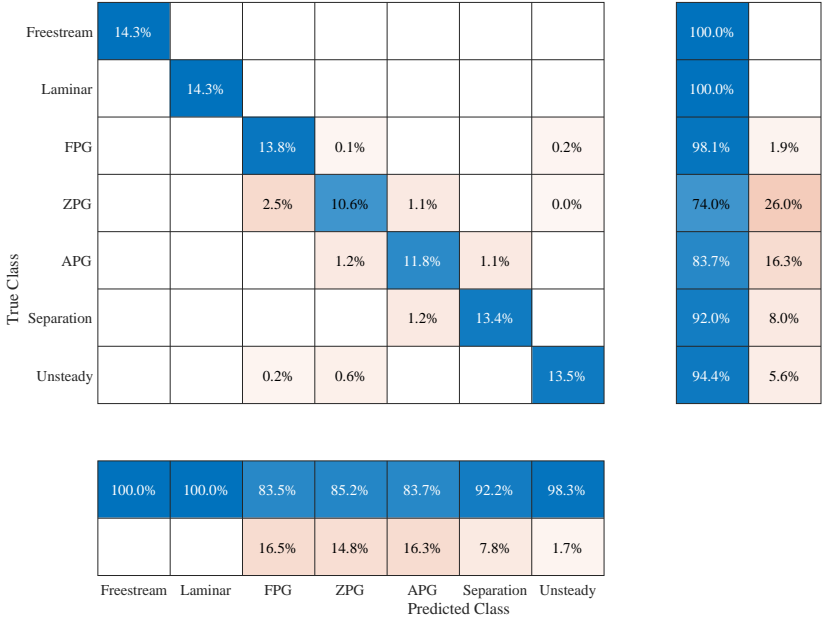}
\caption{ Confusion matrix of the classifier. The categories are
  freestream (Freestream), laminar channel flow (Laminar), favourable
  mean pressure gradient wall turbulence (FPG), zero
  mean pressure gradient wall turbulence (ZPG), adverse
  mean pressure gradient wall turbulence (APG), separated turbulence
  (Separation), and statistically unsteady wall turbulence
  (Unsteady). \label{fig:confusion}}
\end{figure}
    \item[-] \corr{Step 3: Five ANNs are trained, each dedicated to
      predict one building-block flow, i.e. ZPG, APG, FPG, Separation, and
      Unsteady.}  The ANN predictors are obtained via a deep
      feed-forward ANN with 5 hidden layers and 30 neurons per
      layer. The activation functions selected for the hidden layers
      are hyperbolic tangent sigmoid transfer functions and ReLU
      activation transfer functions. \corr{Predictions of the ANN are
        calculated for each building-block category and added together,
        weighted by the probability of belonging to the $i$-th
        category:
\begin{equation}
\frac{\tau_w y_1}{\mu \bar u_1} = \sum_i p_\mathrm{class}^i \frac{\tau_w^i y_1}{\mu \bar u_1}, \quad \gamma_{1w} = \sum_i p_\mathrm{class}^i \gamma_{1w}^i,     
\end{equation}
where $\tau_w^i$ and $\gamma_{1w}^i$ are 
the predictions for the $i$-th category. The process is illustrated in figure \ref{fig:ANN_predictors}.} 
%
\begin{figure}
    \vspace{0.5cm}
  \begin{center}
    \includegraphics[width=0.8\textwidth]{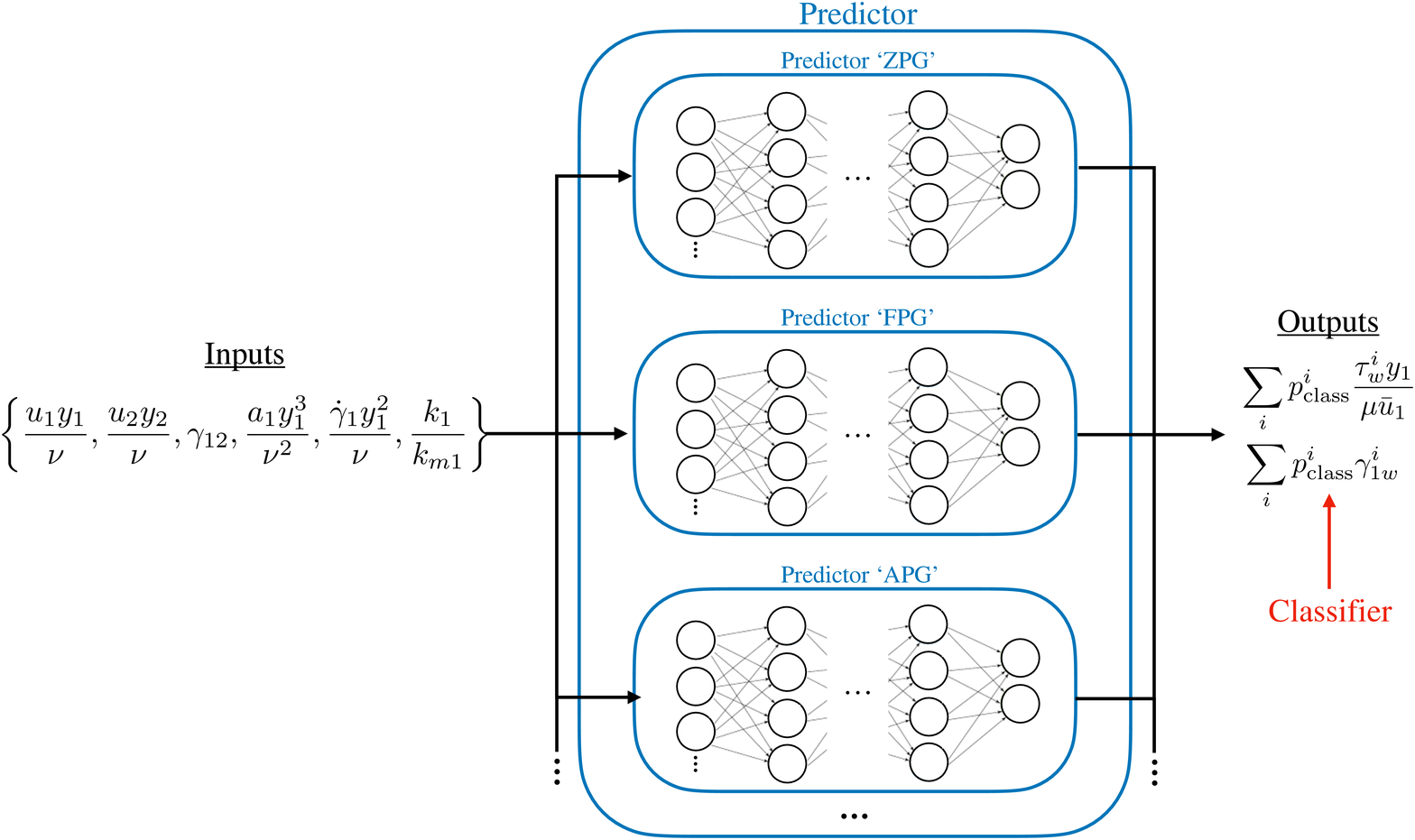}
  \end{center}
\caption{ \corr{Schematic of the predictor. The predictor consists of
    a collection of ANNs/analytical models specialised in predicting
    the wall-shear stress of each building-block flow. The final
    output is the sum of the predictions from each building-block flow
    weighted by the probability of each
    category.} \label{fig:ANN_predictors}}
\end{figure}
%
The ANNs are trained using Bayesian regularisation back propagation by
dividing the training data into two groups: the training set (80\% of
the data) and the test set (20\% of the data). The test set includes
full cases for Reynolds numbers (and $\Pi$ values) that are not part of the
training set. This was found to improve the predictive capability of
the network when interpolating and extrapolating among unseen
cases. The errors from each training sample in the lost function were
weighted to compensate for the different number of snapshots available
for each case. Figure \ref{fig:ANN_correlation} shows the performance
plot of the ANN trained for the Unsteady building-block flow.
%
\begin{figure}
    \vspace{0.5cm}
  \begin{center}
    \subfloat[]{\includegraphics[width=0.43\textwidth]{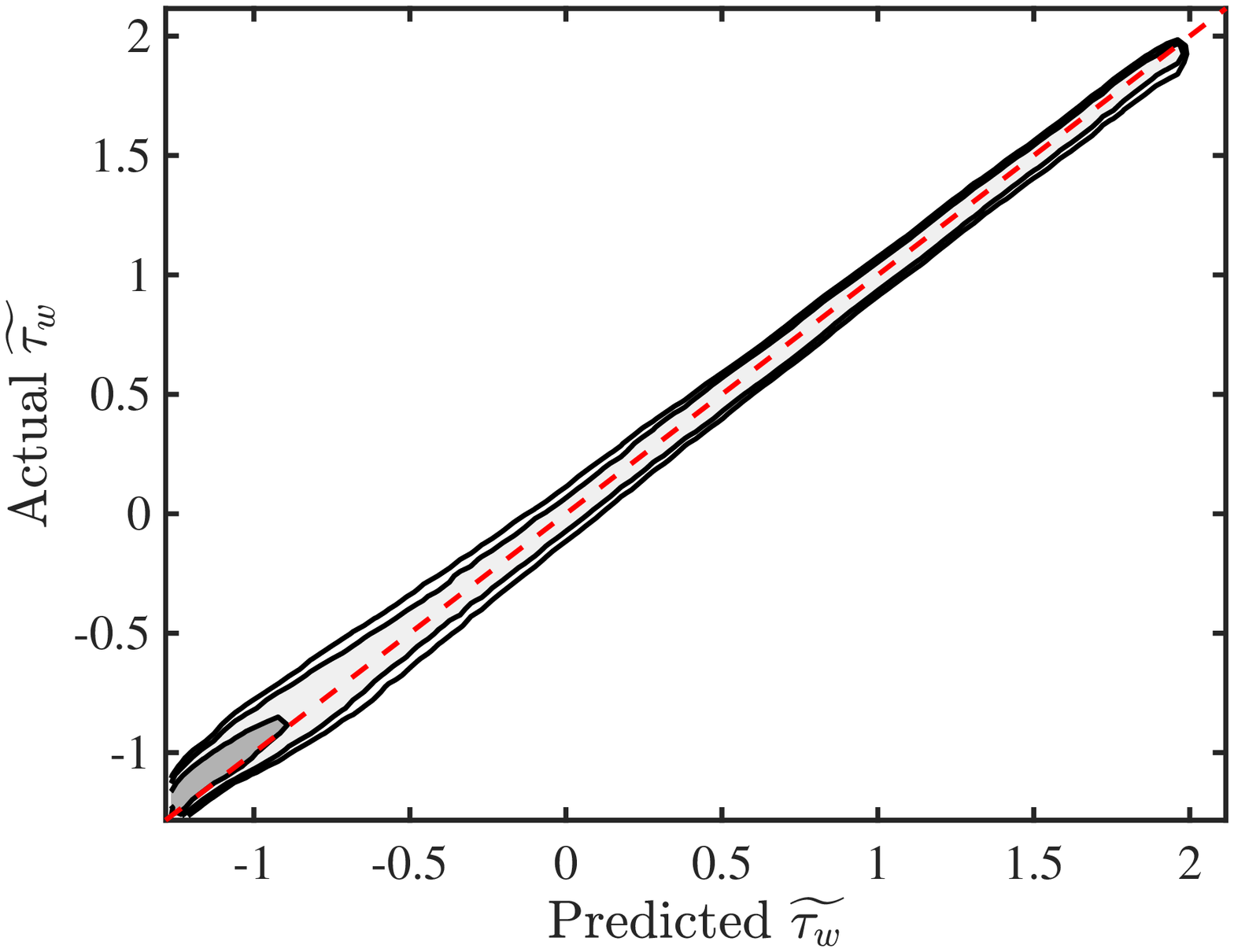}}
    \hspace{0.5cm}
    \subfloat[]{\includegraphics[width=0.43\textwidth]{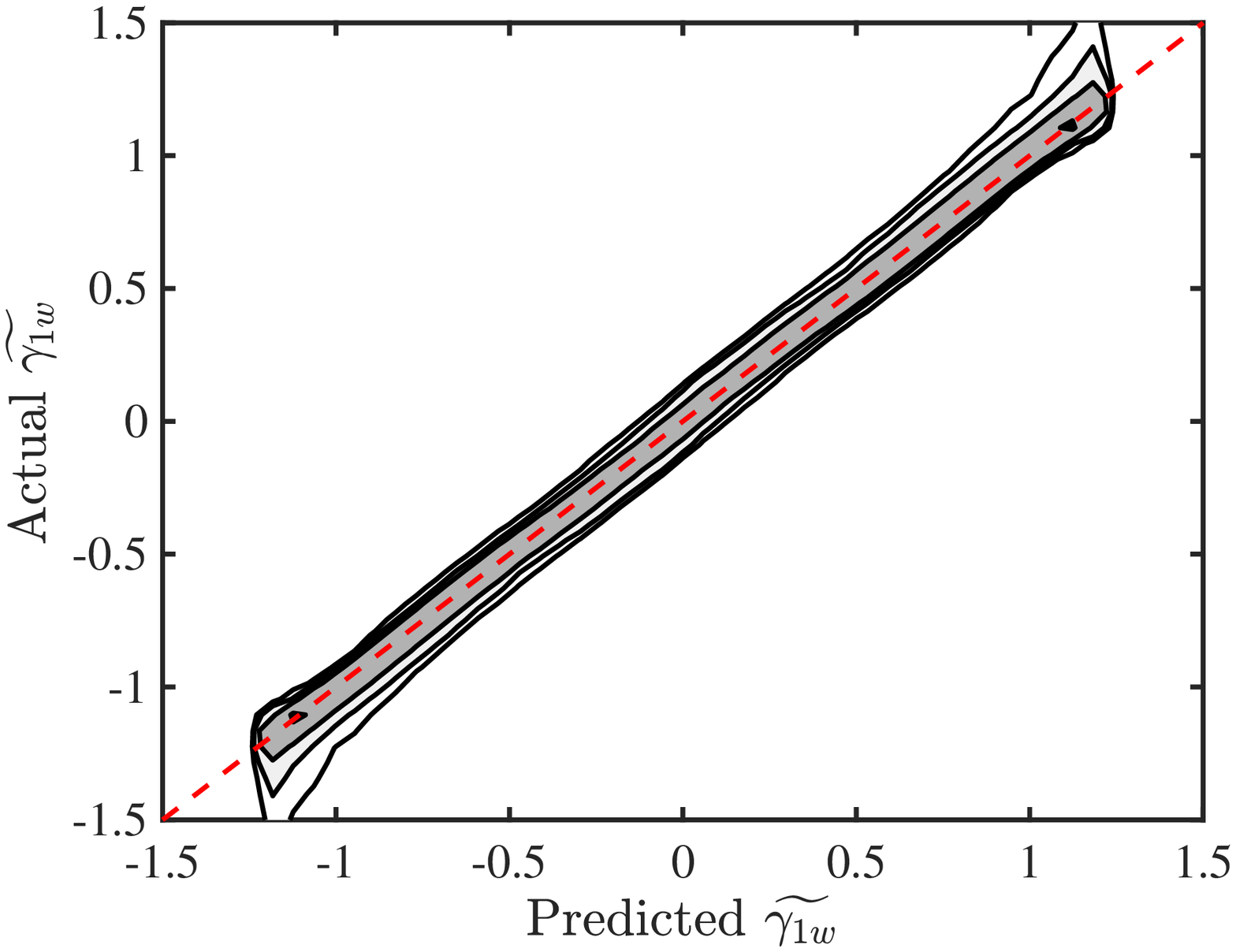}}
  \end{center}
\caption{ Joint p.d.f. of the predicted output
  and actual output for (a)  themagnitude of the wall stress
  $\widetilde{\tau_w}$, and (b) the relative angle
  $\widetilde{\gamma_{1w}}$. The tilde denotes values standardised by
  the mean and standard deviation of the training set. The results are
  for the statistically unsteady turbulent flow
  (Unsteady). \label{fig:ANN_correlation}}
\end{figure}
\end{itemize}


\corr{For both the classifier and predictor, the inputs and outputs
  are standardised using the mean and the standard deviation of the
  training set. These values are stored as part of the model and
  used to standardise the inputs and outputs when the model is
  deployed}.  A parametric study was performed in terms of the number
of layers and neurons per layer and the present layouts were found to
give a fair compromise between neural network complexity and
predictive capabilities. Finally, \corr{it took roughly 12 hours to
  train each ANN using 4 NVIDIA A100 with 40 GB of memory. The
  computational cost of the WMLES solver using the BFWM is between 1.1 and
  1.3 times more expensive than then same WMLES solver using the
  algebraic equilibrium wall model from \S\ref{sub:tra_models}.  The
  particular cost depends on the number of boundary nodes the BFWM is
  applied to. However, no effort was devoted to optimise the current
  implementation of the BFWM.}

\section{Numerical methods}\label{sec:methods}

\subsection{Flow solver and grid generation}

The simulations are conducted with the high-fidelity solver charLES
developed by Cascade Technologies, Inc.~\citep{Bres2018, Fu2021}. The
code integrates the compressible LES equations using a kinetic-energy-conserving, second-order-accurate, finite-volume method.  The
numerical discretisation relies on a flux formulation which is
approximately entropy-preserving in the inviscid limit, thereby
limiting the amount of numerical dissipation added into the
calculation. The time integration is performed with a third-order
Runge--Kutta explicit method.

The mesh generation follows a Voronoi hexagonal close-packed (HCP)
point-seeding method, which automatically builds locally isotropic
meshes for arbitrarily complex geometries. First, the watertight
surface geometry is provided to describe the computational
domain. Second, the coarsest grid resolution in the domain is set to
uniformly seeded HCP points. Additional refinement levels are
specified in the vicinity of the walls if needed.  Ten iterations of
Lloyd's algorithm smooth the transition between layers with
different grid resolutions.

\subsection{Traditional SGS/wall models}
\label{sub:tra_models}

Two SGS models are considered: the dynamic Smagorinsky model
\citep{Germano1991} with the modification by \citet{Lilly1992} (DSM),
and the Vreman model~\citep{Vreman2004} (VRE).
For the wall, we use a traditional equilibrium wall model (EQWM).  The
no-slip boundary condition at the wall is replaced by a wall stress
boundary condition.  The walls are assumed adiabatic, and the wall
stress is obtained by an algebraic equilibrium wall model derived from
the integration of the one-dimensional stress model along the
wall-normal direction~\citep{Deardorff1970, Piomelli1989}:
\begin{equation}
  u_{2}^+(y_\perp^+) =
  \begin{cases}
    y_\perp^+ + a_1 (y_\perp^{+})^2 \text{\, \, \, for $y_\perp^+ < 23$}, \\
    \frac{1}{\kappa}\ln{y_\perp^+} + B \text{\, \, \, \, otherwise}
  \end{cases}
  \label{eq:charles_algwm}
\end{equation}
where $u_{2}$ is the model wall-parallel velocity at the second grid
point off the wall, $y_\perp$ is the wall-normal direction to the
surface, $\kappa=0.41$ is the K\'arm\'an constant, $B = 5.2$ is the
intercept constant, and $a_1$ is computed to ensure $C^1$
continuity. The superscript $+$ denotes inner units defined in terms
of wall friction velocity and $\nu$.

\section{Model validation}\label{sec:validation}

The BFWM is first validated in canonical flows: laminar and turbulent
boundary layers with zero mean pressure gradient, turbulent pipe flow,
turbulent Poiseuille--Couette flows, and statistically unsteady
turbulent channel flow. These cases are intended to test the
predictive capabilities of the BFWM in flows that are comparable (but not
identical) to the training and test sets. The performance of the BFWM to
predict the wall-shear stress in complex scenarios is evaluated in two
realistic aircraft configurations: the NASA Common Research Model
High-lift and the NASA Juncture Flow experiment.

The validation cases are conducted by combining BFWM with two SGS
models: DSM and VRE. The simulations performed with DSM and VRE
together with BFWM are denoted by DSM-BFWM and VRE-BFWM,
respectively. The results are compared with WMLES using DSM and VRE in
conjunction with the equilibrium wall model (EQWM). The last two cases
are referred to as DSM-EQWM and VRE-EQWM, respectively. When possible,
we also include the results for ESGS combined with BFWM (referred to
as ESGS-BFWM). The latter aims at assessing the performance of the
BFWM in the absence of external SGS modelling errors, as discussed in
the following section.

\subsection{External versus internal wall-modelling errors}

Two sources of errors can be identified in a wall
model~\citep{Lozano2022}: errors from the outer LES input data,
referred to as external wall-modelling errors, and errors from the
wall model physical assumptions, referred to as internal wall-modelling
errors. In the former, errors by the SGS model at the matching
locations propagate to the value of $\tau_w$ predicted by the wall
model.  These errors can be labelled as external to the wall model
inasmuch as they are present even if the wall model provides an exact
physical representation of the near-wall region. The second source of
errors represents the intrinsic wall model limitations: even in the
presence of exact values for the input data, the wall model might
incur errors when the physical assumptions of the model do not
hold. In the BFWM, internal errors may come from the inadequacy of the
building-block flows to represent the physics of the near-wall region
(e.g., compressibility effects, heat transfer effects, separation
differing from the physics modelled by Poiseuille--Couette flows). A consequence of internal errors is that WMLES might not
converge to the DNS solution with grid refinements until the grid is
in the DNS-like regime, when the contribution of the wall model is
negligible. The combined external plus internal error is referred to
as total error. We use ESGS-BFWM to isolate the internal
wall-modelling errors when possible.

\subsection{Laminar boundary layer}

The set-up for the laminar boundary layer is the flow over a flat plate
with zero mean pressure gradient and imposed inflow velocity from the
Blasius solution. At the top boundary, the freestream velocity is
$U_\infty$ and the vertical velocity is also calculated from the
Blasius solution. A convective boundary condition is used at the
outflow. The range of Reynolds numbers is from $\Rey_x = U_\infty
L_0/\nu = 10^4$ to $\Rey_x = 10^6$, where $L_0$ is the distance of the
inlet to the leading edge of the plate. The streamwise, wall-normal
and spanwise sizes of the domain are $ 150 L_0 \times 2 L_0 \times 1
L_0$, respectively. The grid resolution is uniform in the three
directions, with $\Delta/\delta_i \approx 1/2$ at the inlet, which
corresponds to $\Delta/\delta_o \approx 1/20$ at the outlet, where
$\delta_i$ and $\delta_o$ are the boundary layer thicknesses based on
99\% of $U_\infty$ at the inlet and outlet, respectively. The
reference solution was computed by DNS of the same set-up.

\begin{figure}
    \vspace{0.5cm}
  \begin{center}
    \subfloat[]{\includegraphics[width=0.9\textwidth]{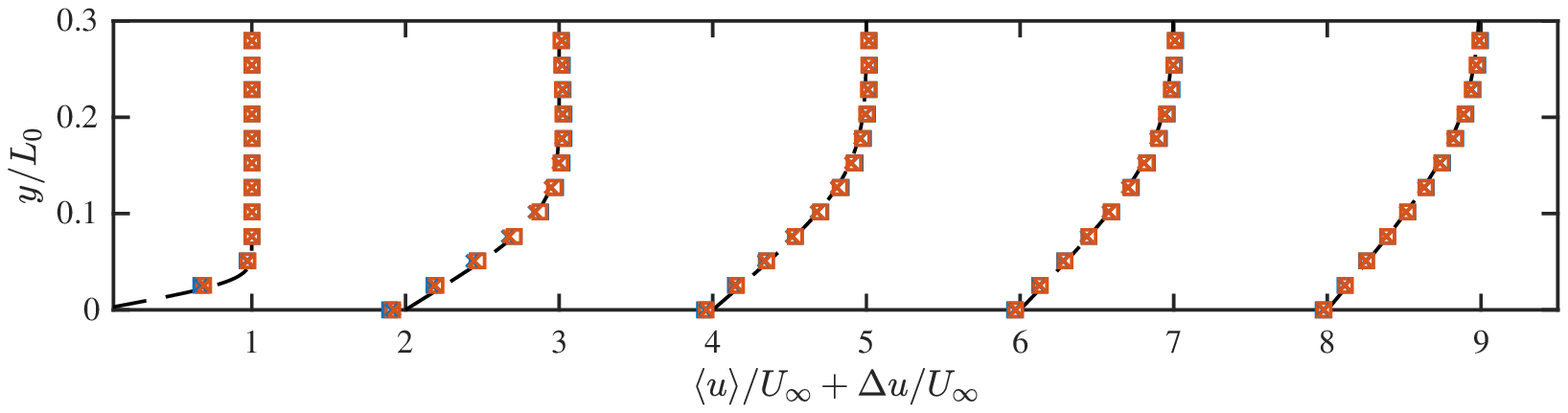}}
  \end{center}
  \begin{center}
    \subfloat[]{\includegraphics[width=0.325\textwidth]{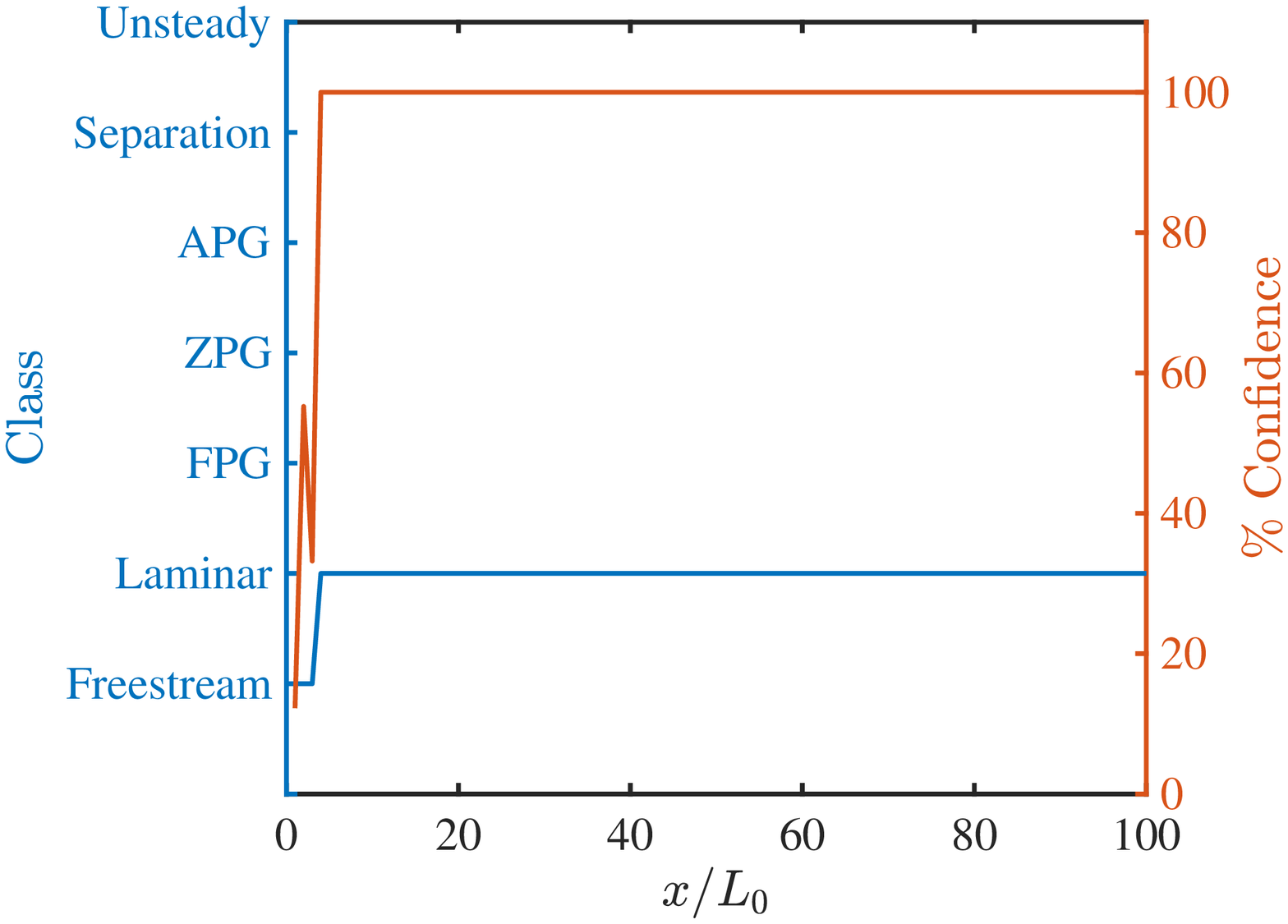}}
    \hspace{0.05cm}
    \subfloat[]{\includegraphics[width=0.305\textwidth]{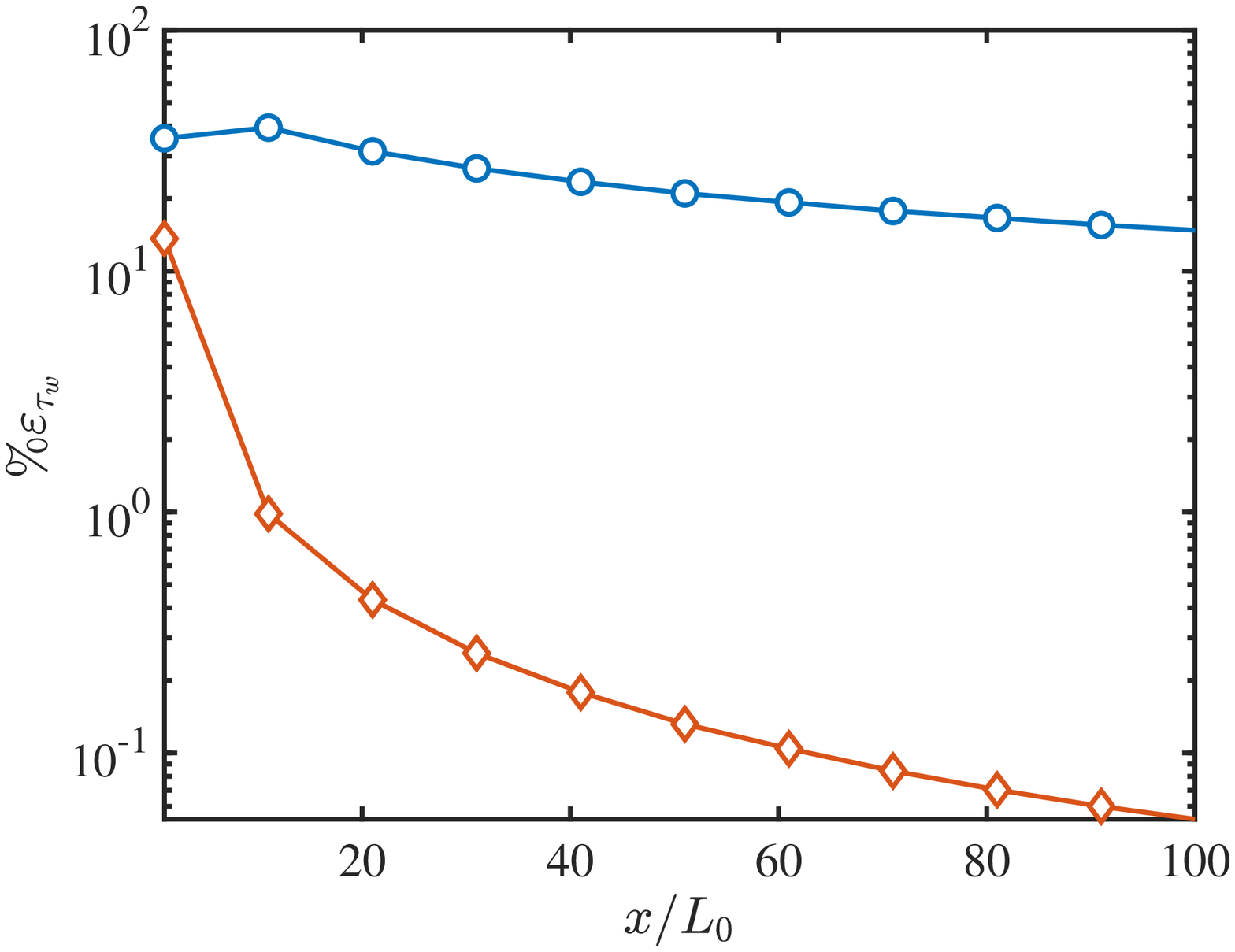}}
    \hspace{0.05cm}
    \subfloat[]{\includegraphics[width=0.305\textwidth]{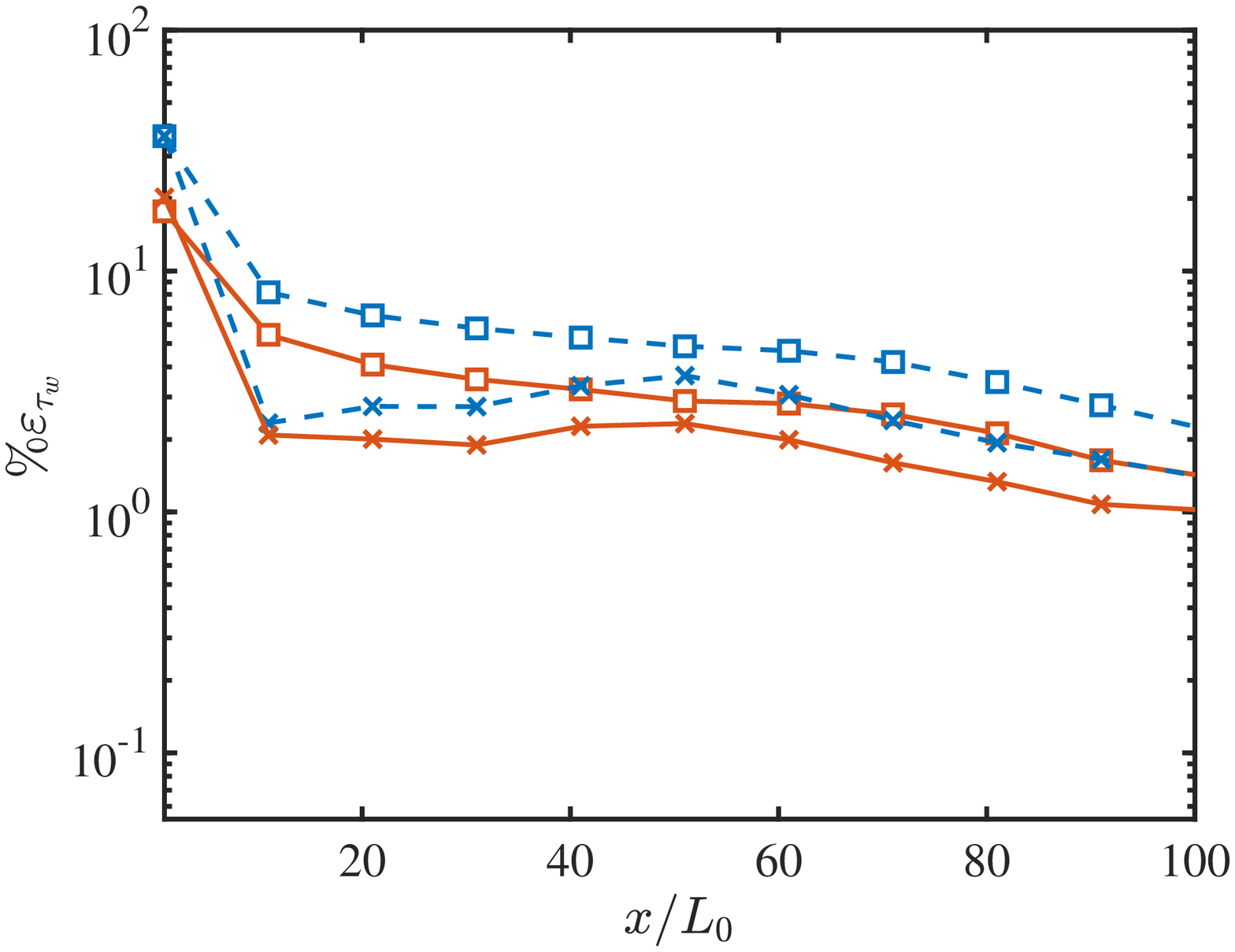}}
  \end{center}
\caption{ Validation case: laminar boundary layer. (a) The streamwise
  mean velocity profiles at $x/L_0 = 1, 11, 21, 31$ and $41$ for
  the DSM-EQWM (blue squares), VRE-EQWM (blue crosses), DSM-BFWM (red
  squares), VRE-BFWM (red crosses), and DNS (solid line). The mean
  velocity profiles are shifted by $\Delta u/U_\infty = (x/L_0-1)/5$. (b) Dominant flow classification (solid blue) and confidence score (solid red) by
  the DSM-BFWM as a function of the streamwise distance. (c) Internal
  wall-modelling error of the wall stress prediction for the ESGS-BFWM
  (diamonds) and ESGS-EQWM (circles) as a function of the streamwise
  distance. (d) Total wall-modelling error for the DSM-EQWM (dashed
  line and squares), DSM-BFWM (solid line and squares), VRE-EQWM (dashed line and crosses), and
  VRE-BFWM (solid line and crosses) as a function of the streamwise
  distance. \label{fig:validation_laminar_BL}}
\end{figure}

The streamwise mean velocity profiles are captured within 5\% error
for DSM-EQWM, VRE-EQWM, DSM-BFWM, and VRE-BFWM as shown in figure
\ref{fig:validation_laminar_BL}(a).  Figure
\ref{fig:validation_laminar_BL}(b) reveals that the flow is correctly
classified as Laminar with confidence near 100\%. The only exception
is a small region close to the inlet, where the flow is labelled as
Freestream with lower confidence owing to the lack of grid resolution
in that region (there are only two points per boundary layer thickness at the
inlet).  The internal wall model errors are included in figure
\ref{fig:validation_laminar_BL}(c), which shows a clear improvement in
performance by the BFWM compared to the EQWM. After the initial transient from
the inflow, errors for the BFWM are below 1\% and decay faster than
errors for the EQWM, which are always above 20\%. The total wall-stress
error is evaluated in figure \ref{fig:validation_laminar_BL}(d). The
performance of the BFWM is still superior to that of the EQWM. However, the
accuracy of the BFWM is severely diminished due to the external
wall-modelling errors from the DSM and VRE. Figure
\ref{fig:validation_laminar_BL}(d) also reveals that the DSM-EQWM and
VRE-EQWM are subjected to internal/external error cancellation (i.e.,
predictions may improve despite the fact that input values differ from
ESGS). The degraded accuracy of the BFWM due to SGS model errors and the
presence of error cancellation using the EQWM are recurrent
observations in the following validation cases.
%

\subsection{Zero mean pressure gradient turbulent boundary layer}

For the zero pressure gradient turbulent boundary layer case, the
range of $\Rey_\theta$ for the turbulent boundary layer is from 800 to
2,000, where $\Rey_\theta$ is the Reynolds number based on $U_\infty$
and the momentum thickness ($\theta$). The length, height and width
of the simulated box are $L_x=1060\theta_\text{avg}$,
$L_y=18\theta_\text{avg}$, and $L_z=35\theta_\text{avg}$, where
$\theta_\mathrm{avg}$ denotes $\theta$ averaged along the streamwise
coordinate.  The boundary conditions at the top plane are $u =
U_\infty$, $w=0$, and $v$ is estimated from the known experimental
growth of the displacement thickness for the corresponding range of
Reynolds numbers as in \citet{Jimenez2010}.  This controls the average
streamwise pressure gradient, whose nominal value is set to zero. The
turbulent inflow is generated by the recycling scheme of
\citet{Lund1998}, in which the velocities from a reference downstream
plane, $x_\text{ref}$, are used to synthesise the incoming
turbulence. The reference plane is located well beyond the end of the
inflow region to avoid spurious feedback~\citep{Nikitin2007,
  Simens2009}. In our case, $x_\text{ref}/\theta_i = 890$, where
$\theta_i$ is the momentum thickness at the inlet.  A convective
boundary condition is applied at the outlet with convective velocity
$U_\infty$~\citep{Pauley1990} and small corrections to enforce global
mass conservation~\citep{Simens2009}. The spanwise direction is
periodic. The grid resolution is uniform in the three
directions. Compared to the boundary layer thickness, the grid size is $\Delta/\delta_i \approx 1/5$  at
the inlet and
$\Delta/\delta_o \approx 1/20$ at the outlet. The reference DNS solution is from
\citet{Towne2022}, which follows an identical numerical set-up.

\begin{figure}
    \vspace{0.5cm}
  \begin{center}
    \subfloat[]{\includegraphics[width=0.9\textwidth]{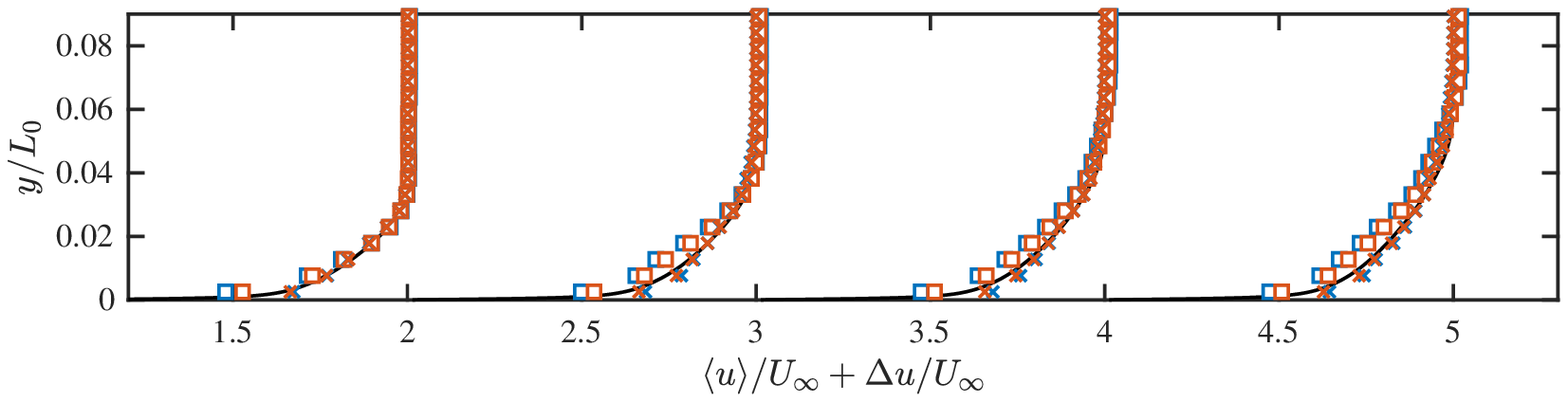}}
    \hspace{0.5cm}
  \end{center}
  \begin{center}
    \subfloat[]{\includegraphics[width=0.325\textwidth]{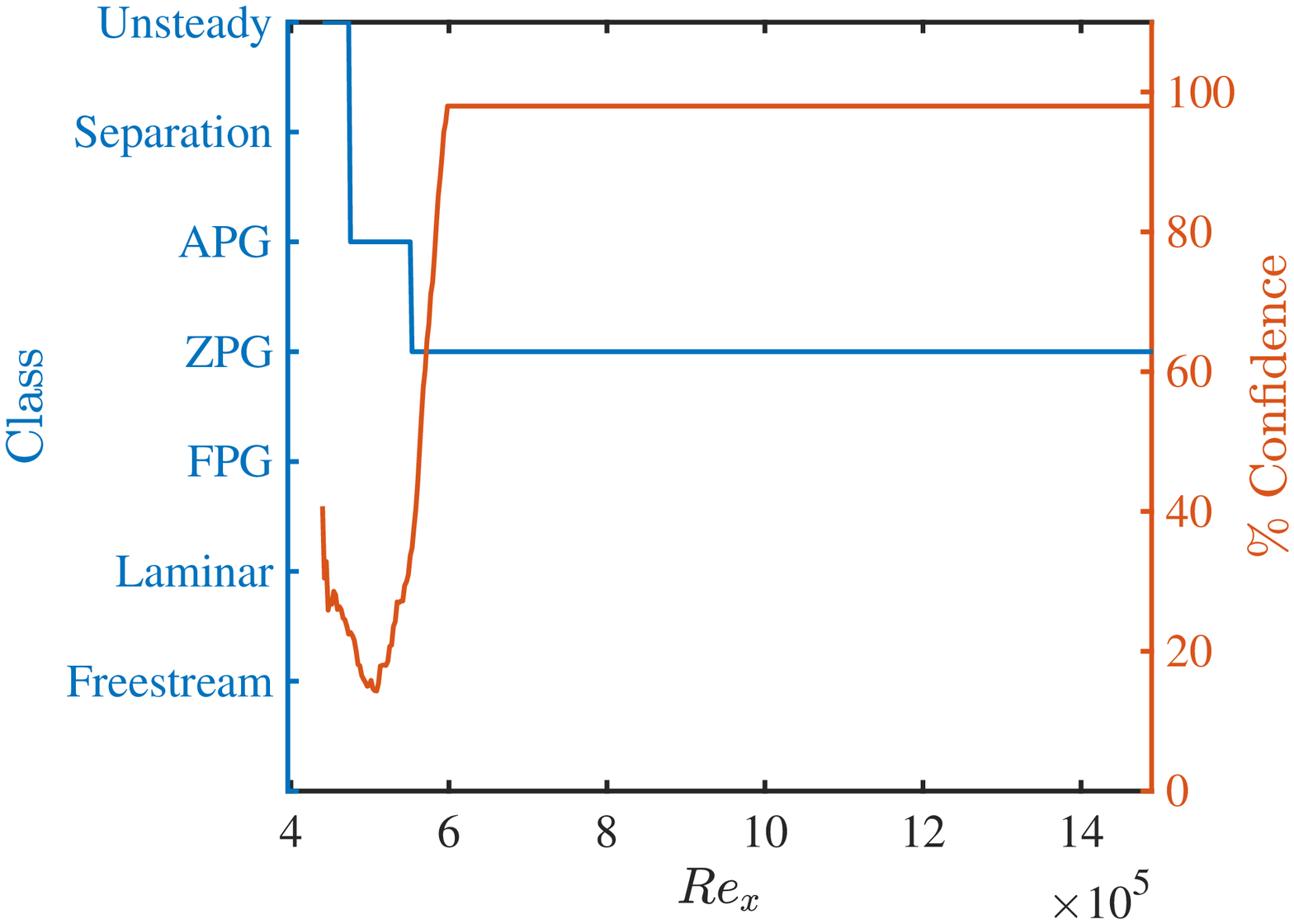}}
    \hspace{0.05cm}
    \subfloat[]{\includegraphics[width=0.303\textwidth]{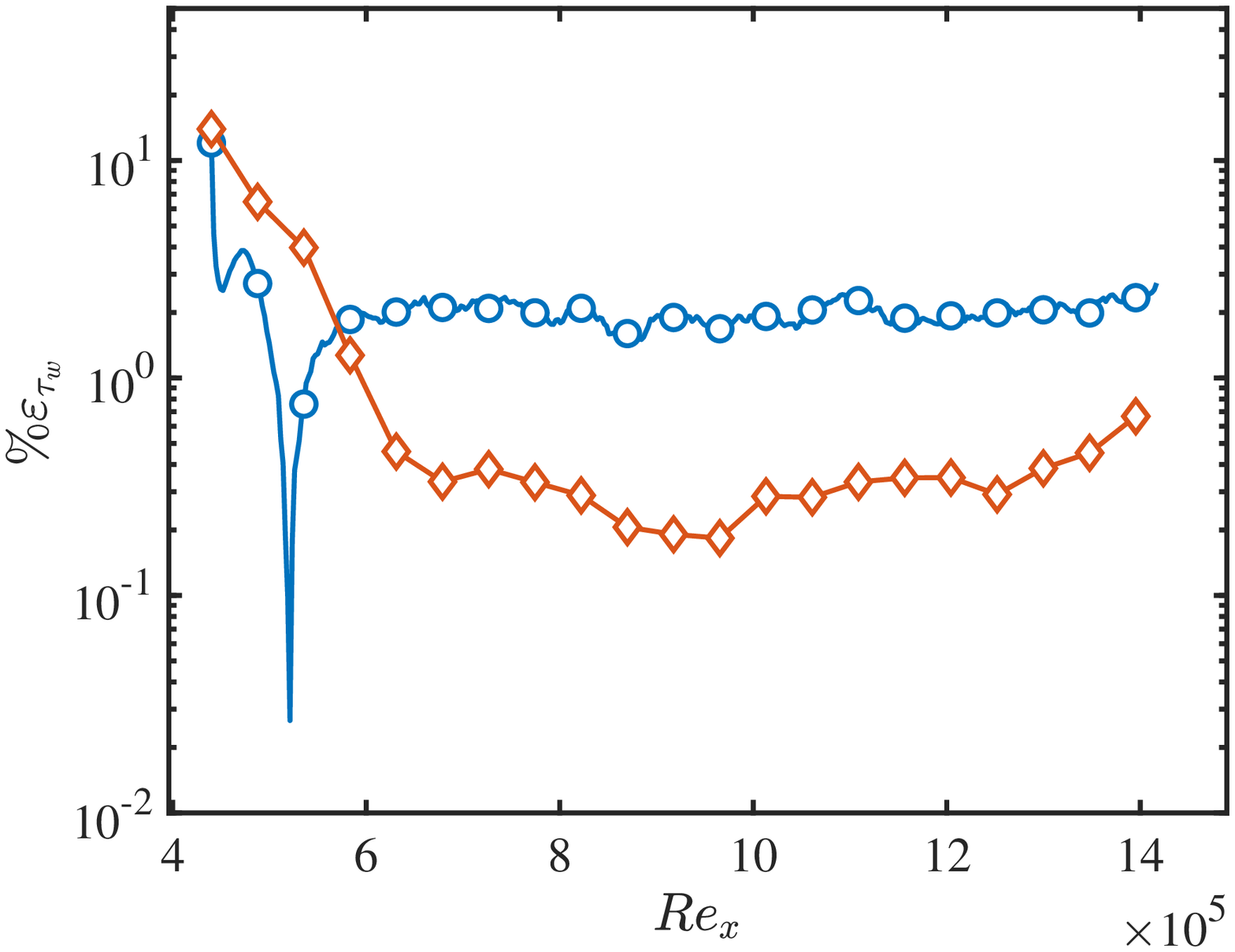}}
    \hspace{0.05cm}
    \subfloat[]{\includegraphics[width=0.303\textwidth]{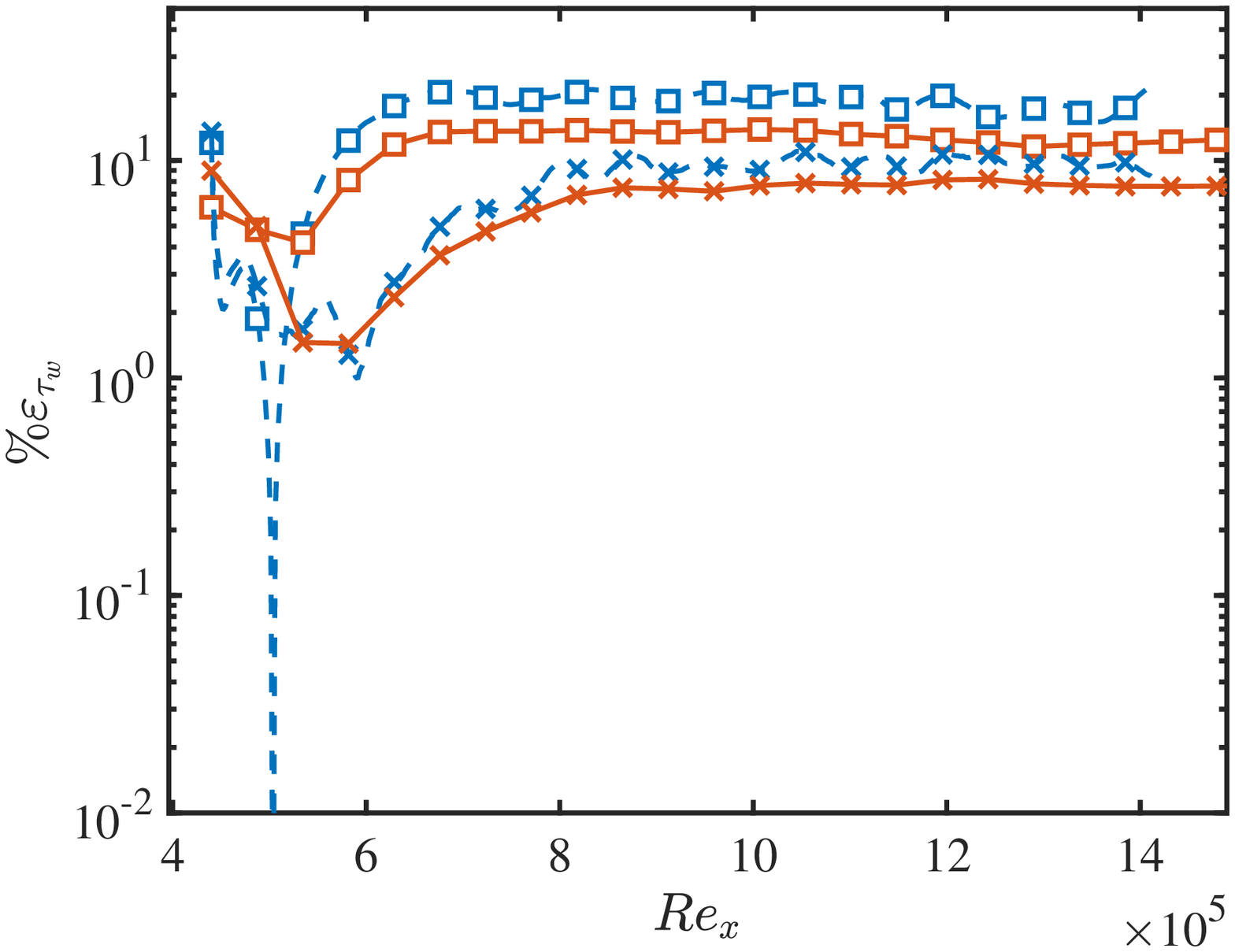}}
  \end{center}
\caption{ Validation case: zero mean pressure gradient turbulent
  boundary layer.  (a) The streamwise mean velocity profiles at
  $\Rey_x = 6\times 10^5, 8\times 10^5, 10\times 10^5$ and $12\times
  10^5$, for the DSM-EQWM (blue squares), VRE-EQWM (blue crosses),
  DSM-BFWM (red squares), VRE-BFWM (red crosses) and DNS (solid
  line). The mean velocity profiles are shifted by $\Delta u/U_\infty =
  (x/L_0-1)/5$. (b) Dominant flow classification (solid blue) and confidence
  score (solid red) by the DSM-BFWM as a function of the streamwise
  distance. (c) Internal wall-modelling error of the wall stress
  prediction for the ESGS-BFWM (diamonds) and ESGS-EQWM (circles) as a
  function of the streamwise distance. (d) Total wall-modelling error
  for the DSM-EQWM (dashed line and squares), DSM-BFWM (solid line and squares), VRE-EQWM
  (dashed line and crosses), and VRE-BFWM (solid line and crosses) as a function of the
  streamwise distance.  \label{fig:validation_TBL}}
\end{figure}

To facilitate the comparison between DNS and WMLES, we use as
independent variables $\Rey_x$ and $x/L_0$, as these are free
from modelling errors. The streamwise mean velocity profiles (figure
\ref{fig:validation_TBL}a) deviate about 10\% to 30\% from the DNS
profile.  The flow is correctly classified by the BFWM as ZPG with
almost 100\% confidence across the vast majority of the domain (figure
\ref{fig:validation_TBL}b). Close to the inlet, the flow is labelled as
Unsteady and APG, probably because it is still recovering from
the artificial inlet boundary condition.  
\corr{From the inlet, the confidence first drops from 40\% to 20\%, and
  then recovers up to 100\%. In particular, the flow deccelerates
  close to the inlet under an adverse mean pressure gradient, as seen
  in the classification of Unsteady and APG.  This condition is not
  explicitly contained in the training set.  Therefore, the drop in
  confidence occurs because the input data to the model move far away
  from the input data in the training set. After the transient, the
  flow approaches equilibrium under zero mean pressure gradient,
  resembling a ZPG turbulent boundary layer. At this point, the flow
  is identified as ZPG and the confidence increases close to 100\%.}
The internal wall-modelling errors are quantified in figure
\ref{fig:validation_TBL}(c), which shows moderate improvements in the
performance of the BFWM compared to the EQWM. In particular, the averaged
wall stress error is below 0.5\% for the BFWM and about 2\% for the EQWM once
the effect of the inlet condition is forgotten. This case demonstrates
that the BFWM is able to successfully interpolate among grid resolutions
and Reynolds numbers.  In terms of total wall stress error (figure
\ref{fig:validation_TBL}d), the BFWM and EQWM exhibit comparable
performance, with errors between 5\% and 15\%. Similar to the previous
case, the accuracy of the BFWM worsens due to the external wall-modelling
errors when the wall model is coupled with the DSM and VRE.
%

\subsection{Turbulent pipe flow at $\Rey_\tau=40,000$}

The next validation case is a turbulent pipe flow at $\Rey_\tau
\approx 40,000$ with reference data obtained from experiments by
\citet{Baidya2019}. The flow is driven by a mean streamwise pressure
gradient that fixes the Reynolds number based on the centreline
velocity to the correct experimental value. The radial coordinate is
$r$. The length of the pipe is $100R$, where $R$ is the radius of the
pipe, and the streamwise direction is periodic. Four grid resolutions
are considered: $\Delta/R \approx 1/5, 1/10, 1/20,$ and $1/40$.

\begin{figure}
    \vspace{0.5cm}
  \begin{center}
    \subfloat[]{\includegraphics[width=0.41\textwidth]{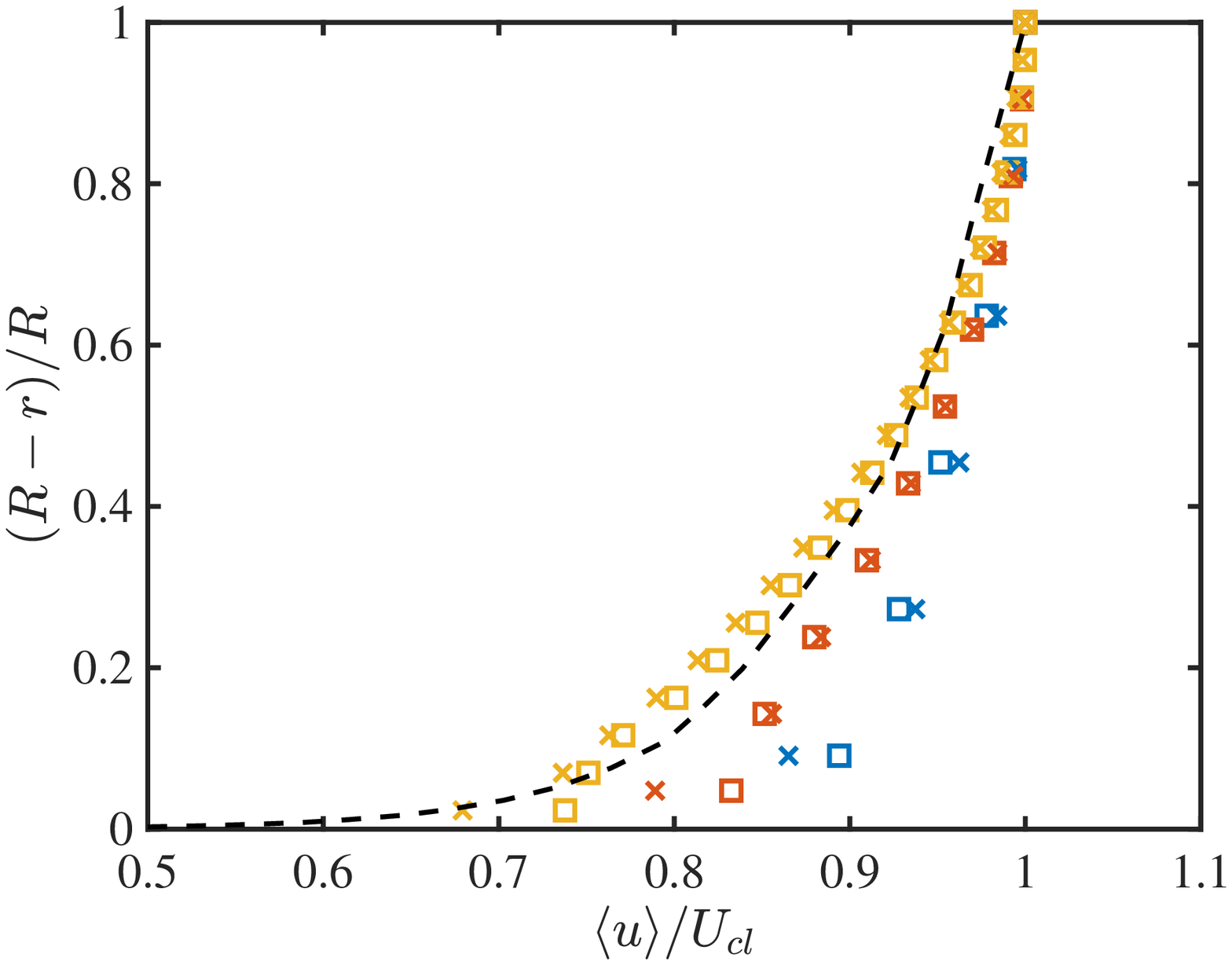}}
    \hspace{0.05cm}
    \subfloat[]{\includegraphics[width=0.45\textwidth]{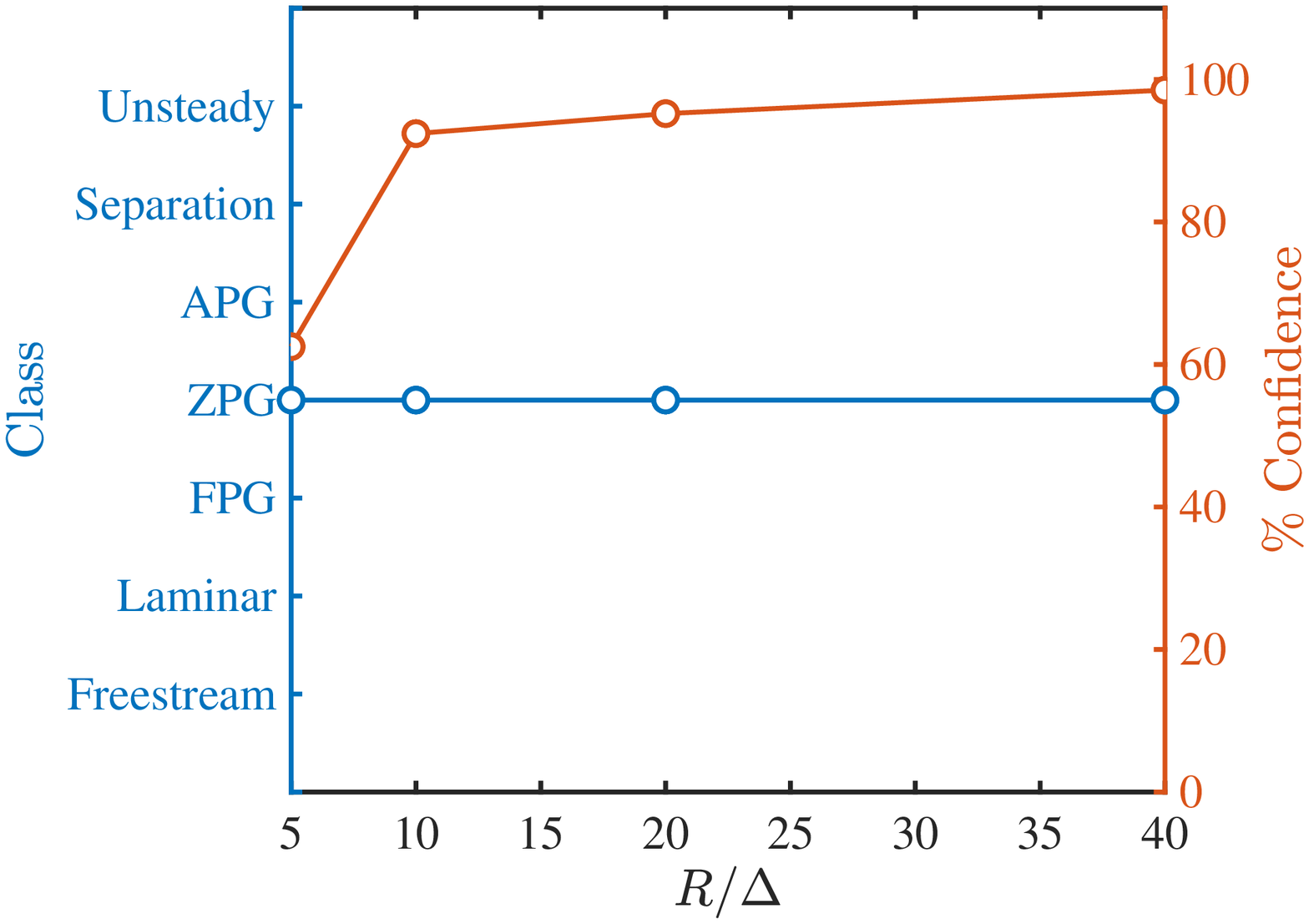}}
  \end{center}
  \begin{center}
    \subfloat[]{\includegraphics[width=0.41\textwidth]{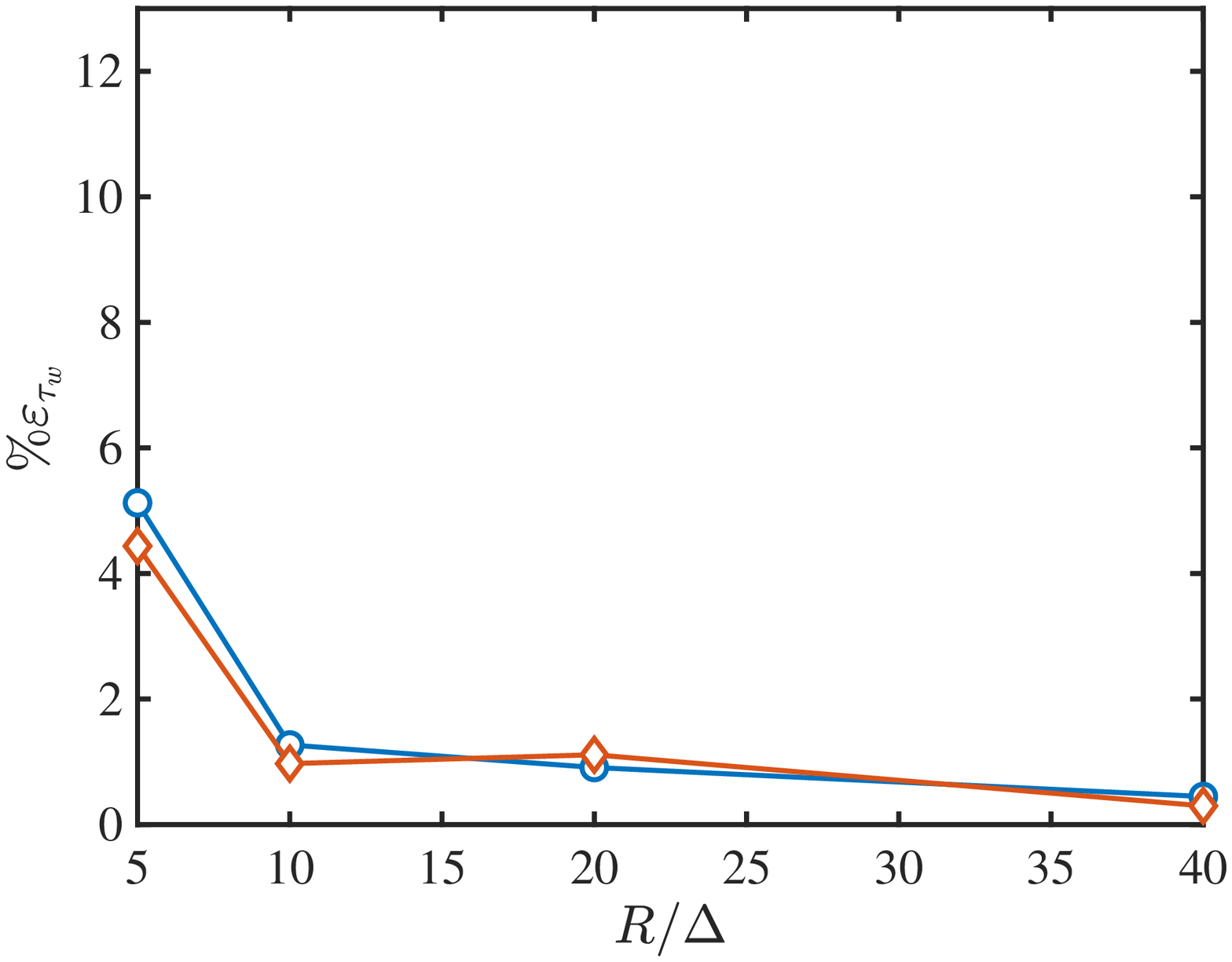}}
    \hspace{0.15cm}
    \subfloat[]{\includegraphics[width=0.41\textwidth]{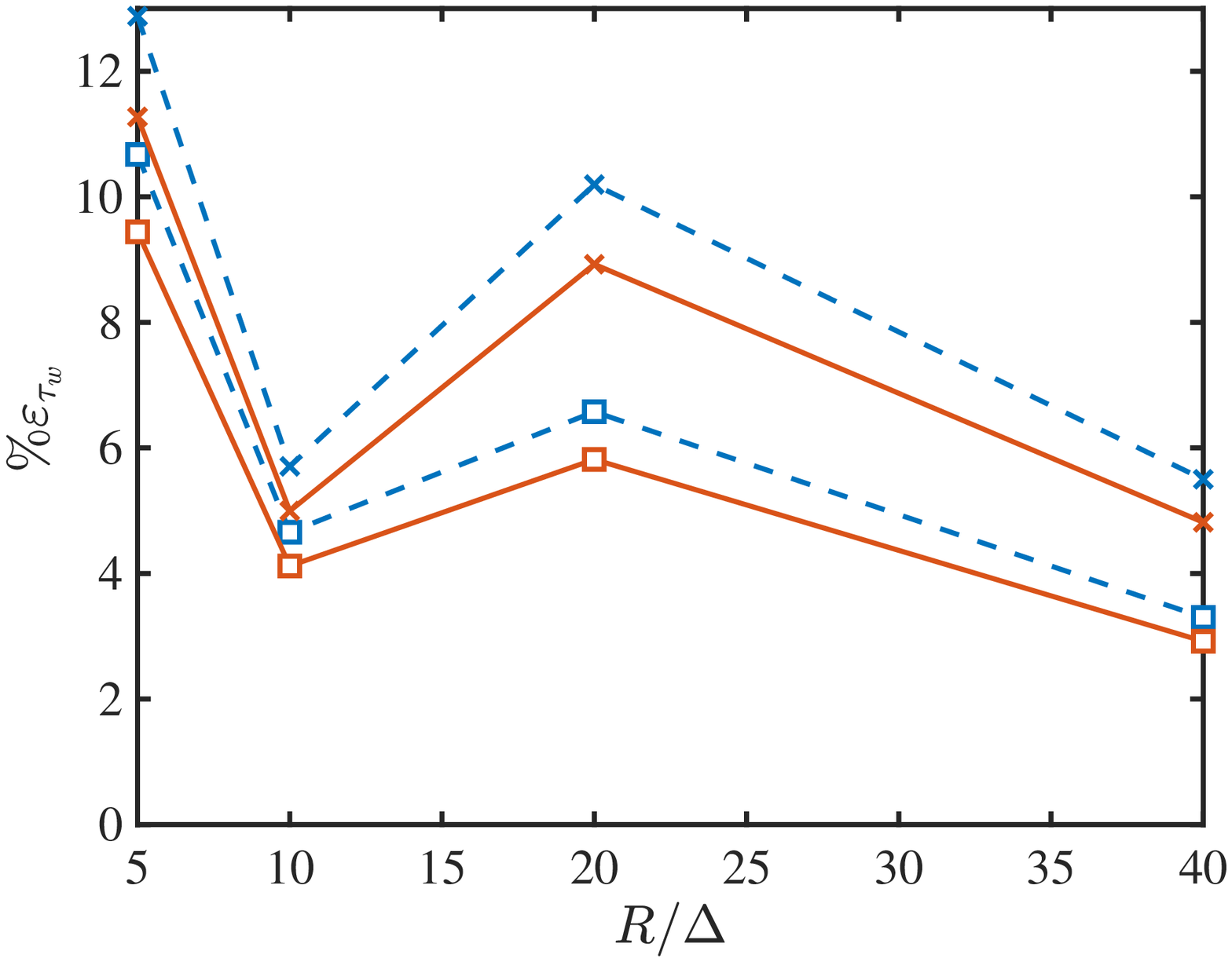}}
  \end{center}
  \caption{ Validation case: turbulent pipe flow at
    $\Rey_\tau=40,000$.  (a) The streamwise mean velocity
    profiles non-dimensionalised by the mean centreline velocity
    ($U_{cl}$) for the DSM-BFWM (squares) and VRE-BFWM (crosses) for
    $R/\Delta \approx$ 5 (blue), 10 (red) and 20 (yellow). The dashed
    line is the experimental mean velocity profile from
    \citet{Baidya2019}.  (b) Dominant flow classification (solid blue) and
    confidence score (solid red) by the DSM-BFWM as a function of the grid
    resolution. (c) Internal wall-modelling error of the wall stress
    prediction for the ESGS-BFWM (diamonds) and ESGS-EQWM (circles) as a
    function of the grid resolution. (d) Total wall-modelling error for
    the DSM-EQWM (dashed line and squares), DSM-BFWM (solid line and squares), VRE-EQWM
    (dashed line and crosses), and VRE-BFWM (solid line and crosses) as a function of
    the grid resolution. \label{fig:validation_pipe}}
\end{figure}

Figure \ref{fig:validation_pipe}(a) shows that the streamwise mean
velocity profiles for the DSM-BFWM and the VRE-BFWM approach the reference
data with grid refinement. The flow is accurately classified as ZPG
for all the grid resolutions (figure \ref{fig:validation_pipe}b).
The BFWM correctly identifies the flow as ZPG with high confidence
($>$95\%) for $R/\Delta>5$. For the coarsest grid
($R/\Delta\approx5$), the flow is also labelled as ZPG but with lower
confidence ($<$60\%).
\corr{The latter is the consequence of the poorer performance of the DSM
  in the coarsest grid resolution, which results in a near-wall mean
  shear lower than expected from the training set.}
The internal wall-modelling errors are shown in figure
\ref{fig:validation_pipe}(c). Overall, both the BFWM and the EQWM perform
satisfactorily, with errors below 5\% even at the coarsest grid
resolution. This validation case demonstrates that the BFWM successfully
extrapolates to high Reynolds numbers for different grid
resolutions. The total wall stress error (figure
\ref{fig:validation_pipe}d) ranges from 4\% to 12\% for the BFWM and the EQWM
regardless of the SGS model, with a mildly improved accuracy for the BFWM.
The results also exhibit the non-monotonic convergence to the true
solution typically observed for WMLES in the range of grid resolution
considered. This effect, that remains a key issue of WMLES, can be
ascribed to the interplay between numerical errors and SGS modelling
errors as discussed in \citet{Lozano2019a} and \citet{Lozano2022}.
%

\subsection{Adverse/favourable mean pressure gradient Poiseuille--Couette turbulence}

Turbulent Poiseuille--Couette flows at $\Rey_U=14,000$ are tested for
adverse mean pressure gradients until separation and beyond for
$\Rey_P\times 10^{-3} =$ -1.40, -1.32, -1.23, -1.13, -1.02, -0.90,
-0.76 and $\Rey_U= 14,000$ and for favourable mean pressure gradients
for $\Rey_P \times 10^{-3}= 1.02, 1.13, 1.23, 1.32, 1.41, 1.49$.  In
all cases, the mean streamwise pressure gradient ensures that the
Reynolds number based on the centreline velocity matches the correct
value. The grid resolution is $h/\Delta = 10$. The reference solutions
were obtained by DNS. It is worth remarking that these validation
cases are not included in the training or testing set discussed in
\S\ref{subsec:training_data}. Hence the current cases assess the
predictive capabilities of the BFWM when both $\Rey_P$ and $\Rey_U$ are
simultaneously varied.

\begin{figure}
    \vspace{0.5cm}
  \begin{center}
    \subfloat[]{\includegraphics[width=0.41\textwidth]{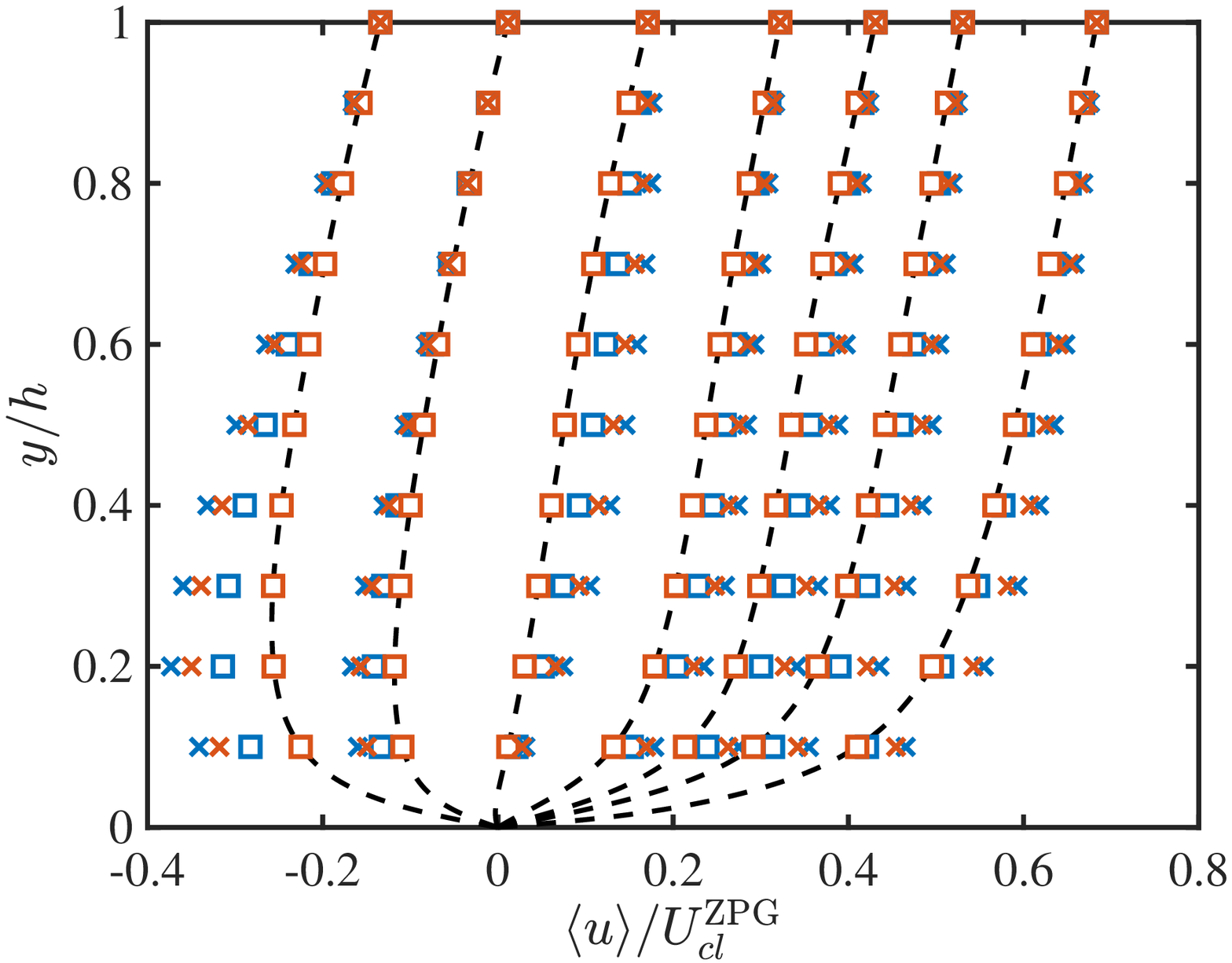}}
    \hspace{0.05cm}
    \subfloat[]{\includegraphics[width=0.45\textwidth]{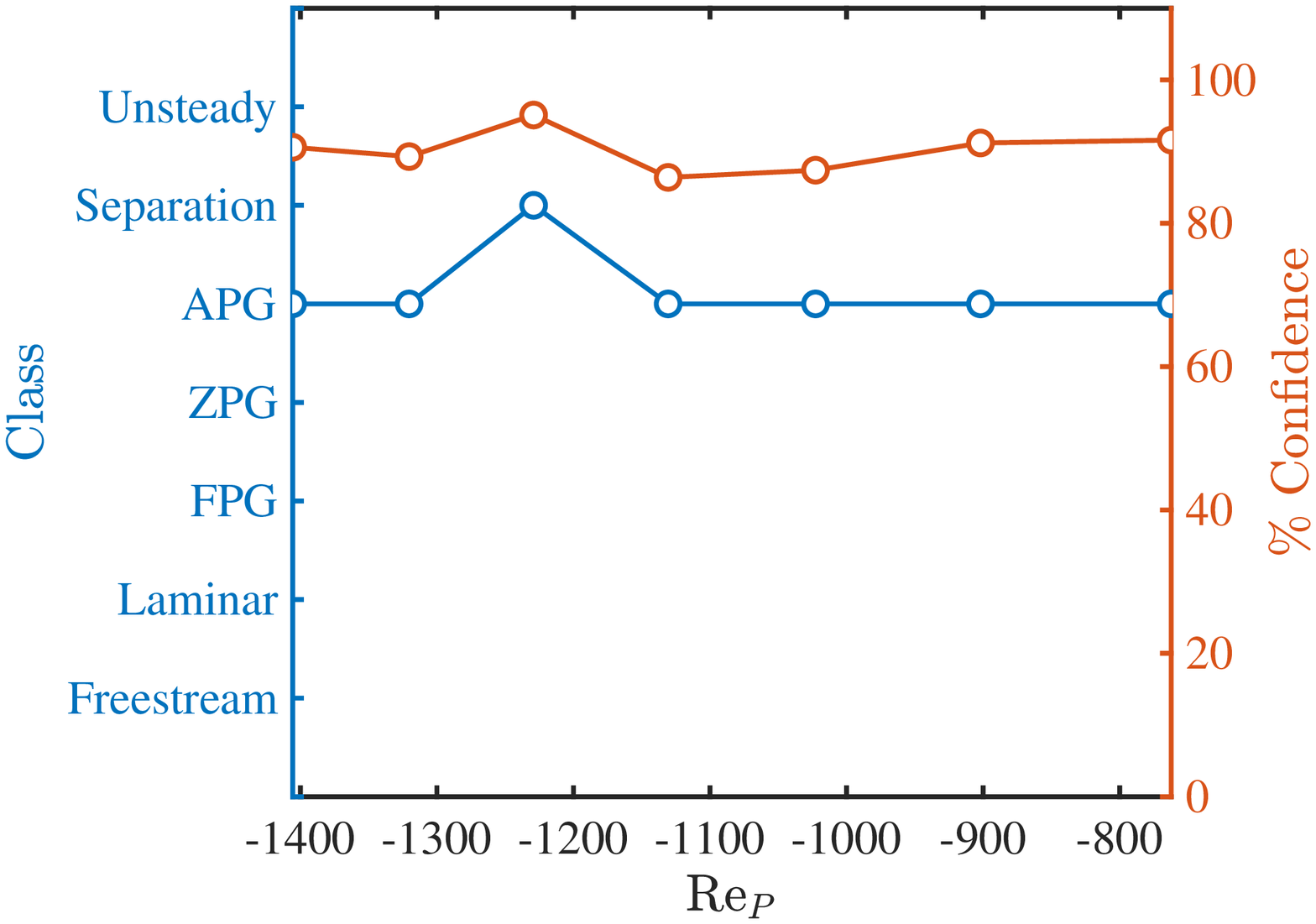}}
  \end{center}
  \begin{center}
    \subfloat[]{\includegraphics[width=0.41\textwidth]{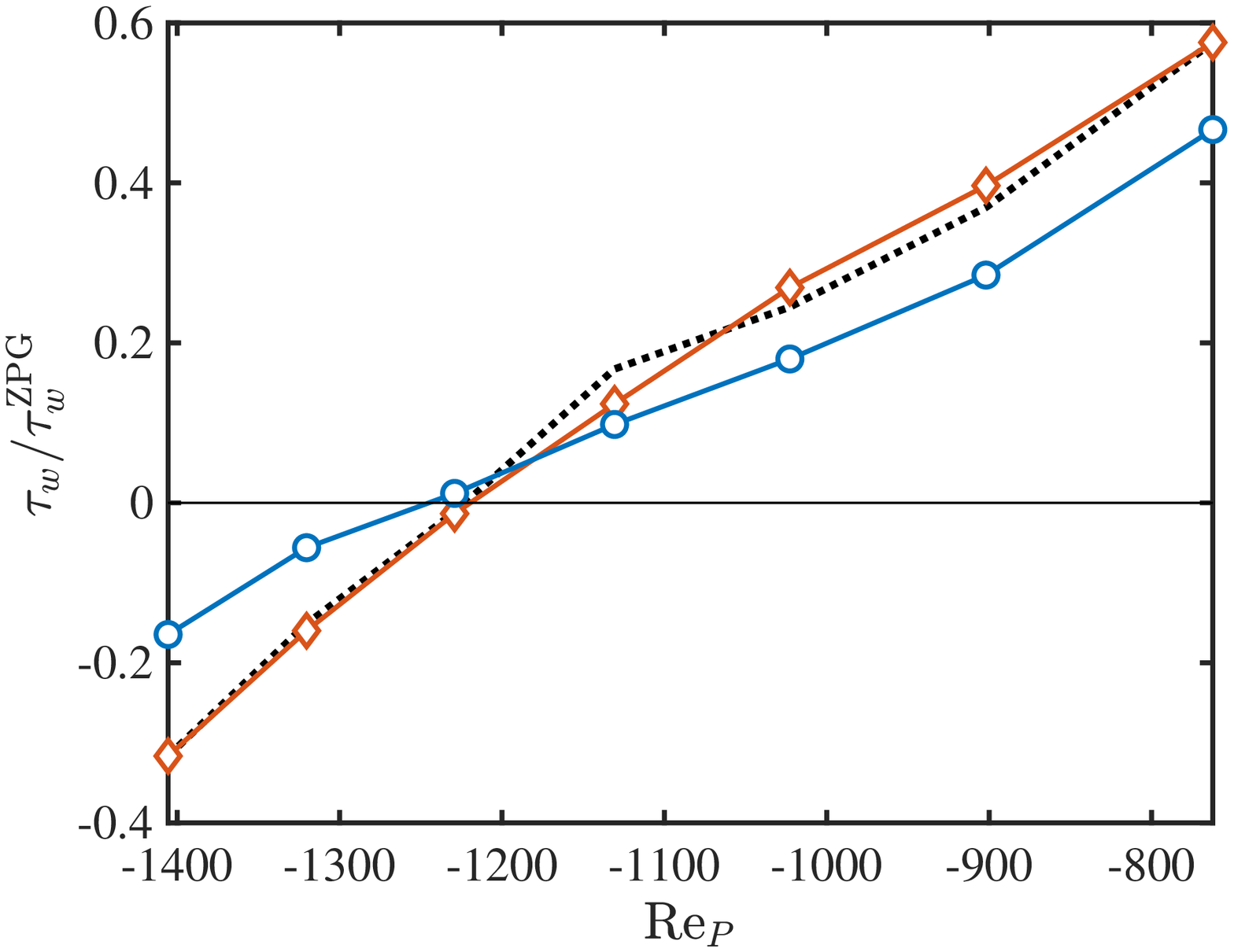}}
    \hspace{0.3cm}
    \subfloat[]{\includegraphics[width=0.41\textwidth]{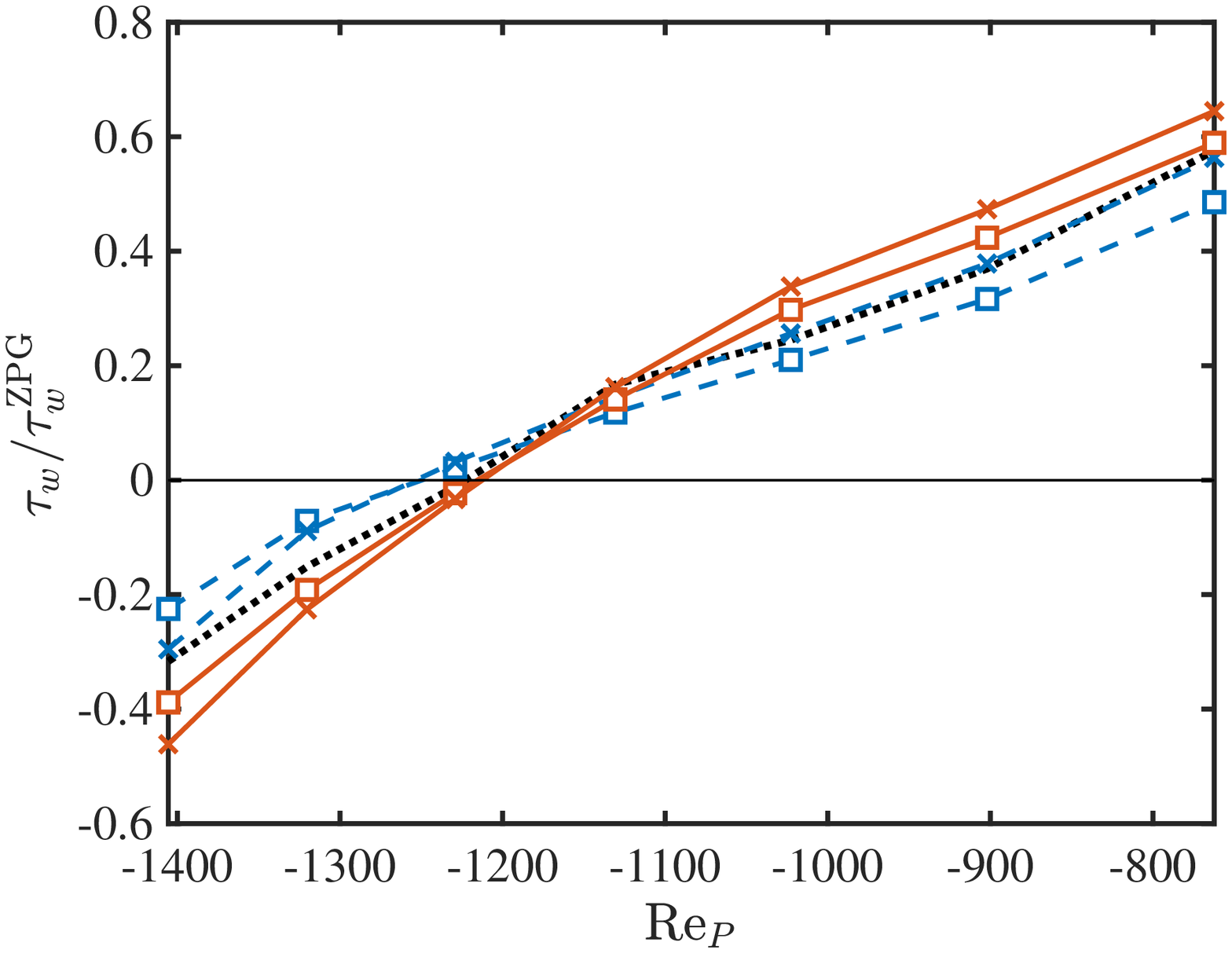}}
  \end{center}
  \caption{ Validation case: turbulent Poiseuille--Couette flow with
    adverse mean pressure gradient.  (a) The streamwise mean velocity
    profiles non-dimensionalised by the mean centreline velocity
    ($U_{cl}^{\mathrm{ZPG}}$) of the case with zero
    mean pressure gradient for DSM-EQWM (blue squares), VRE-EQWM (blue
    crosses), DSM-BFWM (red squares) and VRE-BFWM (red crosses) for
    different $\Rey_P$. The dashed lines are mean velocity profiles from
    DNS.  (b) Dominant flow classification (solid blue) and confidence score
    (solid red) by the DSM-BFWM as a function of the adverse pressure
    gradient $\Rey_P$. (c) Wall stress prediction in the absence of
    external errors for the ESGS-BFWM (diamonds) and ESGS-EQWM (circles)
    as a function of $\Rey_P$. (d) Total wall stress prediction for
    the DSM-EQWM (dashed line and squares), DSM-BFWM (solid line and squares), VRE-EQWM
    (dashed line and crosses), and VRE-BFWM (solid line and crosses) as a function of
    $\Rey_P$. In (c) and (d), the wall stress prediction is
    non-dimensionalised by the DNS wall stress for the case with zero
    mean pressure gradient. \label{fig:validation_APG}}
\end{figure}

\begin{figure}
    \vspace{0.5cm}
  \begin{center}
    \subfloat[]{\includegraphics[width=0.41\textwidth]{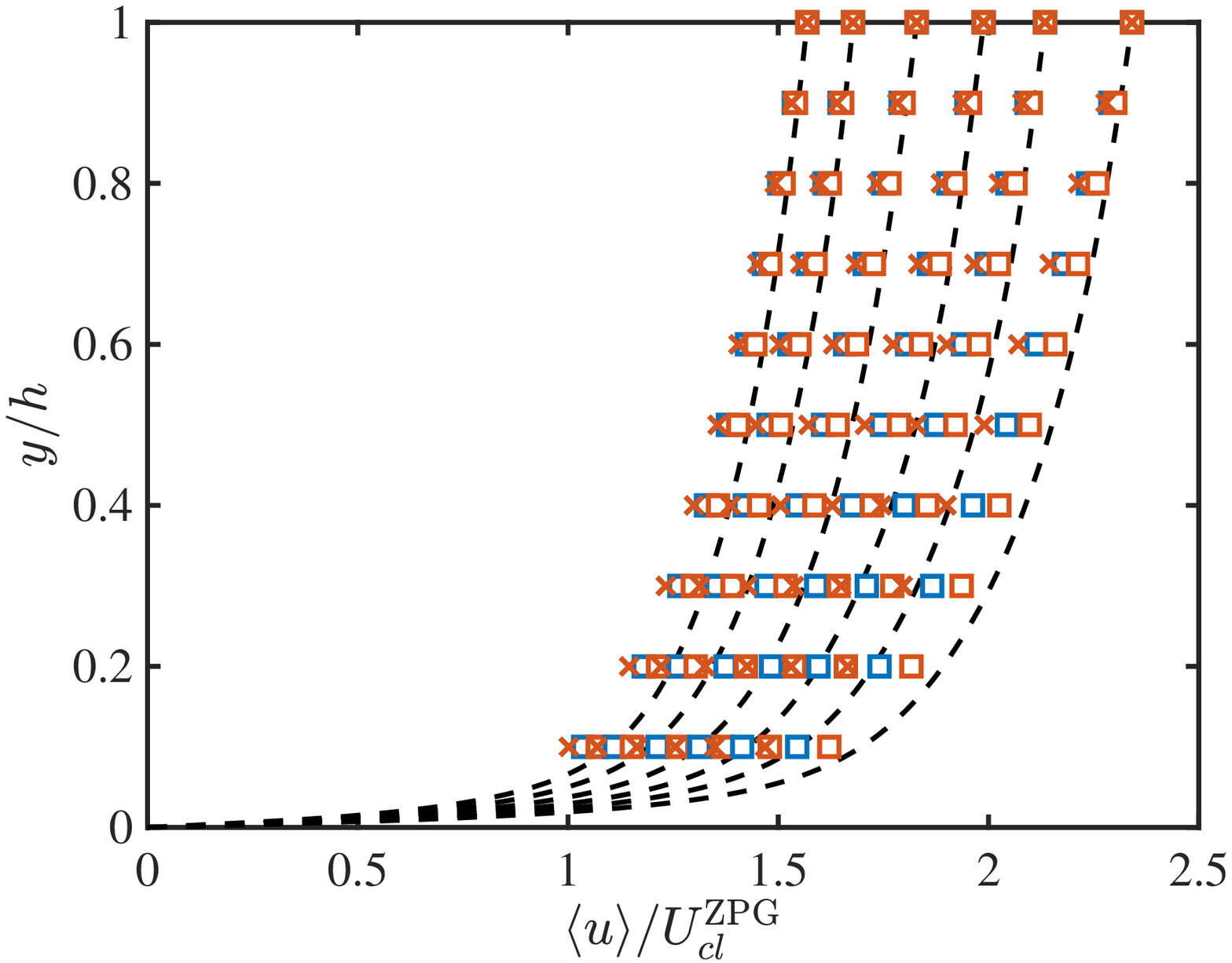}}
    \hspace{0.05cm}     
    \subfloat[]{\includegraphics[width=0.45\textwidth]{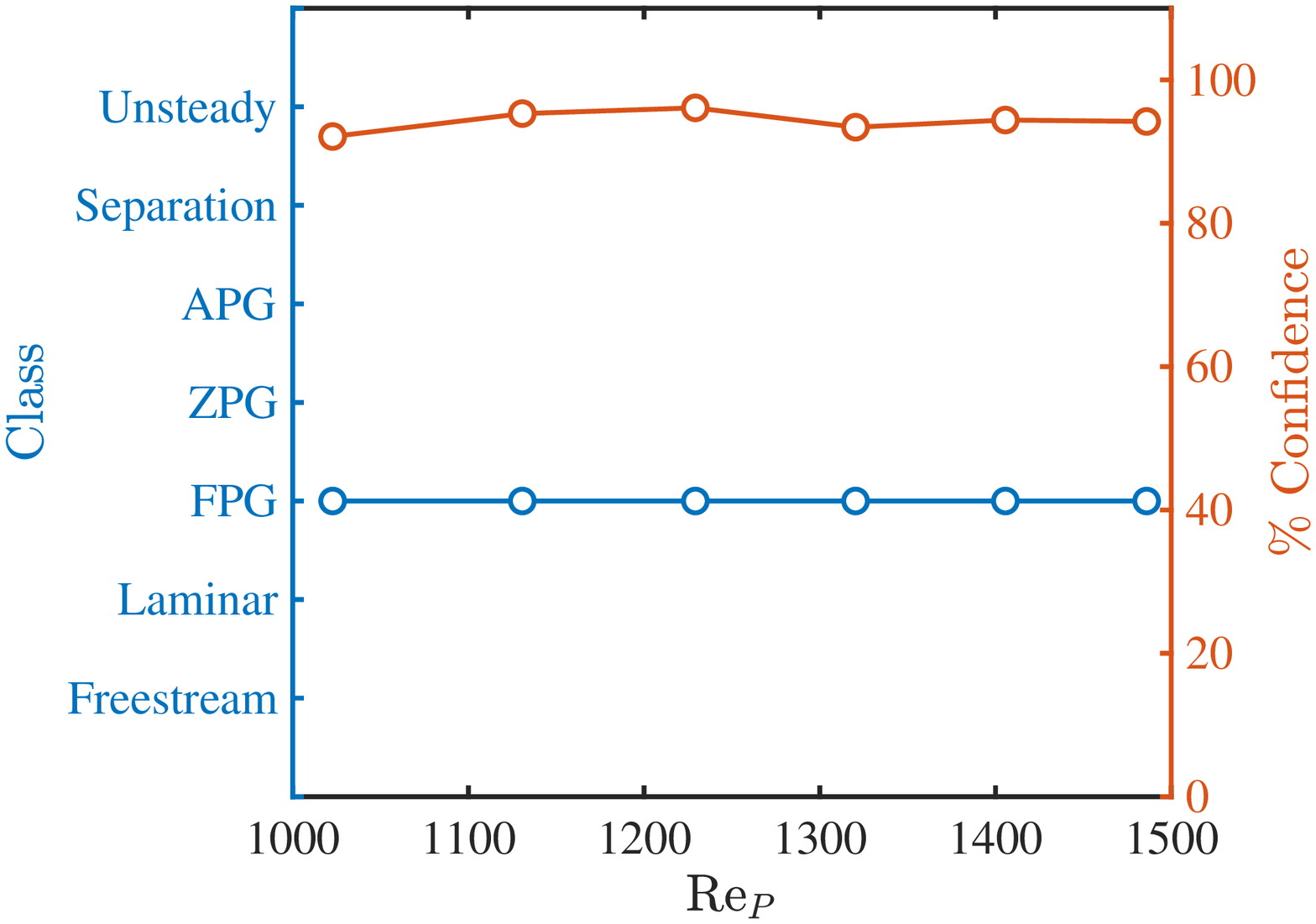}}
  \end{center}
  \begin{center}
    \subfloat[]{\includegraphics[width=0.41\textwidth]{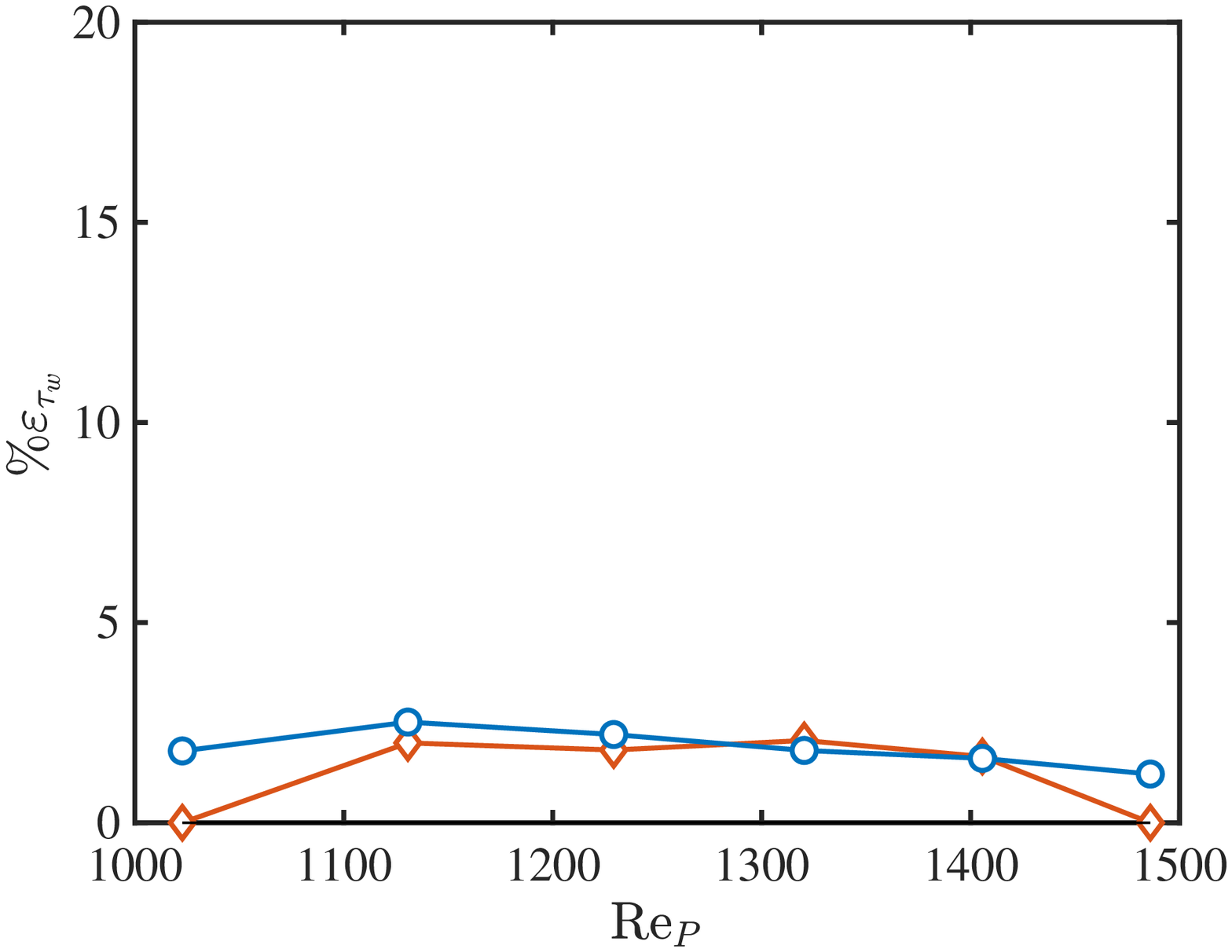}}
    \hspace{0.3cm}
    \subfloat[]{\includegraphics[width=0.41\textwidth]{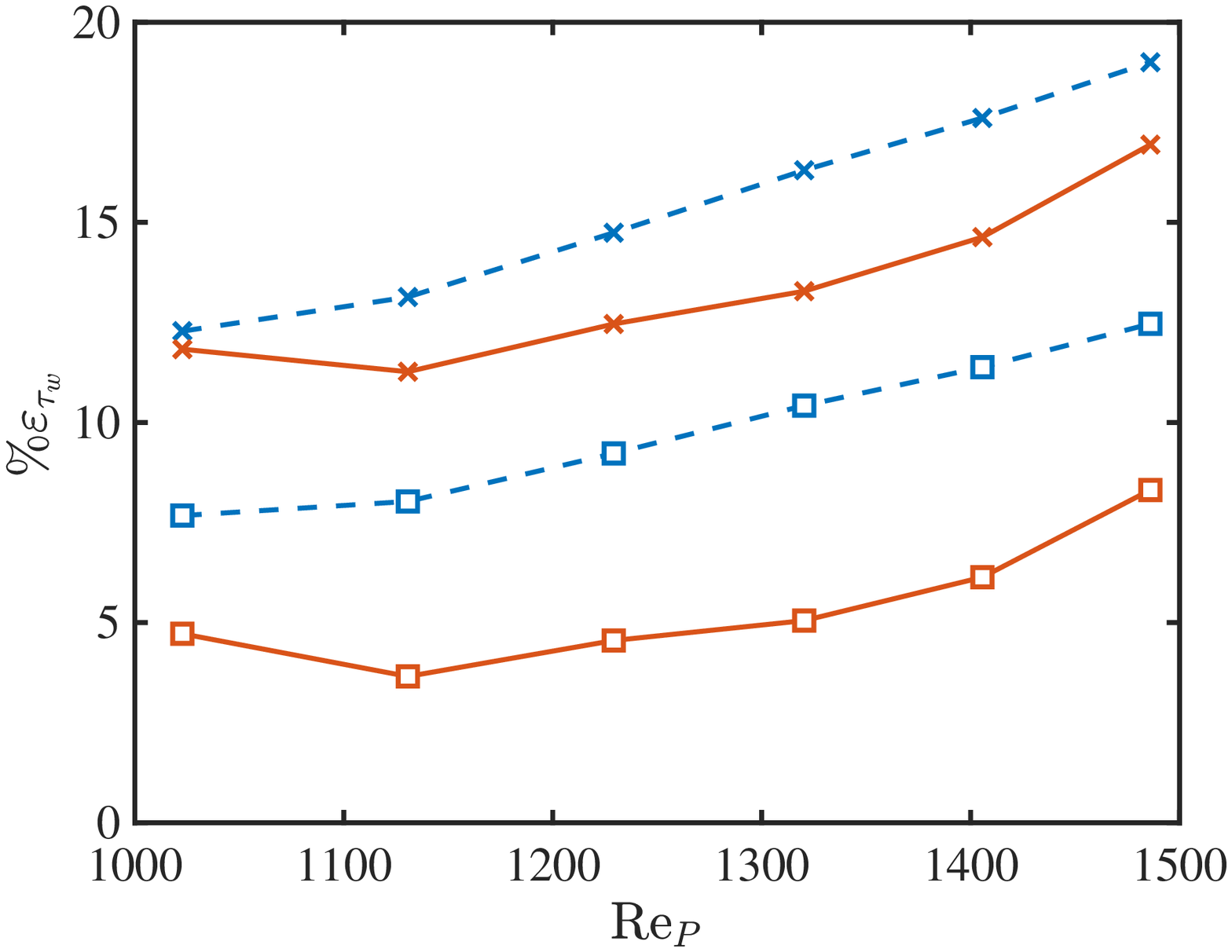}}
  \end{center}
\caption{ Validation case: turbulent Poiseuille--Couette flow with
  favourable mean pressure gradient.  (a) The streamwise mean velocity
  profiles non-dimensionalised by the mean centreline velocity
  ($U_{cl}^{\mathrm{ZPG}}$) of the case with zero
  mean pressure gradient for the DSM-EQWM (blue squares), VRE-EQWM (blue
  crosses), DSM-BFWM (red squares) and VRE-BFWM (red crosses) for
  different $\Rey_P$. The dashed lines are mean velocity profiles from
  DNS.  (b) Dominant flow classification (solid blue) and confidence score
  (solid red) by the DSM-BFWM as a function of $\Rey_P$. (c) Internal
  wall-modelling error of the wall stress prediction for the ESGS-BFWM
  (diamonds) and ESGS-EQWM (circles) as a function $\Rey_P$. (d) Total
  wall-modelling error for the DSM-EQWM (dashed line and squares), DSM-BFWM (solid line and
  squares), VRE-EQWM (dashed line and crosses), and VRE-BFWM (solid line and crosses) as
  a function of $\Rey_P$. \label{fig:validation_FPG}}
\end{figure}

The results for APG/Separation cases and FPG cases are shown in
figures \ref{fig:validation_APG} and \ref{fig:validation_FPG},
respectively. There are noticeable deviations in the prediction of the
mean velocity profiles for both the DSM-EQWM and the VRE-EQWM. Similar results
are obtained for their BFWM counterparts. Nonetheless, the flow
classification (figures \ref{fig:validation_APG}b and
\ref{fig:validation_FPG}b) is accurate, with a high confidence score
($>90\%$).  The internal wall-modelling errors for APG/Separation and
FPG cases are shown in figures \ref{fig:validation_APG}(c) and
\ref{fig:validation_FPG}(c), respectively.  For the APG/Separation
cases, we report the value of the wall-shear stress (rather than the
relative error) to avoid dividing by zero close to separation. The BFWM
outperforms the EQWM for APG/Separation cases, whereas ZPG cases are all
accurately predicted by both wall models, with errors below 3\%. The
total wall-stress error for APG/separation and FPG cases is shown in
figures \ref{fig:validation_APG}(d) and
\ref{fig:validation_FPG}(d), respectively.  For APG/Separation, the EQWM
provides more accurate predictions than the BFWM, but these are subjected
to error cancellation: the EQWM underpredicts the wall-shear stress;
however, the mean velocities close to the wall are overpredicted. The
combination of both effects results in better predictions for the EQWM,
but for the wrong reasons.  For FPG, the total error rises from 5\%
and 20\% for increasing favourable pressure gradient. This trend
applies to both the BFWM and the EQWM, although the BFWM still outperforms the EQWM.
%

\subsection{Turbulent channel flow with the sudden imposition of spanwise pressure gradient}

The last canonical validation case is a turbulent channel flow with
the sudden imposition of a mean spanwise pressure gradient. The ratio of
spanwise to streamwise mean pressure gradient is $\Pi = 80$. The grid
resolution is $h/\Delta = 10$. The simulation is started from a
turbulent channel flow at a statistically steady state with
$\Rey_\tau=700$ for $t<0$ and the lateral pressure gradient is
imposed at time $t=0$.

\begin{figure}
    \vspace{0.5cm}
  \begin{center}
    \subfloat[]{\includegraphics[width=0.30\textwidth]{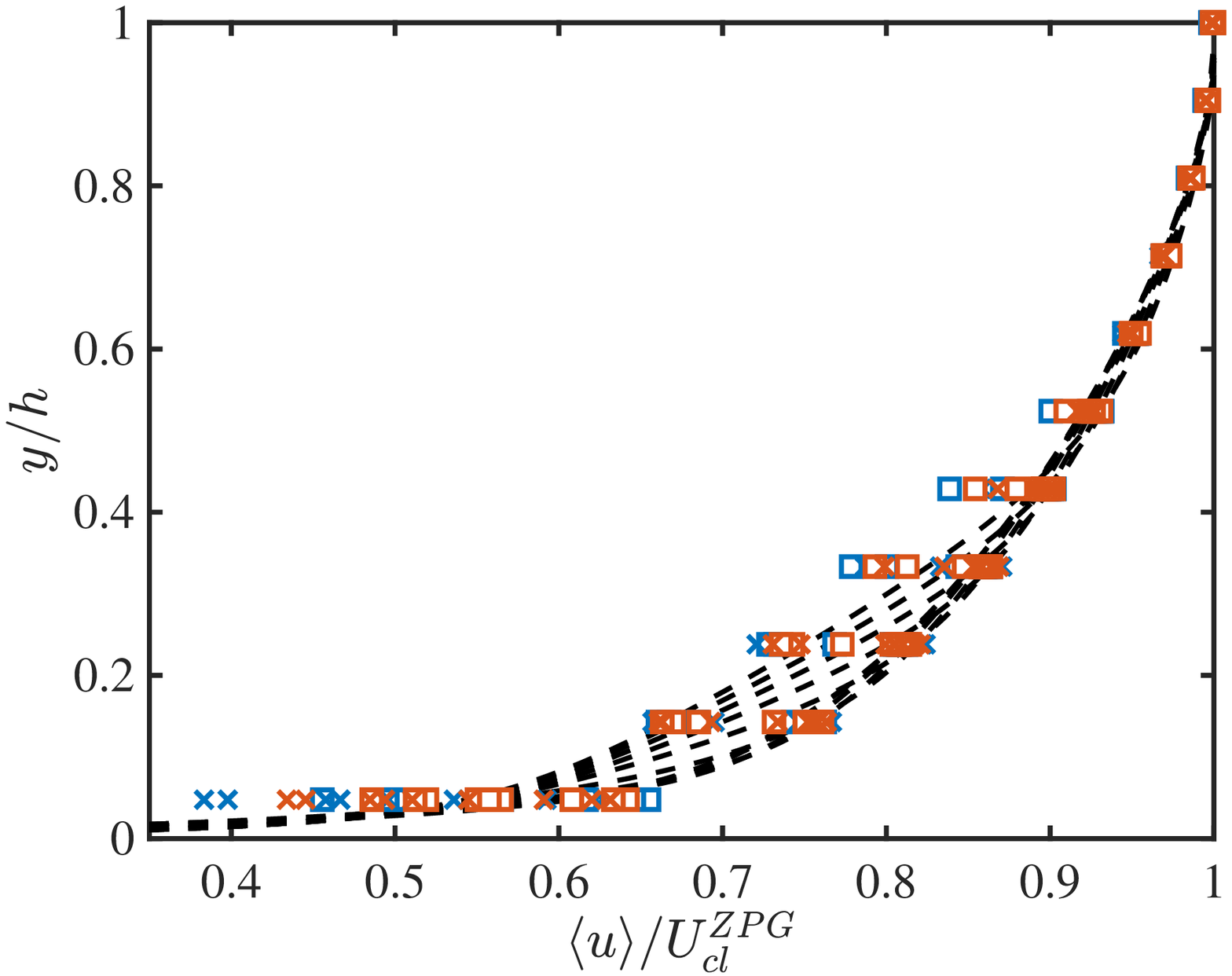}}
    \hspace{0.09cm}
    \subfloat[]{\includegraphics[width=0.30\textwidth]{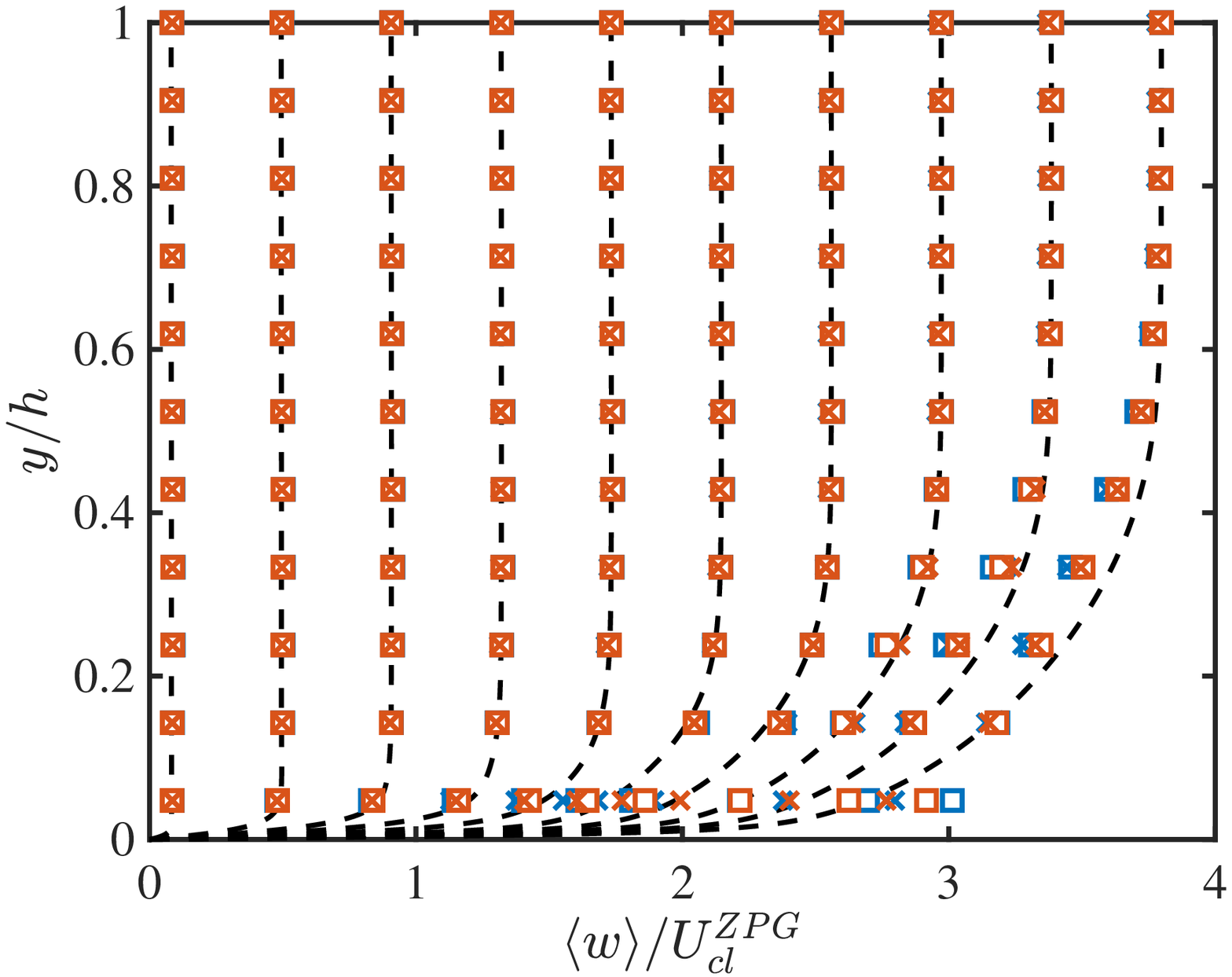}}
    \hspace{0.09cm}
    \subfloat[]{\includegraphics[width=0.33\textwidth]{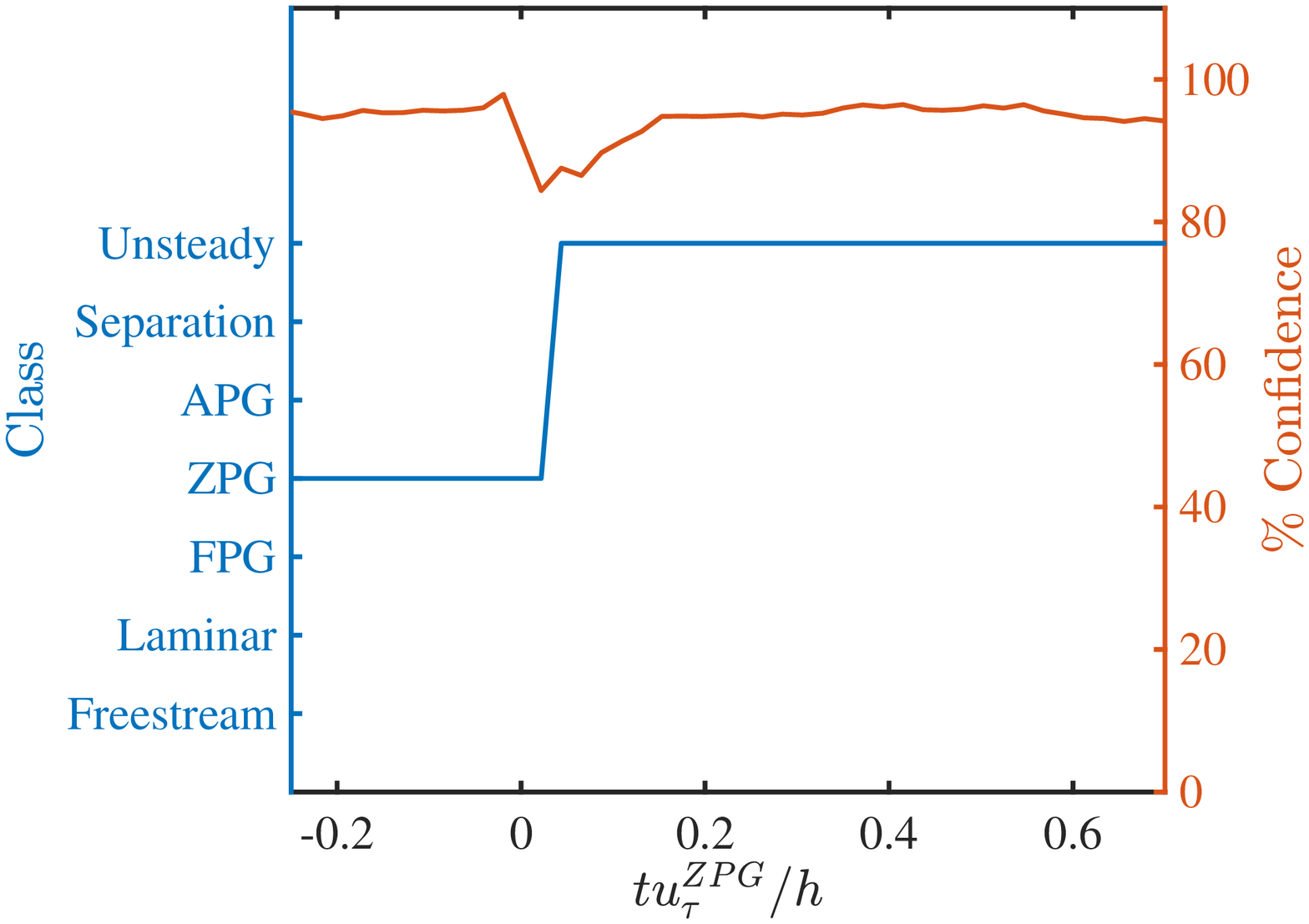}}
  \end{center}
  \begin{center}
    \subfloat[]{\includegraphics[width=0.30\textwidth]{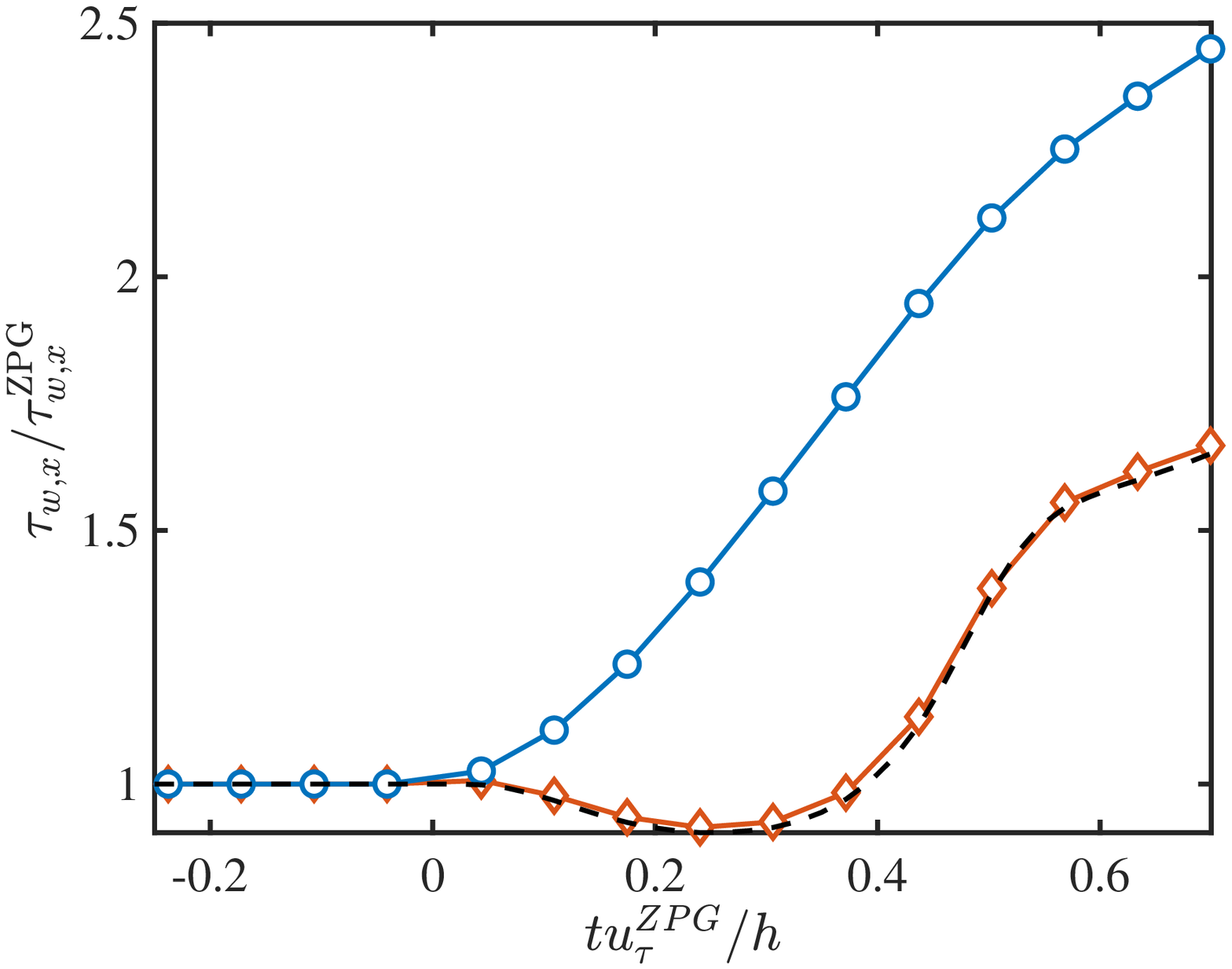}}
    \hspace{0.09cm}
    \subfloat[]{\includegraphics[width=0.30\textwidth]{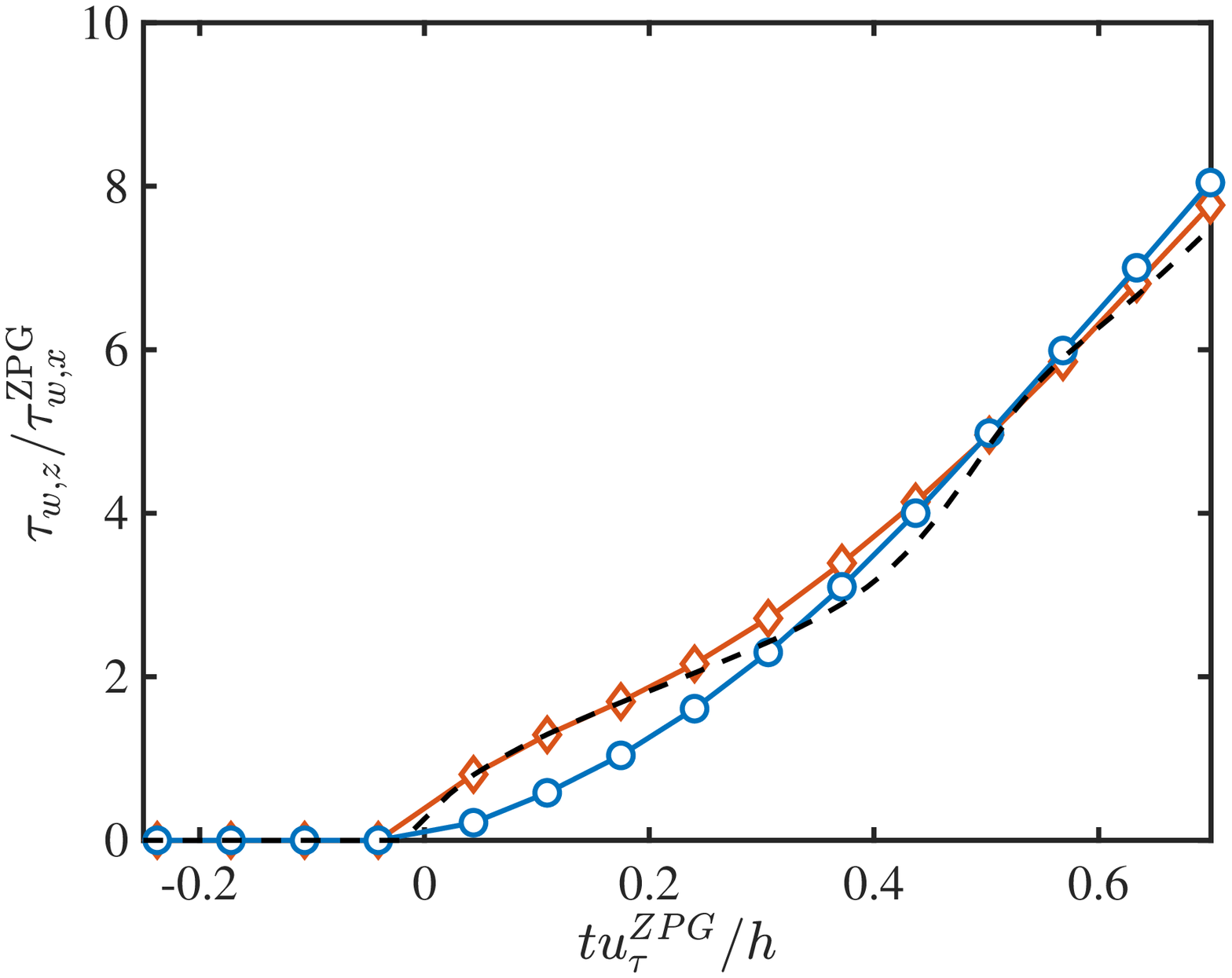}}
    \hspace{0.09cm}
    \subfloat[]{\includegraphics[width=0.30\textwidth]{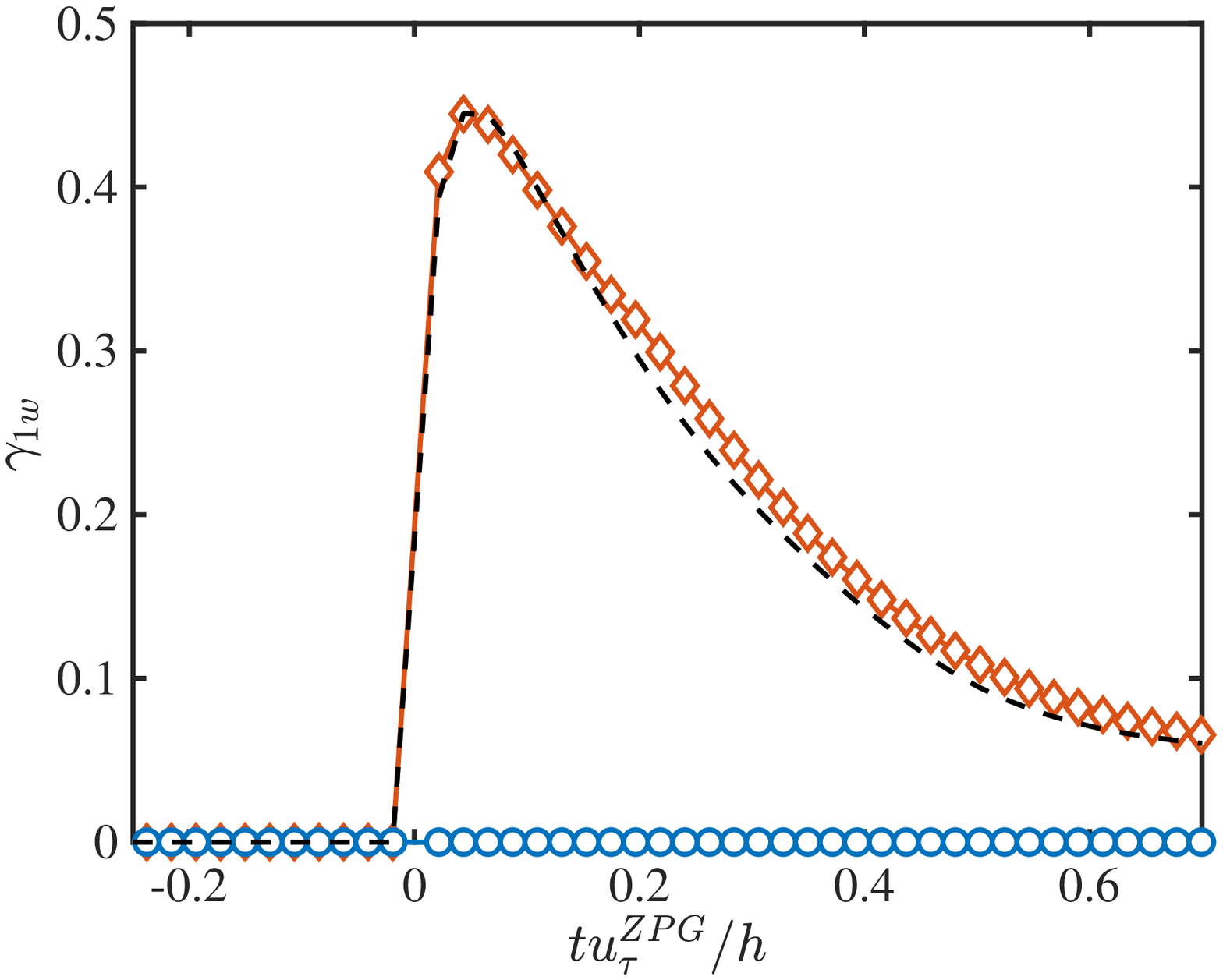}}
  \end{center}
  \begin{center}
    \subfloat[]{\includegraphics[width=0.30\textwidth]{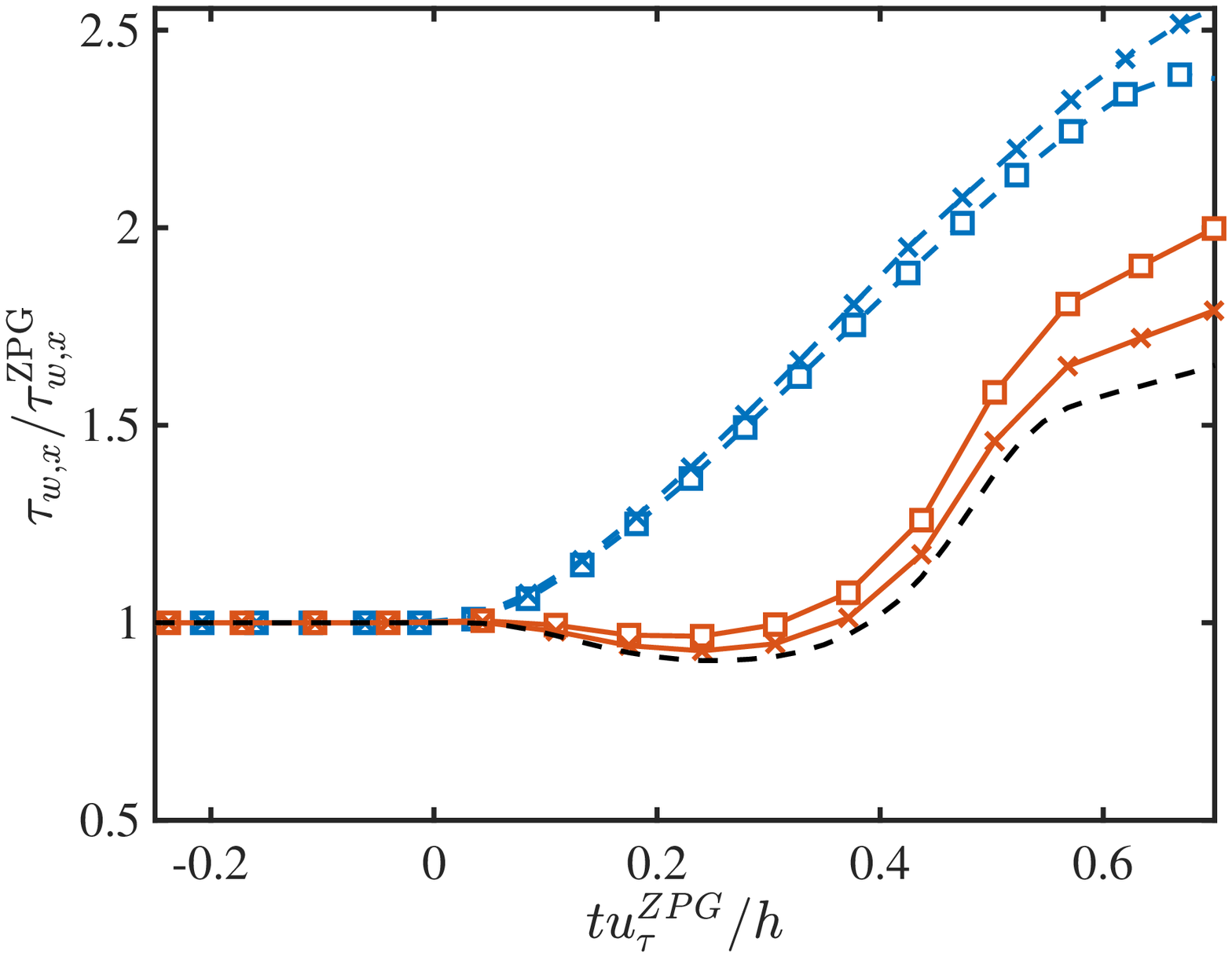}}
    \hspace{0.09cm}
    \subfloat[]{\includegraphics[width=0.30\textwidth]{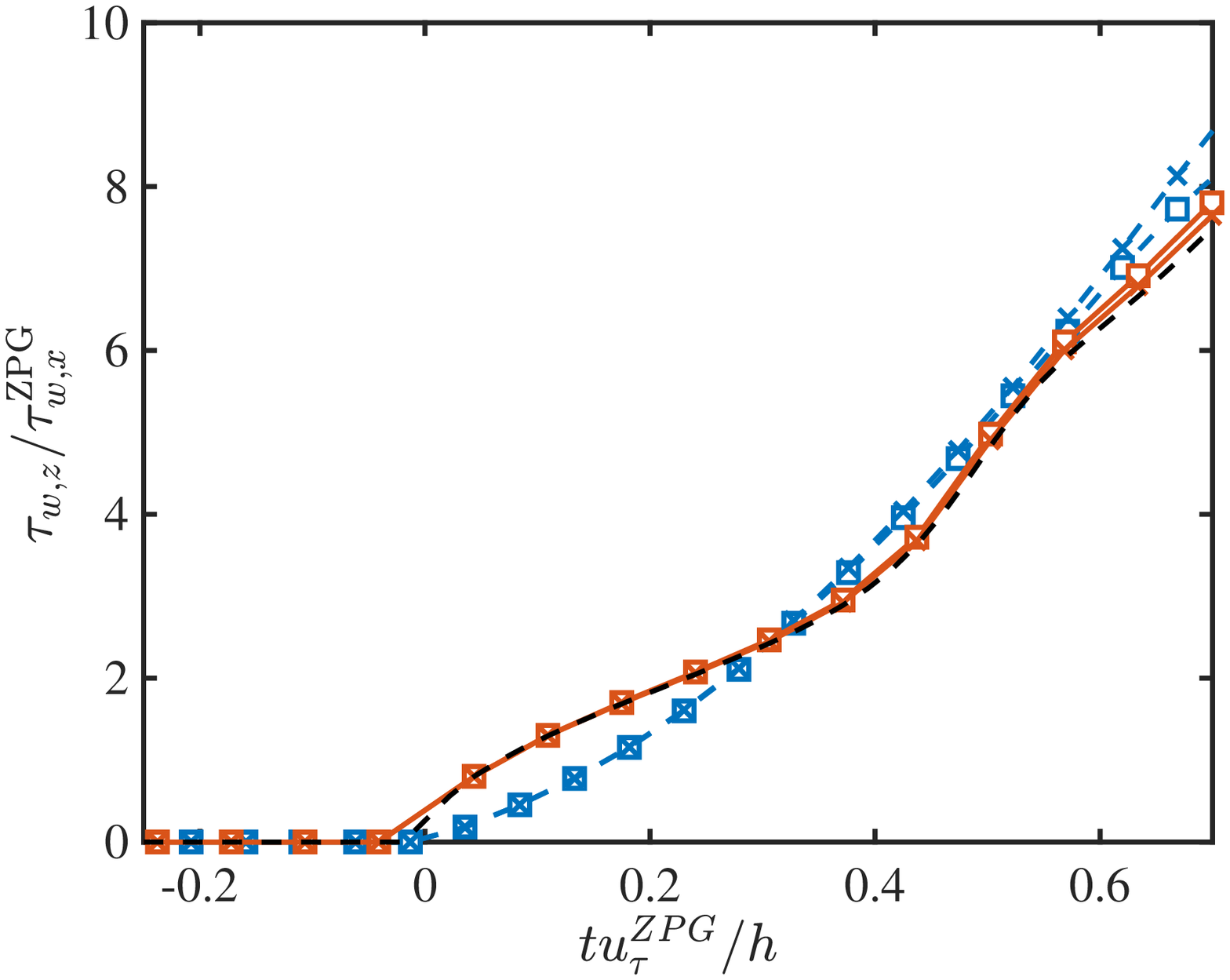}}
    \hspace{0.09cm}
    \subfloat[]{\includegraphics[width=0.30\textwidth]{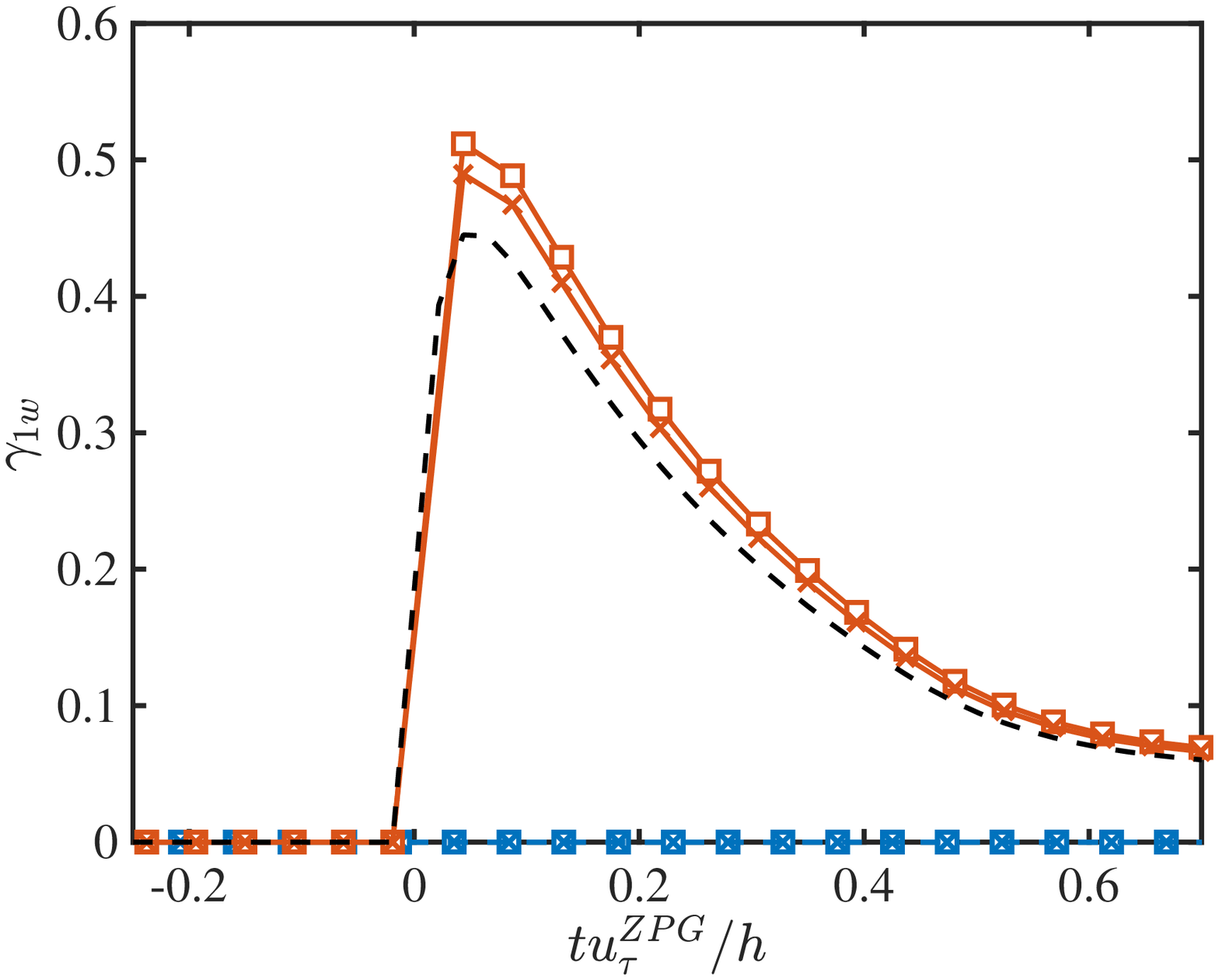}}
  \end{center}
\caption{ Validation case: turbulent channel flow sudden imposition of
  spanwise mean pressure gradient.  (a) The streamwise and (b)
  spanwise mean velocity profiles non-dimensionalised by the
  centreline mean velocity at $t=0$ ($U_{cl}^{\mathrm{ZPG}}$) for
  the DSM-EQWM (blue squares), VRE-EQWM (blue crosses), DSM-BFWM (red
  squares) and VRE-BFWM (red crosses) at $t
  u_\tau^{\mathrm{ZPG}}/h=0.01, 0.2,...,0.9$, where
  $u_\tau^{\mathrm{ZPG}}$ is $u_\tau$ at $t=0$. The dashed lines are
  DNS mean velocity profiles (same for all following panels).  (c)
  Dominant flow classification (solid blue) and confidence score (solid red) by
  the DSM-BFWM as a function of time. Internal wall-modelling error of the
  (d) streamwise and (e) spanwise wall stress prediction and (f)
  relative angle between $\vec{u}_1$ and $\vec{\tau}_w$ for the ESGS-BFWM
  (diamonds) and ESGS-EQWM (circles) as a function of time.  Total
  wall-modelling error for the streamwise wall stress (g), spanwise
  wall stress (h), and (i) relative angle between $\vec{u}_1$ and
  $\vec{\tau}_w$ for the DSM-EQWM (dashed line and squares), DSM-BFWM (solid line and
  squares), VRE-EQWM (dashed line and crosses), and VRE-BFWM (solid line and crosses) as
  a function of time.
  \label{fig:validation_unsteady}}
\end{figure}

Figures \ref{fig:validation_unsteady}(a,b) show the history of
the streamwise and spanwise mean velocity profiles for all
cases. Predictions of the mean flow remain accurate within 5\% error
for the second grid point off the wall. More acute errors are observed
for the first grid point off the wall. The classifier accurately
detects the change of regime from ZPG to Unsteady at $t=0$ with high
confidence (figure \ref{fig:validation_unsteady}c). The internal
wall-modelling errors for the streamwise ($\tau_{w,x}$) and spanwise
($\tau_{w,z}$) wall stresses are shown in figures
\ref{fig:validation_unsteady}(d,e), respectively.  As expected,
the BFWM performs better than the EQWM, with errors below 1\% for all
times. Additionally, the BFWM is able to predict the mild drop in
$\tau_{w,x}$ as a function of time, which is completely missed by
the EQWM. A similar conclusion applies to the relative angle between
$\vec{u}_1$ and $\vec{\tau}_w$ (namely, $\gamma_{1w}$). The BFWM correctly
predicts the increase and decrease of $\gamma_{1w}$, whereas
$\gamma_{1w}$ remains zero for the EQWM essentially by construction of the
model. Finally, the BFWM still improves the predictions compared to the EQWM
in terms of the total wall stress and $\gamma_{1w}$ errors (figures
\ref{fig:validation_unsteady}g, h, and i), but with lower accuracy
than before due to the presence of external wall-modelling errors.
%

\subsection{NASA Common Research Model High-lift}
\label{subsec:CRM}

The NASA Common Research Model High-lift (CRM-HL) is a geometrically
complex aircraft that includes the bracketry associated with deployed
flaps and slats as well as a flow-through nacelle mounted on the
underside of the wing.  The simulations are performed in free air at
Reynolds number of 5.49 million based on the mean aerodynamic
chord and freestream velocity. The freestream Mach number is 0.20. The
results are compared with wind tunnel experimental data from
\citet{Evans2020} corrected for free air conditions. Further details
of the experimental set-up can be found in \citet{Lacy2016}.

\begin{figure}
  \centering
  \vspace{1cm}
  \includegraphics[width=1\textwidth]{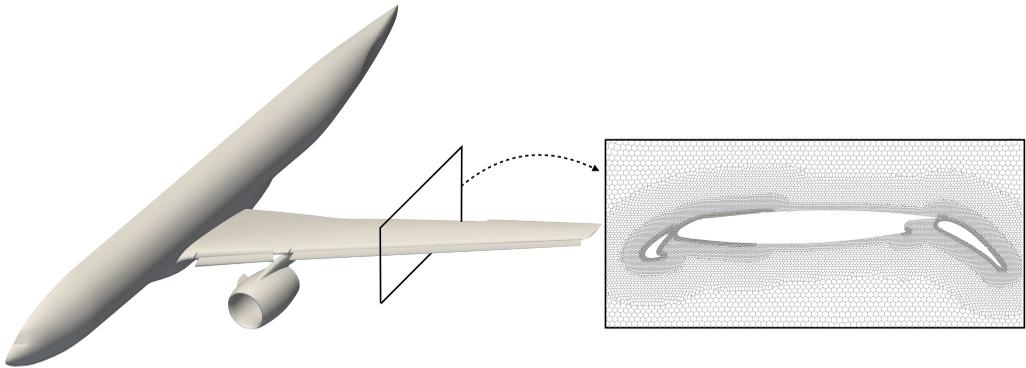}
\caption{ Validation case: NASA Common Research Model
  High-lift. Geometry of the aircraft (left) and an inset of the grid
  across the wing section (right). \label{fig:CRM_grid}}
\end{figure}

We follow the computational set-up from \citet{Goc2021}.  A semi-span
aircraft geometry is simulated, and the symmetry plane is treated with
free-slip and no-penetration boundary conditions.  Uniform plug flow
is used as the inlet. A non-reflecting boundary condition with
specified freestream pressure is imposed at the
outlet~\citep{Poinsot1992}. The total number of grid points is 30
million, and the number of grid points per boundary layer thickness
ranges from 0 to 20. The reader is referred to \citet{Goc2021}
for more details about the gridding strategy. Figure
\ref{fig:CRM_grid} offers one example of the grid topology.  The SGS
model is the DSM, and we perform simulations for the DSM-EQWM and
the DSM-BFWM. No ESGS model is available for comparison in this case.  A
summary of previous WMLES of the CRM-HL can be found in
\citet{Kiris2022}, using traditional SGS and wall models. Here, we
re-run the simulations using the DSM-EQWM, and perform new simulations for
the DSM-BFWM at angles of attack 7 and 19 degrees.
%

\begin{figure}
  \centering
  \vspace{1cm}
  \includegraphics[width=0.32\textwidth]{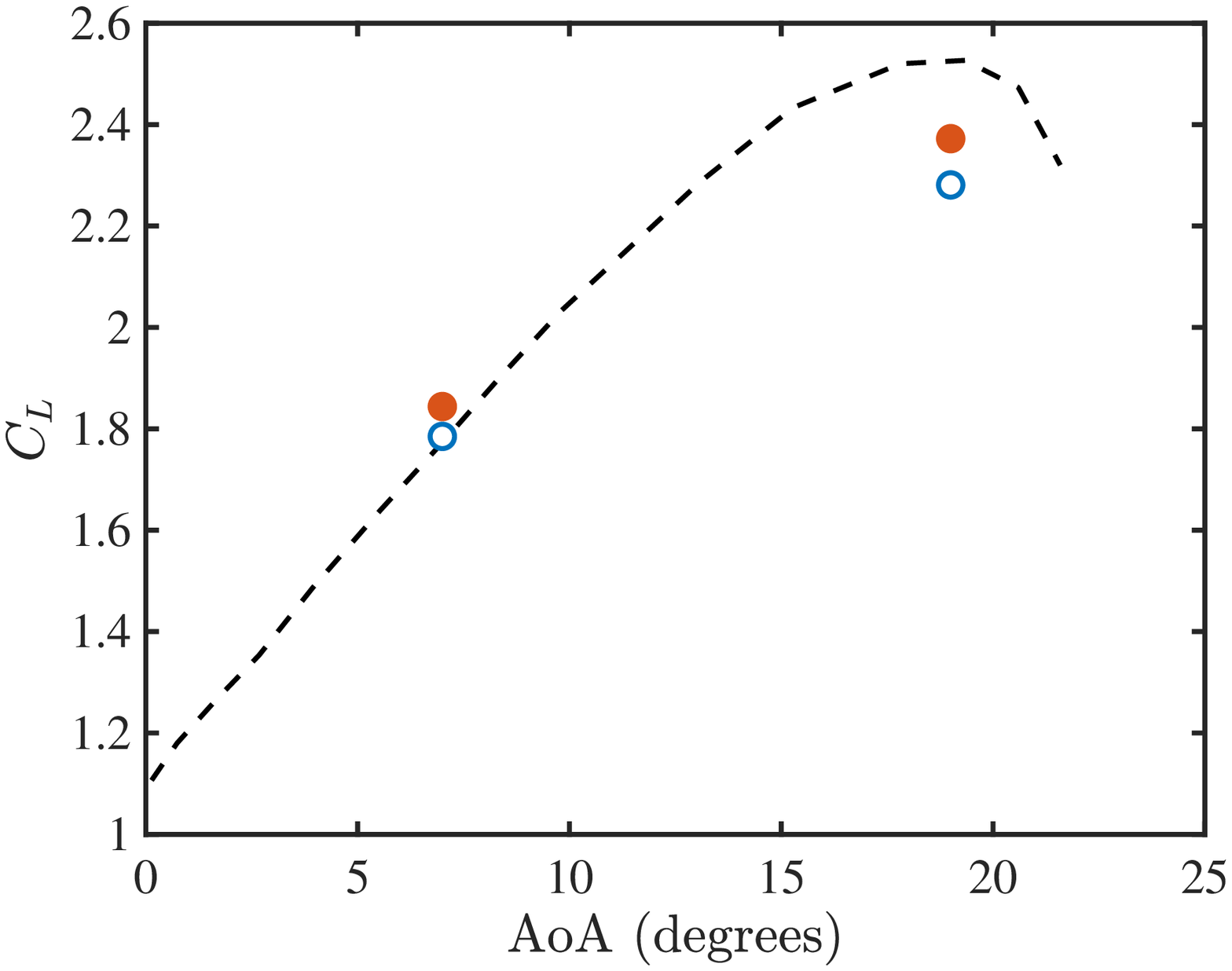}
  \hspace{0.1cm}
  \includegraphics[width=0.32\textwidth]{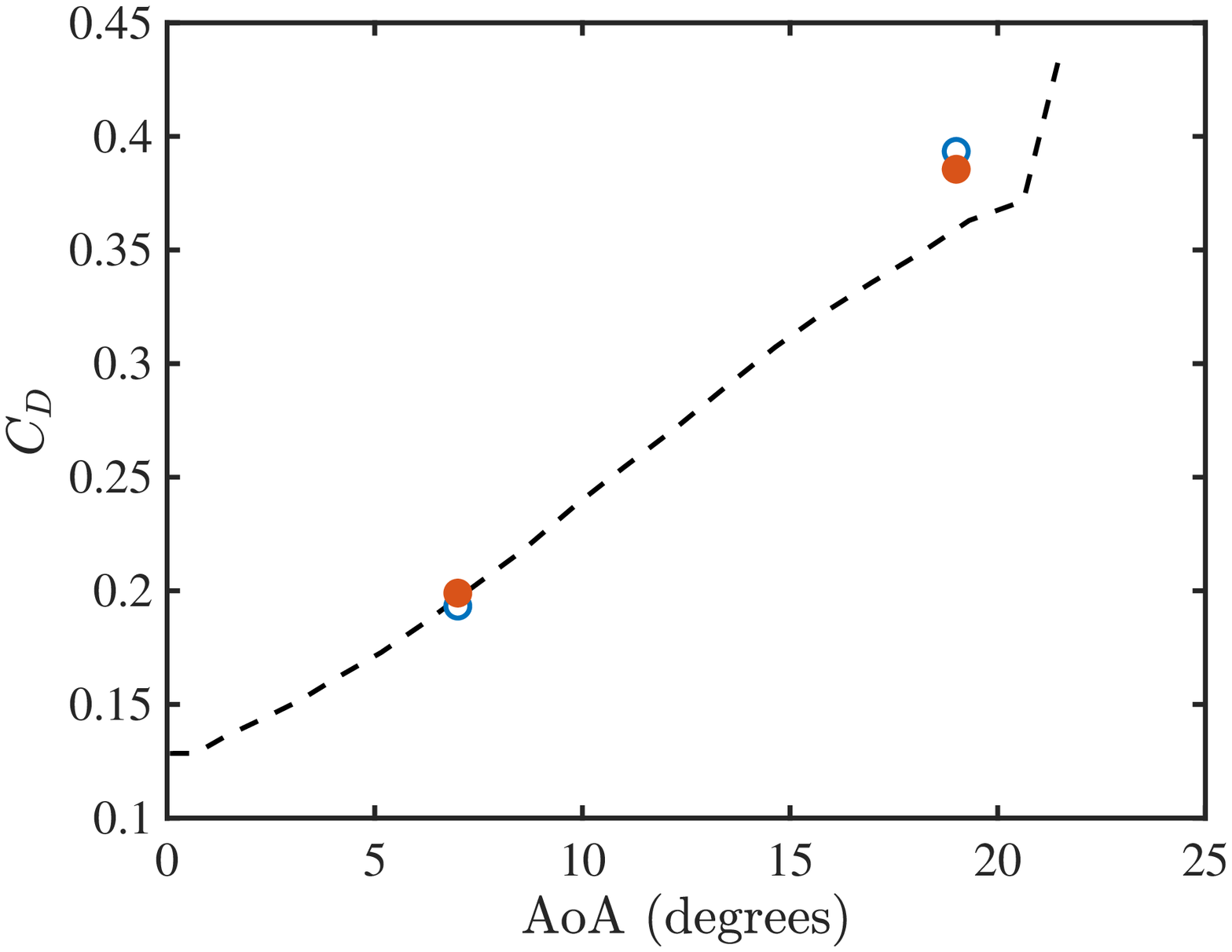}
  \hspace{0.1cm}
  \includegraphics[width=0.32\textwidth]{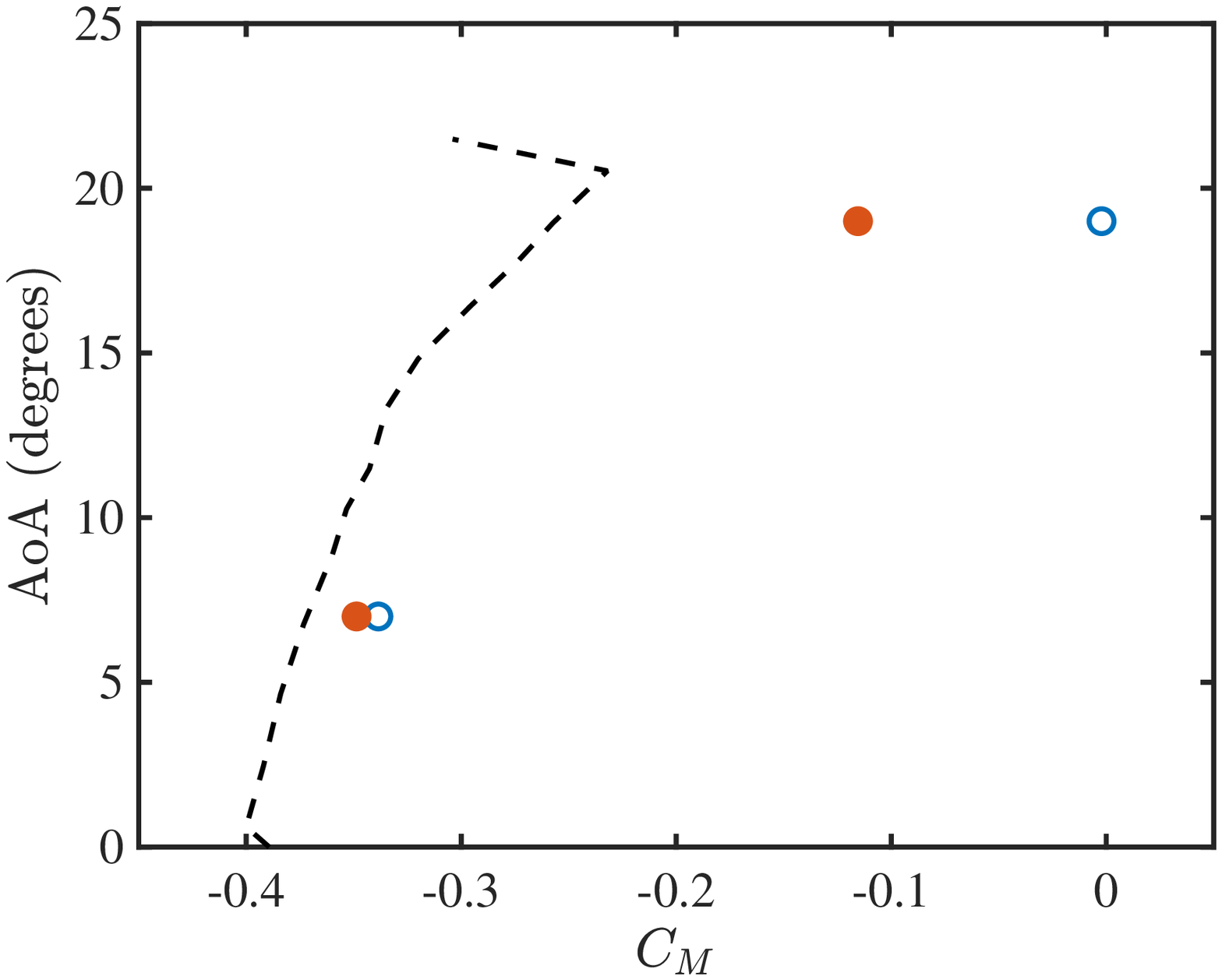}
\caption{Validation case: NASA Common Research Model High-lift. (a)
  The lift, (b) drag, and (c) pitching moment coefficients, as a
  functions of the angle of attack (AoA) for the DSM-BFWM (red circles) and
  DSM-EQWM (blue circles). The dashed lines are experimental
  results. \label{fig:CRM_results}}
\end{figure}

The lift ($C_L$), drag ($C_D)$ and pitching moment ($C_M$)
coefficients are shown in figure \ref{fig:CRM_results}.  It is worth
remarking that the BFWM has never `seen' an aircraft-like flow or been
trained in a case that resembles an aerofoil or a wing. The DSM-BFWM
provides moderate improvements with respect to DSM-EQWM, especially
close to the stall. The improvements are more clearly manifested in
the pitching moment, which is a marker of the accuracy in the force
distribution along the wing. The exception is $C_L$ at low angles of
attack, where the DSM-EQWM outperforms the DSM-BFWM. However, it has been
reported that the good predictions by the DSM-EQWM at low angles of attack
are coincidental and due to error cancellation, so this should not be
taken as a failure of the DSM-BFWM. Despite the improvements by the DSM-BFWM,
the results suggest that a superior wall model alone does not suffice
to further boost the performance of WMLES unless the SGS model is also
improved accordingly.
%

\begin{figure}
  \centering
  \vspace{1cm}
  \includegraphics[width=1\textwidth]{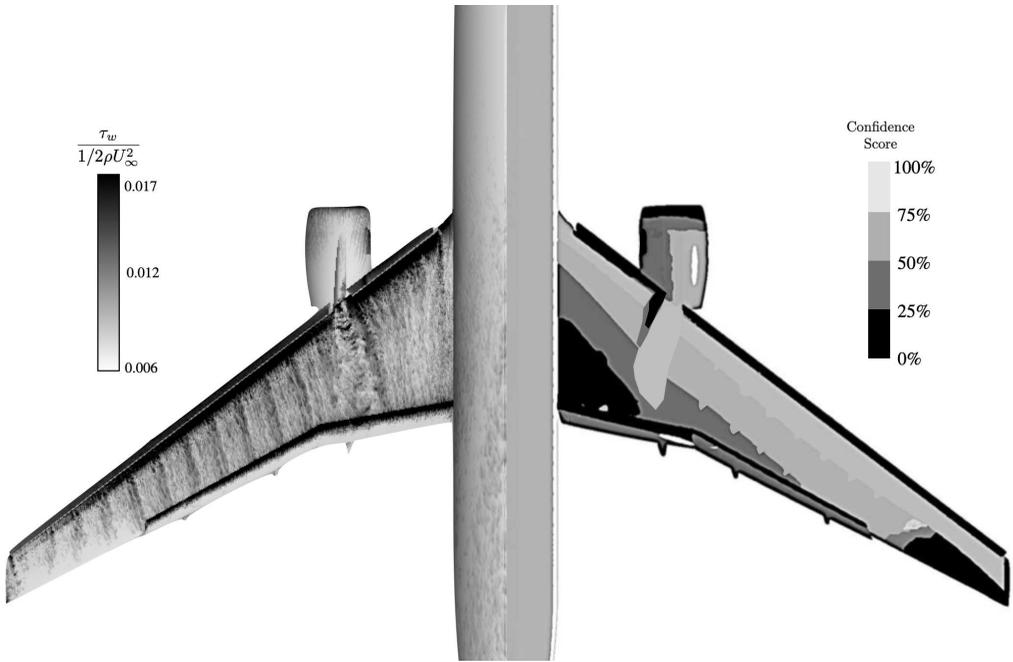}
\caption{ Validation case: NASA Common Research Model
  High-lift. Visualisation of the instantaneous wall-shear stress
  (left) and confidence map (right) by the DSM-BFWM.
  \label{fig:CRM_confidence}}
\end{figure}

Figure \ref{fig:CRM_confidence} shows the confidence map of BFWM over
the surface of the aircraft. Two types of low-confidence ($<$20\%)
regions are identified. The first region is located at the leading edges of
the nacelle, slats and flaps, and is probably related to the lack of
grid resolution in those locations (i.e., zero points per boundary
layer thickness). The second low-confidence region is related to
separation at the wing tip and close to the wing root. This may hint
at a lack of grid resolution, the inadequacy of the building-block
flows to model separation, or a combination of both. It is worth
mentioning that confidence values of 20\% still indicate that the flow
features detected resemble the building blocks up to some reasonable
extent. For example, the presence of completely unknown flow
(i.e. random values of the model inputs) will result in confidence
scores of 0\%. The confidence map is a key feature of the BFWM which
enables further evaluation of the model performance.  As such, the
information from figure \ref{fig:CRM_confidence} might aid grid
refinement in poor-confidence regions and/or inform future
developments of the model, for example, by suggesting alternative
building-block flows or the necessity for new ones.
%

\subsection{NASA Juncture Flow experiment}

The NASA Juncture Flow experiment consists of a full-span
wing-fuselage body configured with truncated DLR-F6 wings and has been
tested in the Langley 14 feet by 22 feet Subsonic Tunnel.  The case has
been proposed as a validation experiment for generic wing--fuselage
junctions at subsonic conditions~\citep{Rumsey2019}. The Reynolds
number is $\Rey=L U_\infty/\nu=2.4$ million, where $L$ is the chord at
the Yehudi break. We consider an angle of attack of 5 degrees. The
experimental dataset comprises a collection of local-in-space
time-averaged measurements~\citep{Kegerise2019}, such as velocity
profiles and Reynolds stresses, which aid the validation of the BFWM in
more detail. The frame of reference is such that the fuselage nose is
located at $x = 0$, the $x$-axis is aligned with the fuselage
centreline, the $y$-axis denotes the spanwise direction, and the
$z$-axis is the vertical direction.  The associated instantaneous
velocities are denoted by $u$, $v$ and $w$.

\begin{figure}
  \centering
  \vspace{1cm}
\includegraphics[width=1\textwidth]{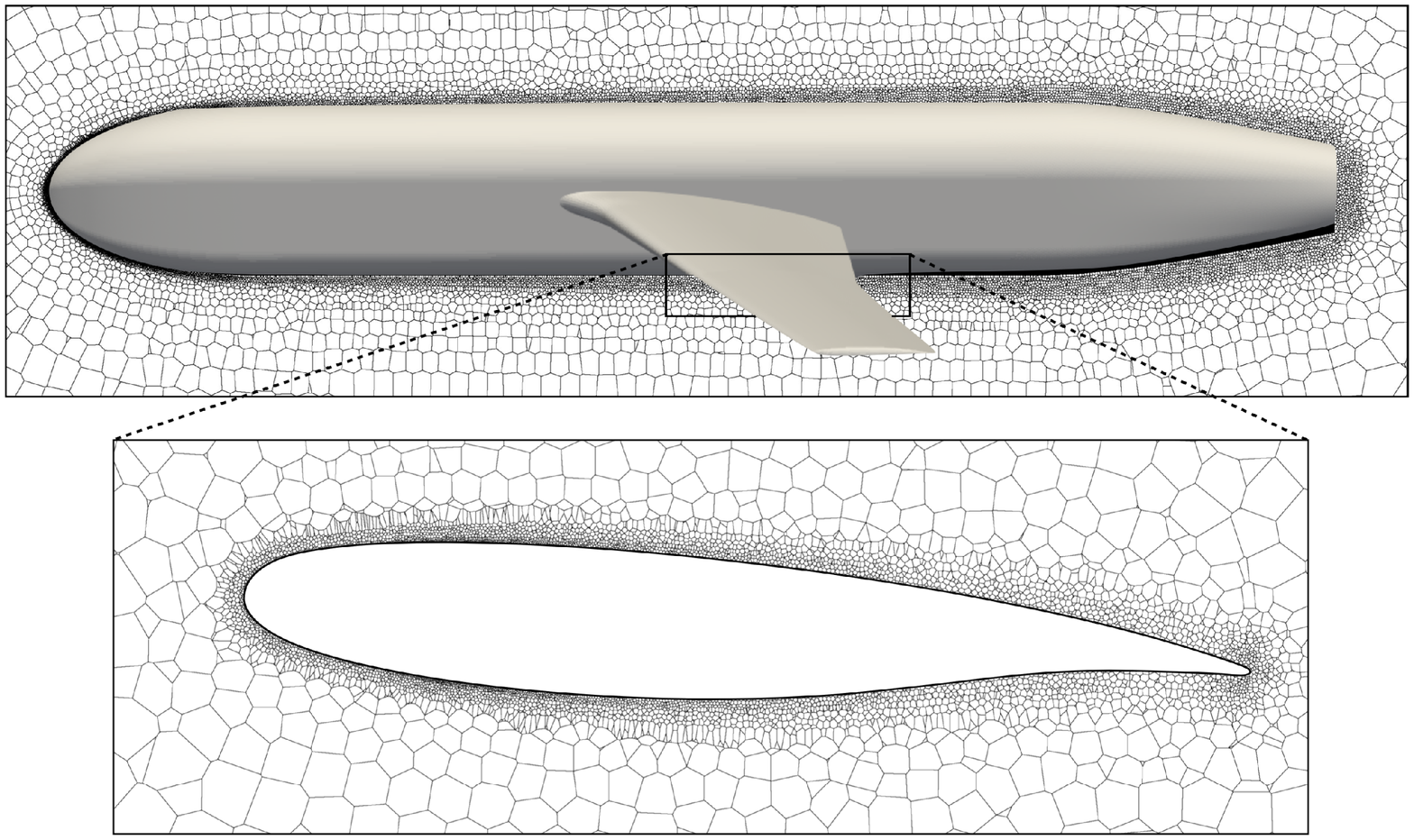}
\caption{Validation case: NASA Juncture flow experiment.  Visualisation of
  Voronoi control volumes for boundary-layer-conforming grid with five
  points per boundary layer thickness. \label{fig:grid_JF}}
\end{figure}

The SGS model is the DSM, and we perform simulations for the DSM-EQWM and
the DSM-BFWM.  We use a boundary-layer-conforming grid~\citep{Lozano2022},
such that the control volumes are distributed to achieve five points
per boundary layer thickness. Figure \ref{fig:grid_JF} contains a
visualisation of Voronoi control volumes for the
boundary-layer-conforming grid. The reader is referred to
\citet{Lozano2022} for an in-depth description of the computational
set-up, numerical methods and grid generation.
%
%

\begin{figure}
  \centering
  \vspace{1cm}
\includegraphics[width=1\textwidth]{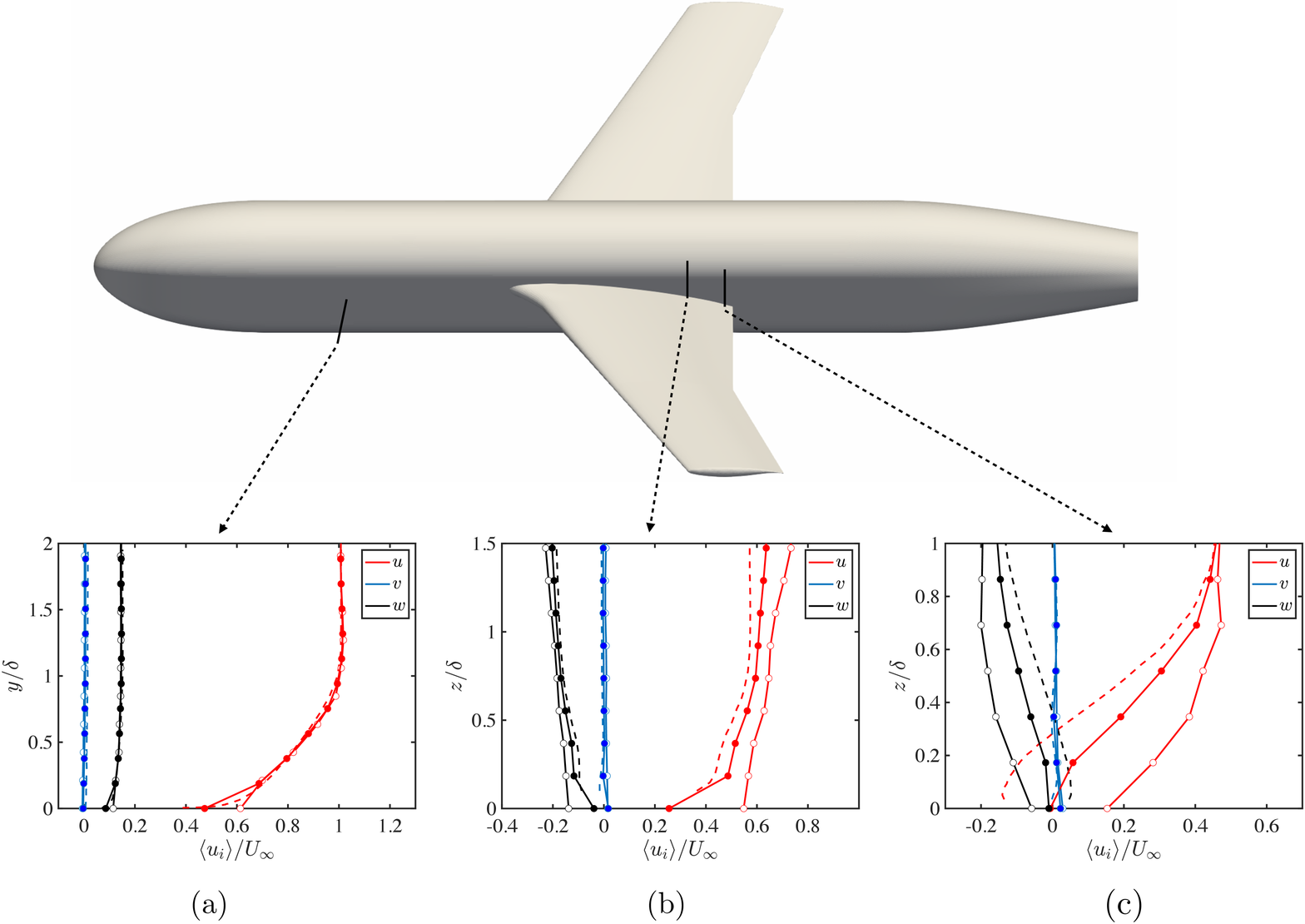}
\caption{ Validation case: NASA Juncture flow experiment. The mean velocity
  profiles at (a) the upstream region of the fuselage at $x=1168.4$~mm and
  $z=0$~mm, (b) the wing--body juncture at $x=2747.6$~mm and $y=239.1$~mm,
  and (c) the wing--body juncture close to the trailing edge at
  $x=2922.6$~mm and $y=239.1$~mm. Solid lines with open symbols are
  for the DSM-EQWM, and closed symbols are for the DSM-BFWM. Dashed lines are
  experiments. Colours denote different velocity components. The
  distances $y$ and $z$ are normalised by the local boundary layer
  thickness $\delta$ at each location.\label{fig:results_JF}}
\end{figure}

The prediction of the mean velocity profiles is shown in
figure~\ref{fig:results_JF} and compared with experimental
measurements.  Three locations are considered: the upstream region of
the fuselage, the wing--body juncture, and the wing--body juncture close
to the trailing edge. Table~\ref{table:class_JF} contains information
about the flow classification and confidence in the solution at each
location. In the first region, the flow resembles a ZPG turbulent
boundary layer. Hence the DSM-EQWM and the DSM-BFWM perform accordingly
(i.e., errors below 2\%). The flow is identified by the BFWM as 88\% ZPG
and 12\% FPG with $98\%$ confidence.  The outcome is expected as the BFWM
was trained in turbulent channel flows and the boundary layer at the
fuselage resembles a ZPG.

There is a decline in accuracy by the DSM-EQWM in the wing--body juncture
and trailing-edge region, which are dominated by adverse pressure
gradient effects in the corner and flow separation at the trailing
edge. \citet{Lozano2022} have shown that not only does the DSM-EQWM exhibit
larger errors in the wing--body juncture and trailing-edge region, but the rate of convergence of the DSM-EQWM by merely refining the grid
is too slow to compensate for the modelling deficiencies at a
reasonable computational cost.

The DSM-BFWM provides better predictions in the juncture location
compared to the DSM-EQWM. The main reason for the improved results comes
from the augmentation of the wall stress compared to the EQWM. This
effect accumulates along the streamwise direction, decelerating the
flow and alleviating the overprediction of $u$ and $w$ observed with
the EQWM. The flow in the juncture is classified as 52\% APG and 41\%
ZPG, which shows that the increase in accuracy is mostly due to the
use of the APG building-block flow. The confidence in the prediction
is above 80\%.

The performance of the DSM-BFWM in the separated region is also
improved compared to the DSM-EQWM, but it is still far from
satisfactory. The DSM-BFWM classifies the flow as a mix of ZPG, APG
and Separation, the latter being the most dominant class (56\%). The
improvements with respect to the EQWM are caused not only by the use of
the APG and Separation building-block flows, but also the improved
streamwise history of the flow discussed above.  Interestingly, the
model prompts a warning about its potential poor performance, which
is evidenced by the low confidence ($\sim$20\%) in the wall-stress
prediction.  The deficient performance of the DSM-BFWM and the DSM-EQWM is
not surprising if we note that the separation zone has wall-normal
thickness 0.3$\delta$ (with $\delta$ the boundary layer
thickness), whereas the WMLES grid size is $\Delta\approx
0.2\delta$. Thus there is only one grid point across the separation
bubble. Despite the fact that the model was trained in separated
flows, the external SGS model errors dominate the LES solution in
this case. These errors hinder the capability of the BFWM to correctly
classify and predict the flow. Nonetheless, the improvements from
BFWM and its ability to assess the confidence in the prediction is a
competitive advantage with respect to traditional wall models.

Simulations of the NASA Juncture Flow experiment were also performed by
\cite{Lozano_brief_2020} using a preliminary version of the BFWM trained
with filtered DNS data and a smaller collection of building-block
flows. The current version of the BFWM provides higher accuracy in the
predictions compared to \cite{Lozano_brief_2020} because of the
enhanced collection of building-block flows and the consistency with
the numerical scheme of the flow solver.
%
%
\begin{table}
  \centering
\begin{tabular}{llrrrl}
                                   &                                                                                                            & Location (a)                                                & Location (b)                                                & Location (c)                                                 &  \\ \hline
\multicolumn{1}{c}{Classification} & \multicolumn{1}{c}{\begin{tabular}[c]{@{}c@{}}Freestream\\ Laminar\\ FPG \\ ZPG \\ APG \\ Separation \\  Unsteady \end{tabular}} & \begin{tabular}[c]{@{}r@{}} 0\% \\  0\% \\ 12\% \\ 88\% \\ 0\% \\ 0\% \\ 0\% \end{tabular} & \begin{tabular}[c]{@{}r@{}} 0\% \\ 0\% \\ 0\% \\ 41\% \\ 52\% \\ 7\% \\ 0\% \end{tabular} & \begin{tabular}[c]{@{}r@{}} 0\% \\ 0\% \\ 0\% \\ 13\% \\ 31\% \\56\% \\ 0\% \end{tabular} &  \\ \hline
\multicolumn{1}{c}{Confidence}     & \multicolumn{1}{c}{}                                                                                       & 98\%                                                        & 82\%                                                        & 22\%                                                         &  \\
&                                                                                                            & \multicolumn{1}{l}{}                                        & \multicolumn{1}{l}{}                                        & \multicolumn{1}{l}{}                                         &  \\ \hline 
\end{tabular} 
\caption{Validation case: NASA Juncture flow. The flow classification
  and confidence in the solution by BFWM.  The locations are (a)
  upstream region of the fuselage, (b) wing-body juncture, and (c)
  wing-body juncture close to the trailing
  edge.\label{table:class_JF}}
\end{table}

\section{Model limitations}\label{sec:limitations}

We close this work by discussing some of the limitations of the BFWM and potential improvements.
\begin{itemize}
  \item First, the identification of meaningful building-block flows
    and the minimum number of blocks required to make accurate
    predictions remains an open question. The present version of the
    model uses seven building blocks, which is far from being
    representative of the rich flow physics that might occur in all
    complex scenarios (e.g. compressibility effects, shock waves,
    chemical reactions, multiphase flows, different mean flow
    three-dimensionalities and separation patterns). Additionally, there is no specific mechanism in the BFWM to
    faithfully capture the laminar-to-turbulent transition, which is
    of paramount importance in many applications.
 \item The confidence score offered by the BLWM is based on the
   similarities between the input flow features and the building-block
   flows used in the training set. This implies that low confidence
   scores may still result in accurate wall stress predictions and
   vice versa. Moreover, the confidence score does not provide an
   error bound on the value of the prediction, which would be more
   representative of the model performance. \corr{In that vein, future
     developments of the model could incorporate uncertainty
     quantification frameworks to estimate error bounds for the
     quantities of interest~\citep[e.g.,][]{Sun2020, Maulik2020,
       Morimoto2022, Lozano2022_info, Reza2022}.}
 \item Consistency between the model and the numerical/gridding
   schemes is solver-dependent. As such, the BFWM must be re-trained to
   yield accurate predictions in different flow solvers. However, this
   inconvenience seems inevitable considering the intimate
   relationship between numerics, gridding and modelling in
   implicitly filtered LES.
 \item A key conclusion of this work is that the main limiting factor
   in the accuracy of the BFWM predictions originates from external
   modelling errors due to the poor performance of SGS models. This
   hints at the necessity for a unified SGS/wall model approach where
   the building-block flow model is also used to devise a numerically
   consistent SGS model (referred to as a building-block flow model,
   BFM). Steps in that direction are already been undertaken by our
   group, and \citet{Ling2022} have shown that the prediction of the
   lift, drag and pitching moment coefficients for the CRM-HL in
   \S\ref{subsec:CRM} are greatly improved in the first version of
   the BFM.
 \item The prediction and classification tasks in the BFWM rely on
   ANNs. In this work, it was found that feedforward ANNs with 5
   hidden layers and roughly 20--30 neurons per layer enabled
   accurate results. However, the neural network architectures
   utilised here may be far from optimal. Moreover, there was no
   attempt to optimise the computational efficiency of the ANNs, for
   example via GPU implementation, and further work should be devoted
   in this direction.
   \item \corr{Variables from past time steps were not included in the
     model input. Adding such information has the potential of
     improving the accuracy of the model predictions, as shown in
     previous studies in the context of turbulence forecasting via
     ANNs~\citep{Lozano2020_cau,Srinivasan2019,Nakamura2021}.}
\end{itemize}

\section{Conclusions}\label{sec:conclusions}

The prediction of aerodynamic quantities of interest remains among the
most pressing challenges for computational fluid dynamics, and it has
been highlighted as a Critical Flow Phenomena in the NASA CFD Vision
2030~\citep{Slotnick2014}. The aircraft aerodynamics are inherently
turbulent with mean flow three-dimensionality, often accompanied by
laminar-to-turbulent transition, flow separation, secondary flow
motions at corners, and shock wave formation, to name a few. However,
the most widespread wall models are built upon the assumption of
canonical flat-plate statistically-in-equilibrium wall turbulence, and
do not faithfully account for the wide variety of flow conditions
described above.  This raises the question of how to devise models
capable of accounting for such a vast and rich collection of flow
physics in a feasible manner.

In this work, we have developed a wall model for large-eddy simulation
by devising the flow as a collection of building blocks whose
information enables the prediction of the wall stress. The concept was
first introduced by \citet{Lozano_brief_2020}, and it is referred to as
the building-block-flow wall model (BFWM).  The model relies on the
assumption that simple flows contain the essential flow physics to
formulate generalisable wall models. Seven types of building-block
units were used to train the model accounting for
zero/favourable/adverse mean pressure gradient wall turbulence,
separation, statistically unsteady turbulence with three-dimensional mean flow, and
laminar flow.  The approach is implemented using two types of
artificial neural networks: a classifier, which identifies the
contribution of each building block in the flow; and a predictor,
which estimates the wall stress via a combinations of building-block
units.  The output of the model is accompanied by a confidence
score. The latter value aids the detection of areas where the model
underperforms, such as flow regions that are not representative of the
building blocks used to train the model.  This a critical component of
the BFWM and is considered a key step for developing reliable models. For
example, the present model will provide a low confidence score in the
presence of a flow that it has never seen before (e.g. a shock wave), or
when the wall model input is located in the freestream. Finally, the
model is designed to guarantee consistency with the numerical scheme
and the gridding strategy. This was attained by training the model
using WMLES data obtained with an optimised SGS model that accounts
for the numerical errors of the solver.

The BFWM has been validated in laminar and turbulent boundary layers,
turbulent pipe flow, turbulent Poiseuille--Couette flows mimicking
favourable/adverse mean pressure gradient effects, and statistically
unsteady turbulent channel flow with 3-D mean flow.  The performance
of the BFWM in complex scenarios was evaluated in two realistic aircraft
configurations: the NASA Common Research Model High-lift and the NASA
Juncture Flow experiment. The validation cases were conducted by
combining the BFWM with two SGS models: the dynamic Smagorinsky model and
the Vreman model. The BFWM outperforms the traditional equilibrium wall model
in canonical cases and the two realistic aircraft scenarios. The only
exceptions are the coincidental accurate predictions by the
equilibrium wall model due to error cancellation. Despite the improved
performance of the BFWM, its accuracy deteriorates considerably owing to
external errors from the SGS model. This suggests that additional
wall model improvements will not suffice to further boost the
performance of WMLES unless the SGS model is also improved.

The modelling capabilities demonstrated by the BFWM, and the ability to
extend the model to a richer collection of flow conditions, makes the
building-block flow approach a realistic contender to overcome the key
deficiencies of current CFD methodologies.  We have argued that truly
revolutionary improvements in WMLES will encompass advancements in
numerics, grid generation and wall/SGS modelling.  Here, we have
focused on wall modelling and its consistency with the numerics and
the grid. However, work remains to be done on the SGS modelling
front. In a recent work by our group, \citet{Ling2022} extended the
building-block flow methodology to the SGS model and showed that the
prediction of the lift, drag and pitching moment coefficients in the
CRM-HL are greatly improved compared to the BFWM combined with traditional
SGS models. There is also a data science component to the problem,
such as the need for efficient and reliable machine learning
techniques for data classification and regression. However, in our
experience, the model performance is mainly controlled by the physical
assumptions rather than by the details of the neural network
architecture at hand.

\section{Acknowledgements}

A.L.-D. acknowledges the support of NASA under grant No. NNX15AU93A
from a preliminary version of this work. We thank Konrad Goc for
providing the grids for the CRM-HL. The authors acknowledge the MIT
SuperCloud and Lincoln Laboratory Supercomputing Center for providing
HPC resources that have contributed to the research results reported
within this paper.

\bibliographystyle{jfm}
\bibliography{references}

\end{document}